\documentclass[preprint,  aps,  superscriptaddress,  nofootinbib,  showkeys,  10pt]{revtex4}
\usepackage{caption}
\usepackage{float}
\usepackage{subfigure}
\usepackage[a4paper, left=1.9cm, right=1.9cm, top=2.5cm, bottom=2.5cm]{geometry}

\usepackage{caption}
\DeclareCaptionStyle{centered_sub}{
    justification = centering,
    singlelinecheck = true
}
\usepackage{hyperref}
\usepackage{physics}
\usepackage{simpler-wick}
\usepackage{annotate-equations}
\usepackage{wrapfig}
\usepackage{slashed}
\usepackage{subcaption}
\usepackage{amsmath,amssymb}
\usepackage{mathtools}
\usepackage{cancel}
\usepackage{graphicx}
\usepackage{xcolor}
\usepackage{nicefrac}
\usepackage{hyperref}
\usepackage{subfigure}
\hypersetup{colorlinks=true, linkcolor=blue, citecolor=blue, filecolor=blue, urlcolor=blue}
\usepackage{makecell}
\usepackage[framemethod=TikZ]{mdframed}
\usetikzlibrary{decorations.markings, arrows.meta}
\mdfdefinestyle{MyFrame}{%
	linecolor=black,
	outerlinewidth=1pt,
	roundcorner=2pt,
	innertopmargin=\baselineskip,
	innerbottommargin=\baselineskip,
	innerrightmargin=0pt,
	innerleftmargin=0pt,
	backgroundcolor=gray!10!white}

\newcommand{\nt}{\notag\\}

\allowdisplaybreaks

\begin{document}
\title{ Cosmological correlators from the Inflation end to CMB sky via reheating\\
}
\author{Chandra Prakash}
\email{chandra.pp@alumni.iitg.ac.in}
\author{Debaprasad Maity}
\email{debu@iitg.ac.in}
\affiliation{Department of Physics, Indian Institute of Technology, Guwahati, 
Assam, India}

\date{\today}

\begin{abstract}
We investigate the imprint of post-inflationary evolution on primordial cosmological correlators by relaxing the standard assumption of instantaneous reheating. We compute the power spectrum and bispectrum for a conformally coupled and non-minimally coupled ($\xi \neq \frac{1}{6},0$) scalar field with cubic self-interaction, across a finite reheating epoch characterized by an effective equation of state $w$, and reheating temperature $T_{\rm reh}$. We find that for a conformally coupled scalar field, the signatures of reheating, such as the modification to oscillatory features, are strictly confined to the subhorizon regime. On superhorizon scales, the correlators receive negligible corrections at leading order. In sharp contrast, for a non-minimally coupled field, we identify a distinct phenomenology where non-trivial values of $\xi$ and reheating equation of state $w$, can induce a tachyonic enhancement of the field modes on superhorizon scales. This enhancement modifies both the power spectrum and bispectrum, marking a sharp deviation from the instantaneous transition limit. Our results demonstrate that to the leading order while conformal coupling shields large-scale correlations from the expansion history, non-minimal coupling breaks this degeneracy and opens an observable window into the physics of reheating.
\end{abstract}

\maketitle

\section{Introduction}
Cosmology is famously an observational rather than an experimental science. No experimentalists were present in the early universe, and no terrestrial experiment can reproduce the enormous energies and extreme conditions of the primordial epoch. Our knowledge of the early Universe is therefore indirect: it is inferred from the statistical properties of late-time cosmological structures. Observations of the cosmic microwave background (CMB) anisotropies and the large-scale distribution of matter encode precise information of correlation functions that act as the fossil record of the primordial universe \cite{brandenberger2004lectures,maldacena2003non,Dalal:2007cu,bartolo2004non,aghanim2020planck,ross2017clustering}. Any theoretical reconstruction of the earliest moments must explain the origin, amplitude, scale dependence, and non-Gaussian structure of these correlators.

One such theoretical framework is Inflation. Inflation offers the leading paradigm for this reconstruction. A brief period of accelerated expansion dynamically resolves the horizon and flatness problems\cite{guth1981inflation,guth2005inflationary, linde1982new, albrecht1982cosmology}, while amplifying quantum fluctuations into macroscopic curvature perturbations that seed the formation of structure \cite{mukhanov1985gravitational,sasaki1986large,mukhanov1988quantum}. In single-field slow-roll inflation, these perturbations evolve into a conserved quantity on superhorizon scales, called curvature perturbation denoted by $\zeta$ \cite{lyth2003conserved,weinberg2003adiabatic}. This conservation law implies a powerful simplification: correlators computed at the end of inflation can be directly propagated to late times without modification via transfer function. Under this assumption, the primordial power spectrum and bispectrum measured today reflect the predictions of the inflationary era with minimal contamination from subsequent dynamics. This assumption caries with itself a very stringent constraint that the correlations become classical in nature and  interaction is turned off at the end of inflation. This has been shown to be qualitatively true in the case of instantaneous reheating but may not be a fundamentally consistent assumption to make. Furthermore, due to inherent quantum mechanical nature of reheating process transporting correlations across reheating should be more involved. 

Furthermore, the robustness of $\zeta$ conservation is far from guaranteed. A wide range of physical effects can violate this conservation and alter cosmological correlators after inflation. In single-field inflation, $\zeta$ generically becomes ill-defined at instants where $\dot\varphi=0$ during reheating, leading to divergences which are problematic for its conservation~\cite{algan2015breakdown}. On the other hand, multi-field inflation generically excites entropy (isocurvature) modes that feed back into the adiabatic curvature perturbation, leading to superhorizon evolution \cite{gordon2000adiabatic,bartolo2004non, chen2010primordial}. Sudden transition, non-attractor phases, or quasi-single-field models can introduce non-adiabatic stresses and generate additional correlation structures beyond those predicted by the minimal scenario \cite{Chen:2009we,Namjoo:2012aa,achucarro2011features}. Even in nominally single-field models, the transition from inflation to the hot Big Bang is neither instantaneous nor trivial.

During inflation, particle production occurs as quantum fields evolve in the rapidly expanding background spacetime. Fluctuations of the inflaton field are amplified and, upon exiting the Hubble radius, become effectively classical. In a quasi–de Sitter regime, the field perturbations behave as independent harmonic oscillators with a time-dependent frequency, and their quantum state can be well approximated by a Gaussian wave functional. This Bogoliubov particle production due to expanding background does not require any direct interaction with other matter fields \cite{zago2019quantum,parker1968particle,kolb2024cosmological}. The nontrivial particle content arises purely from the time dependence of the classical gravitational background, and the probability distribution of field amplitudes remains Gaussian. It is only in the presence of interaction with other fields, or with itself, that the associated probability distribution acquires non-gaussianity \cite{arkani2015cosmological,lee2016non,racco2024gravitational}. The resulting curvature perturbation is typically assumed to remain conserved on superhorizon scales until horizon re-entry in the post-inflationary Universe \cite{lyth2005general,senatore2013constancy,weinberg2003adiabatic}. This implies that the observable spectrum of density fluctuations is insensitive to the detailed dynamics occurring between the end of inflation and horizon entry, including the processes associated with reheating. 

However, the primordial spectrum can acquire additional features that encode information about the spacetime transition between the end of inflation and the onset of the radiation-dominated era. These features arise from the non-adiabatic dynamics of the transition and may leave observable imprints on the curvature perturbation spectrum. Studying the evolution of scalar fields during this period, both free and with simple interaction terms, provides a controlled framework to investigate how these transitions affect particle production and associated various correlation functions in the early Universe. The linear evolution of perturbation in the field does not capture the effects originating from the decay channels taking place during that era.
\begin{figure}[h]
		\centering
		\includegraphics[width=0.9\linewidth]{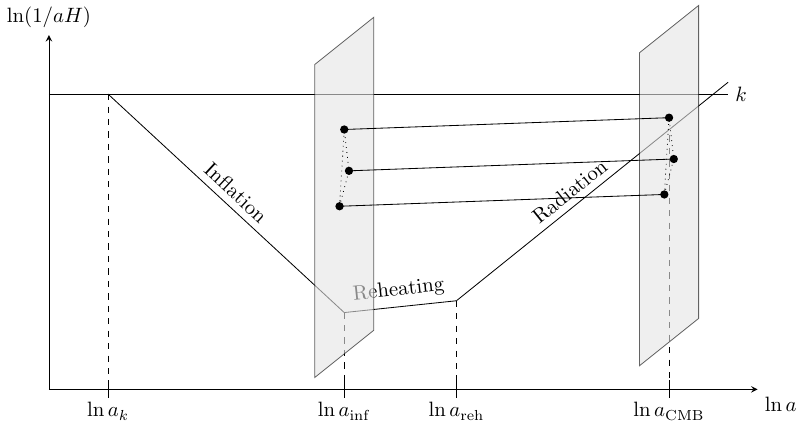}
        \caption{Schematic illustration emphasizing that while cosmological correlators are often evaluated at the end of inflation, observables at later times can differ due to subsequent dynamical evolution.}
        \label{fig:schema}
	\end{figure}
Reheating and preheating involve nonlinear, non-perturbative dynamics driven by parametric resonance, explosive particle production, mode coupling, and backreaction \cite{bassett2006inflation,kofman1997towards, Lozanov:2019lecturesReheating}. These processes are generically out of equilibrium and can modify the curvature perturbation, especially on intermediate or small scales, as demonstrated in analytical studies and lattice simulations \cite{ChambersRajantie:2008latticeNG,Jedamzik:2010collapse,Imrith:2019deltaN_lattice}. Preheating can also generate substantial non-Gaussianity or introduce corrections to frozen correlators \cite{Enqvist:2005NGfromPreheating,Bond:2009chaoticBilliards}. In addition, the quantum-to-classical transition of cosmological perturbations, often assumed to be completed by the end of inflation, can be environment-dependent, and decoherence may leave subtle imprints in correlation functions \cite{KieferPolarski:1998classicality,DaddiHammouBartolo:2023decoherence}. More speculative possibilities arise in the presence of trans-Planckian physics, modified initial states, or quantum-gravity effects, each of which can imprint deviations in the primordial correlators \cite{yin2024cosmological,GozziniVidotto:2021QGfluctuations}.

Given that cosmological correlators represent our primary observational window into physics at energy scales far beyond those accessible to us, it is essential to understand the conditions under which inflationary predictions survive unaltered. Assuming that correlators computed at the end of inflation evolve adiabatically through reheating is a strong assumption. A rigorous assessment requires tracking the quantum mechanical evolution of perturbations, including their correlations, through the entire post-inflationary epoch, encompassing  reheating, and the onset of radiation dominated era. The specific sequence of these stages and timeline is depicted in Fig.~\ref{fig:schema}.

In this work, we revisit the computation of cosmological correlators treating the fields being evolved quantum mechanically through out the entire cosmological periods, and systematically examining the circumstances under which their post-inflationary evolution can generate measurable deviations. Our goal is to determine the extent to which late-time observables can be robustly interpreted as direct signatures of inflation and to identify scenarios in which reheating-era dynamics leave observable imprints that challenge the conventional and naive quantum-to-classical transition pictures  of various correlators at the end of inflation. 

The paper is organized as follows: 
\begin{itemize} 
    \item Section \ref{section 1} provides the foundational setup for a conformally coupled spectator field during a sudden transition to radiation era. 
    \item Section \ref{section 2} explores the dynamical evolution of the field and its three-point statistics with the inclusion of prolonged reheating era. 
    \item Section \ref{section 3} addresses the more general case of non-minimal coupling, developing the Bogoliubov formalism required for multi-phase cosmic histories. 
\end{itemize}
The spectator field correlators computed in this work are of direct 
physical interest as the fundamental ingredients from which observable 
isocurvature statistics are built, as outlined in Sec.~\ref{sec:isocurv}. 
Rather than immediately connecting to observables, which introduces 
additional model-dependent features such as the choice of mass range, 
non-minimal coupling, and reheating temperature, our primary goal is 
to first understand how these fundamental objects themselves are affected 
by the reheating epoch. To this end, we study the propagation of the 
spectator field correlators from inflation through reheating, adopting 
the instantaneous transition approximation as the simplest reheating 
model, and use the reheating temperature $T_{\rm reh}$ to parametrize 
the duration of reheating. 
\section{Physical Setup and Theoretical Framework}\label{section 1}
\subsection{Instantaneous Reheating and Background Evolution}
Following the inflationary epoch, the cosmic background transitions from a quasi-de Sitter phase to a radiation-dominated (RD) era. This transition is mediated by a reheating phase where inflaton decay triggers efficient particle production, thermalizing the Universe into a hot, interacting plasma. For the first part of our work, we consider this process in the limit of instantaneous reheating. We assume that at the conformal time $\eta = \eta_{\rm inf}$, the inflaton $\varphi$ decays instantaneously into radiation. Consequently, the subsequent dynamics are governed by a radiation fluid, with initial conditions for the perturbations established at the transition boundary. Under these assumptions, the background scale factor $a(\eta)$ can be piecewise defined to capture the transition:
\begin{align}\label{sudden-transition_scalefactor}
	a(\eta)=
	\begin{cases}
		-\frac{1}{H_{\inf}\eta},	&\text{when } \eta<\eta_{\text{inf}}\\
		\frac{1}{H_{\inf}\eta^2_{\text{inf}}}(\eta-2\eta_{\text{inf}})=\frac{1}{H_{\inf}\eta^2_{\text{inf}}}(\eta+2|\eta_{\text{inf}}|), &\text{when } \eta>\eta_{\text{inf}}
	\end{cases}
\end{align}
\begin{figure}[t]
    \centering
    \includegraphics[width=0.9\linewidth]{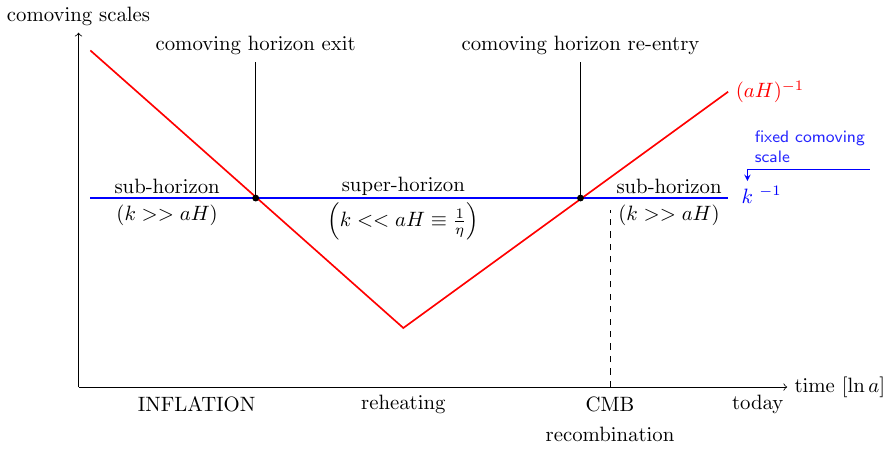}
    \caption{Schematic illustration of the evolution of the comoving Hubble radius across different cosmological eras.}
\end{figure}
This parametrization ensures that both the scale factor $a(\eta)$ and Hubble parameter $H = \nicefrac{\dot{a}}{a}$ are continuous across the transition hypersurface.

\subsection{Field Dynamics}
\noindent
In the first part we restrict our attention to the dynamics of a spectator conformally coupled scalar field whose action is given by:
\begin{equation}
    S = -\int d^4x \sqrt{|g|} \left[ \frac{1}{2} g^{\mu\nu} \partial_\mu \phi \, \partial_\nu \phi+\frac{1}{2}m_{\phi}\phi^2 + \frac{1}{12} \mathcal{R} \phi^2 +\frac{1}{3!}g\phi^3\right].
\end{equation}
We consider the case of $m_{\phi}=0$, where $\mathcal{R}$ denotes the Ricci scalar. The choice of conformal coupling $\xi = 1/6$ ensures invariance under conformal transformations. In this case, the mode functions reduce to plane waves, the vacuum remains unchanged throughout the expansion, and consequently particle production will be prohibited \cite{garani2025particle,kolb2024cosmological}. Throughout, we assume the metric signature to be $(-+++)$ with the metric $ds^2 = a(\eta)^2(-d\eta^2 + dx^2 + dy^2 + dz^2)$, where $\eta$ and $a(\eta)$ are the conformal time and scale factor for Friedmann-Lemaitre-Robertson-Walker (FLRW) metric. With this we now decompose the scalar field as
\begin{equation}
    \phi(\vec x,\eta)=\int \frac{d^3k}{(2\pi)^3}\phi_k e^{i\vec k\cdot\vec x}= \int \frac{d^3k}{(2\pi)^3}(f_k(\eta) a_k+f^*_k(\eta)a_{-k}^{\dagger})e^{i\vec k\cdot\vec x}.
\end{equation}
Since the spatial coordinates $x$ are comoving, the Fourier conjugate $k$ represents the comoving momentum, which are evaluated relative to the present-day scale factor $a_0$ (we will keep this implicit unless stated otherwise). The Equation of Motion (EoM) for the mode function in the free field limit is given by:
\begin{align}
    \pdv[2]{(a f_k)}{\eta}+\left(k^2-\frac{\partial_{\eta}^2a}{a}+\frac{a^2\mathcal{R}}{6}\right)(a f_k) &=0.
\end{align}
To solve the equations during de Sitter inflation and the subsequent radiation-dominated era, we require the following expressions:
\begin{align}
    \frac{\partial_\eta^2 a}{a} =
    \begin{cases}
        \dfrac{2}{\eta^2}, & \text{for } \eta < \eta_{\rm inf}, \\[2mm]
        0, & \text{for } \eta > \eta_{\rm inf},
    \end{cases}~~~~~~~~;~~~~~~~~~
    \mathcal{R} =
    \begin{cases}
        12 H^2_{\inf}, & \text{for } \eta < \eta_{\rm inf}, \\[1mm]
        0, & \text{for } \eta > \eta_{\rm inf}.
    \end{cases}
\end{align}
Considering the well known Bunch-Davis vacuum condition, during inflation, the mode equation reads:

\begin{align}
    \left( \frac{\partial^2}{\partial \eta^2} + k^2 - \frac{2}{\eta^2} + \frac{2 H^2_{\inf}}{H^2_{\inf}\eta^2} \right) a f_k = 0
    \quad \implies \quad
 f_k(\eta) = - \frac{H_{\inf} \eta}{\sqrt{2k}} e^{-i k \eta}.
\end{align}
During the radiation-dominated era on the other hand, the conformally coupled scalar field mode satisfies:

\begin{align}
    \left( \frac{\partial^2}{\partial \eta^2} + k^2 \right) a f_k = 0 
    \quad \implies \quad 
    f_k(\eta) = c_1 \frac{H_{\inf} \eta_{\rm inf}^2}{\sqrt{2k}\, (\eta - 2 \eta_{\rm inf})} e^{-i k \eta} 
    + c_2 \frac{H_{\inf} \eta_{\rm inf}^2}{\sqrt{2k}\, (\eta - 2 \eta_{\rm inf})} e^{i k \eta}.
\end{align}
For a conformally coupled field, the mode function $f_k(\eta)$ possesses a distinct physical structure: it is scaled by the inverse of the scale factor,
\begin{equation}
    f_k(\eta) =\frac{X_k(\eta)}{a(\eta)},
\end{equation}
where $X_k(\eta)$ represents a normalized plane wave. This $1/a(\eta)$ factor accounts for the geometric dilution of the field amplitude as the Universe expands, ensuring that the energy density scales correctly with the increasing volume.

To ensure the continuity of field and its conjugate momenta across the transition, we require the mode functions to be matched at the boundary $\eta = \eta_{\mathrm{inf}}$. We impose the junction conditions:
\begin{equation}
    f_{k}(\eta_{\rm inf}^{-}) = f_k(\eta_{\rm inf}^{+}), \quad f'_{k}(\eta_{\rm inf}^{-}) = f'_k(\eta_{\rm inf}^{+}).
\end{equation}
Applying these conditions to our specific background yields the coefficients $c_1 = 1$ and $c_2 = 0$. Since $a(\eta)$ and $a(\eta)'$ is already continuous, the mode function across both eras takes the form:
\begin{align}
	f_k(\eta)=\frac{1}{a(\eta)}\frac{e^{-ik\eta}}{\sqrt{2k}} =
	\begin{cases}
		 -\frac{H_{\inf}\eta}{\sqrt{2k}}e^{-ik\eta}	\qquad&\text{for }\eta<\eta_{\text{inf}}	\\\\
		\frac{H_{\inf}\eta_{\text{inf}}^2}{\sqrt{2k}(\eta-2\eta_{\text{inf}})}e^{-ik\eta}   \qquad&\text{for }\eta>\eta_{\text{inf}} .
	\end{cases}
\end{align}
The mode functions described above allow for a straightforward definition of the Wightman propagator for the conformally coupled scalar field. In our diagrammatic representation, this propagator is represented by a single line connecting two spacetime points. By convention, the first argument of the propagator corresponds to the left vertex, while the second argument corresponds to the right vertex within the Feynman diagram:

\begin{align}
    W_{k}(\eta,\eta')&=~~\begin{gathered}
 \begin{tikzpicture}
\fill (0,0) circle (2pt);
  \fill (1,0) circle (2pt);
 \draw (0,0)--(1,0);
 \node at (0,0.2) {$\eta$};
 \node at (1,0.2) {$\eta'$};
  \end{tikzpicture}
 \end{gathered}~~
	=f_{k}(\eta)f_{k}^*(\eta')\nt
    &=\left[\theta(\eta_{\rm inf}-\eta)\frac{H_{\inf}|\eta|e^{-ik\eta}}{\sqrt{2k}}+\theta(\eta-\eta_{\rm inf})\frac{H_{\inf}\eta_{\text{inf}}^2e^{-ik\eta}}{\sqrt{2k}(\eta-2\eta_{\text{inf}})}  \right]\nt
    &\qquad\qquad\times
    \left[\theta(\eta_{\rm inf}-\eta')\frac{H_{\inf}|\eta'|e^{ik\eta'}}{\sqrt{2k}}+\theta(\eta'-\eta_{\rm inf})\frac{H_{\inf}\eta_{\text{inf}}^2e^{ik\eta'}}{\sqrt{2k}(\eta'-2\eta_{\text{inf}})} \right],\nt
    &=\left[\theta(\eta_{\rm inf}-\eta)\frac{1}{a^{\rm inf}(\eta)}+\theta(\eta-\eta_{\rm inf})\frac{1}{a^{\rm rad}(\eta)}\right]\left[\theta(\eta_{\rm inf}-\eta')\frac{1}{a^{\rm inf}(\eta')}+\theta(\eta'-\eta_{\rm inf})\frac{1}{a^{\rm rad}(\eta')}\right]\frac{e^{-ik(\eta-\eta')}}{2k},\nt
	&=\frac{1}{2k}\frac{1}{a(\eta)a(\eta')}e^{-ik(\eta-\eta')}=\frac{1}{a(\eta)a(\eta')}W^{(\text{flat})}_{k}(\eta,\eta')  .
\end{align}
At tree-level, the diagrams consist entirely of external legs, meaning the Feynman propagator does not enter the calculation directly.
Where $W_{k}^{(\text{flat})}$ denotes the propagator in Minkowski spacetime.
Using these definitions, the leading order two-point correlation function of the field $\phi$ in Fourier space is given by:
\begin{align}
\bra{\Omega}\phi(\vec{x},\eta)\phi(\vec{x}',\eta)\ket{\Omega}=\bra{0}\phi(\vec{x},\eta)\phi(\vec{x}',\eta)\ket{0}
    &=\int \frac{d^3k}{(2\pi)^3}\underbrace{f_{k}f_{k}^*}_{\mathcal{P}_k}e^{i\vec k\cdot(\vec x-\vec x')}.
\end{align}
where $\ket{\Omega}$ is the interacting theory vacuum and $\ket{0}$ is the free theory vacuum. From this expression, we extract the dimensionless power spectrum, $\Delta_{\phi}(k, \eta)$, which characterizes the variance of the field fluctuations per logarithmic $k$-interval:
\begin{align}
    \Delta_\phi & =
    \frac{k^3}{2\pi^2}W_k(\eta,\eta) =\frac{k^3}{2\pi^2}\mathcal{P}_k= \begin{cases}
       \dfrac{H_{\inf}^2(k\eta)^2}{4\pi^2}, & \eta<\eta_{\text{inf}}<0 \\[8pt]
       \dfrac{H_{\inf}^2(k/k_{\text{inf}})^2}{4\pi^2}
       \left(\dfrac{k/k_{\text{inf}}}{k\eta+2k/k_{\text{inf}}}\right)^2,
       & \eta>\eta_{\text{inf}}
\end{cases}\label{dimless: conform}
\end{align}
where we have identified the transition scale $k_{\text{inf}} = a(\eta_{\text{inf}}) H_{\text{inf}} = -1/\eta_{\text{inf}}$.

\begin{figure}
\centering
\includegraphics[width=0.8\linewidth]{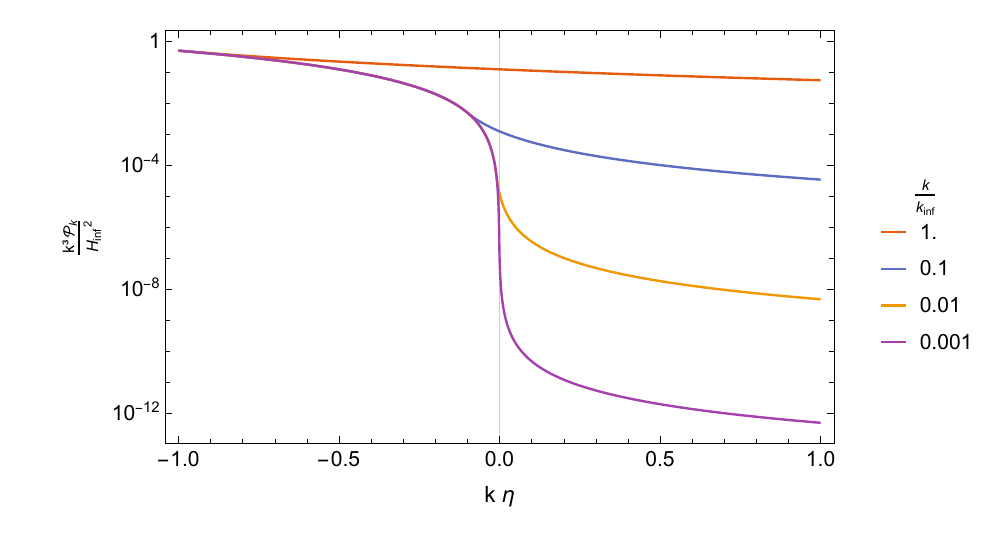}
\caption{Evolution of the dimensionless power spectrum $k^3\mathcal{P}_k$ across the inflationary and radiation-dominated epochs. During inflation ($\eta < \eta_{\text{inf}}<0$), all modes follow a universal power-law decay, highlighting the absence of scale invariance for a conformally coupled scalar. In the subsequent radiation-dominated era ($\eta > \eta_{\text{inf}}$), the spectral evolution becomes mode-dependent, where the degree of suppression is dictated by the horizon-exit scale $k/k_{\text{inf}}$.}
\label{fig:power_spectrum_conf}
\end{figure}
The analytical results obtained in Eq. \eqref{dimless: conform} highlight a crucial departure from the standard inflationary paradigm. The time-evolution of the total energy density for a conformally coupled scalar is directly mapped onto its power spectrum. The conformal coupling basically gives the field a large effective mass $(m_{\rm eff}^2=m^2+\xi\mathcal{R})$ and as a consequence the fluctuations do not ``freeze" at the horizon. Instead, they continue to evolve and decay. This decaying behavior is tied to the background equation of state and the evolution of the scale factor $a(\eta)$. This is demonstrated in Fig.~\ref{fig:power_spectrum_conf}, where the power spectrum exhibits a sustained decay across both regimes. 
\paragraph{Inflationary Era ($\eta < \eta_{\text{inf}} $):} 
The power spectrum exhibits a characteristic power-law decay $\Delta \propto (k\eta)^2$. This departure from scale invariance is a direct consequence of the conformal coupling. Furthermore, the physical field $\phi$ is suppressed by the scale factor $a^{-1}$, resulting in a strongly ``blue'' spectrum where power is predominantly shifted toward small scales (high $k$).
\paragraph{Radiation-Dominated Era ($\eta > \eta_{\text{inf}}$):} 
Following the inflationary transition, the modes undergo secondary evolution governed by the expansion rate and the vanishing of the Ricci scalar ($R \to 0$). The suppression of the power spectrum persists as modes re-enter the horizon, with the specific decay rate determined by the mode's exit time relative to $\eta_{\text{inf}}$. An interesting behavior is observed for modes deep inside the horizon $\frac{k}{k_{\rm inf}}\to\infty$. Rather than freezing, these modes track the instantaneous vacuum solution, with the power spectrum scaling as:
\begin{equation}
    \Delta_\phi\approx\frac{H_{\rm inf}^2(k/k_{\rm inf})^2}{16\pi^2}.
\end{equation}
This $k^2$ scaling is characteristic of the vacuum fluctuations of a conformally coupled field.

\subsection{Bispectrum Dynamics}
\begin{figure}[t]
    \centering
    \includegraphics[width=0.8\linewidth]{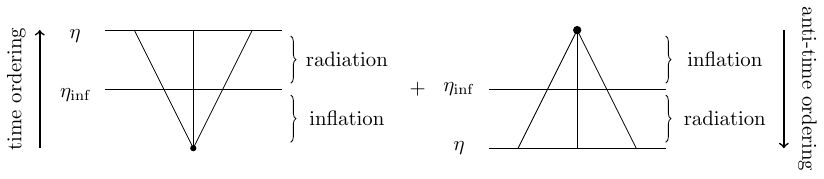}
    \caption{Diagrammatic representation of $\overline{\ev{\phi_{k_1}\phi_{k_2}\phi_{k_3}}}$ during instantaneous reheating. Unlike standard treatments, the correlator is evaluated at a time $\eta>\eta_{\rm inf}$, incorporating the matching conditions at the transition.}
    \label{fig:feynman_diag}
\end{figure}

The first signature of non-Gaussianity, whether arising from interaction-induced particle production or from particle interactions themselves, is encoded in the bispectrum, i.e., the three-point correlator. The form of the three-point function is highly constrained by the conformal invariance properties of the de Sitter boundary \cite{arkani2020cosmological,baumann2024snowmass} defined at $\eta \rightarrow 0$. However, such constraints do not capture the physics that unfolds after the end of inflation. Once inflation ends, the background spacetime undergoes non-trivial evolution such as reheating, 
and hence, to properly study the evolution of non-Gaussianity after the transition, one must adopt a bulk perspective. Before incorporating the reheating phase into our analysis, we first examine the bispectrum for the simpler case of an instantaneous transition to radiation dominance. The relevant interaction term is given by:
$$\mathcal{L}_{\rm int}=-\frac{1}{3!}g\phi^3,$$
with the coupling $g$. We can find the associated hamiltonian:
\begin{align}
	H_{\text{int}}(\eta) = -\int d^3x \, a^3(\eta) \mathcal{L}_{\text{int}}
	= \frac{g}{3!} a^3(\eta) \int d^3x \int \frac{d^3p_1}{(2\pi)^3} \frac{d^3p_2}{(2\pi)^3} \frac{d^3p_3}{(2\pi)^3} \phi_{p_1}(\eta) \phi_{p_2}(\eta) \phi_{p_3}(\eta) e^{-i(p_1 + p_2 + p_3) \cdot x}.
\end{align}
Using this Hamiltonian, the bispectrum can be computed within the in-in formalism as \cite{Chen:2017ryl,werthlecture,pajer2024field}:
\begin{equation}
    \bra{\Omega} \phi_{k_1}(\eta) \phi_{k_2}(\eta) \phi_{k_3}(\eta) \ket{\Omega}= 2\,\text{Im} \left( \int_{-\infty(1-i\epsilon)}^\eta d\eta' a(\eta')\langle 0|\phi_{k_1}(\eta) \phi_{k_2}(\eta) \phi_{k_3}(\eta) H_{\text{int}}(\eta') | 0 \rangle \right) .
\end{equation}
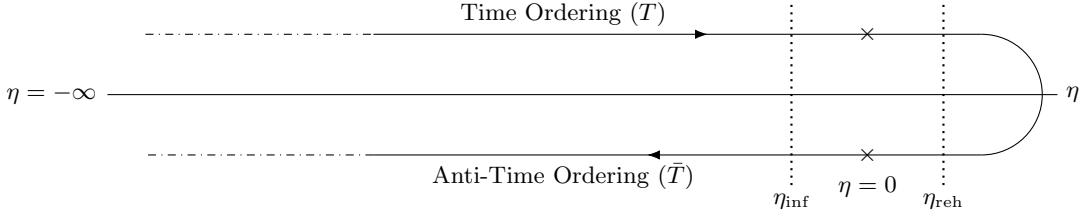
\begin{figure}[t]
    \centering
	\begin{tikzpicture}[>=Latex]
		\def\vsep{0.8} 
		\def\turnaround{4}
		\def\etainf{1.5}
		\def\xmark{2.5}
		\def\etareh{3.5}
        \def\tt{-1.5}
		
		\def\startsolid{-4}
		\def\startinf{-7}
		
		\draw[-] (\startinf - 0.5, 0) -- (\turnaround + 1, 0) node[right] {$\eta$};
		
		\draw[dash dot] (\startinf, \vsep) -- (\startsolid, \vsep);
		\draw[postaction={decorate, decoration={markings, mark=at position 0.55 with {\arrow{>}}}}] 
		(\startsolid, \vsep) -- (\turnaround, \vsep);
        
		\node[above] at (\tt, \vsep) {\small Time Ordering ($T$)};
		
		\draw (\turnaround, \vsep) arc (90:-90:\vsep);
		
		\draw[postaction={decorate, decoration={markings, mark=at position 0.55 with {\arrow{>}}}}] 
		(\turnaround, -\vsep) -- (\startsolid, -\vsep);
		\draw[dash dot] (\startsolid, -\vsep) -- (\startinf, -\vsep);
		
		\node[below] at (\tt, -\vsep) {\small Anti-Time Ordering ($\bar{T}$)};
		
		\node[font=\bfseries] at (\xmark, \vsep) {$\times$};
		\node[font=\bfseries] at (\xmark, -\vsep) {$\times$};
		
		\node[left=15pt] at (\startinf, 0) {$\eta = -\infty$};
		
		\node[below=5pt] at (\xmark, -\vsep) {$\eta = 0$};
		
		\node[below] at (\etareh, -1.5*\vsep) {$\eta_{\rm reh}$};
		\draw[dotted, thick] (\etareh, -1.5*\vsep) -- (\etareh, 1.5*\vsep);
        
		\draw[dotted, thick] (\etainf, -1.5*\vsep) -- (\etainf, 1.5*\vsep);
		\node[below] at (\etainf, -1.5*\vsep) {$\eta_{\text{inf}}$};
	\end{tikzpicture}
    \caption{Graphical illustration of the in-in contour extended beyond $\eta=\eta_{\text{inf}}$ into some instant $\eta$ during the radiation era. The contour has small imaginary parts at early times which relates the interacting vacuum to the free vacuum.}
    \label{fig:in-in-contour}
\end{figure}
The $i\epsilon$ prescription regularizes the in-in integral at early times as well as relate the interacting vacuum ($\ket{\Omega}$) to free Bunch Davies vacuum ($\ket{0}$) and it will be kept implicit throughout the paper.\cite{Christeas:2022ewg} Since the zeroth order term for bispectra $\bra{0} \phi_{k_1}(\eta) \phi_{k_2}(\eta) \phi_{k_3}(\eta) \ket{0}$ vanishes, the leading order contribution comes from interaction channel. The next step is therefore to evaluate the expectation value above and extract the resulting non-Gaussian contribution.

\begin{align}
    &\langle \phi_{k_1}(\eta) \phi_{k_2}(\eta) \phi_{k_3}(\eta) \rangle\equiv\bra{\Omega} \phi_{k_1}(\eta) \phi_{k_2}(\eta) \phi_{k_3}(\eta) \ket{\Omega}\nt
    &\qquad=\frac{2g}{3!}\int \prod_{j=1}^3\frac{d^3p_j}{(2\pi)^3} \delta^3(\vec p_1 + \vec p_2 + \vec p_3)\Im\bigg\{\int_{-\infty}^\eta d\eta'a^4(\eta')\langle 0 | \phi_{k_1}(\eta) \phi_{k_2}(\eta) \phi_{k_3}(\eta) \underbrace{\phi_{p_1}(\eta') \phi_{p_2}(\eta') \phi_{p_3}(\eta')}_{\mathclap{\text{from }H_{\text{int}}}} | 0 \rangle\bigg\},\nt
    &\qquad=\frac{2g}{3!}\int \ \prod_{j=1}^3\frac{d^3p_j}{(2\pi)^3}  \delta^3(\vec p_1 + \vec p_2 + \vec p_3)\Im\bigg\{\int_{-\infty}^\eta d\eta'a^4(\eta')\bra{0}\wick{\c1\phi_{k_1}(\eta)\c2 \phi_{k_2}(\eta) \c3\phi_{k_3}(\eta) \c3\phi_{p_1}(\eta') \c2\phi_{p_2}(\eta') \c1\phi_{p_3}(\eta')}\ket{0}\bigg\}\nt
    &\qquad\qquad+\text{all possible contraction},\nt
    &\qquad=2g\Im{\int_{-\infty}^{\eta}d\eta'a(\eta')^4 W_{k_1}(\eta, \eta')W_{k_2}(\eta,\eta')W_{k_3}(\eta,\eta')}\delta^3(\vec k_1 + \vec k_2 + \vec k_3)\label{ev} .
\end{align}
Each contraction gives us:
\begin{align}
    \wick{\c\phi_k(\eta)\c\phi_{p}(\eta')} = W(\eta,\eta')\delta^3(\vec k+\vec p) =W_k^*(\eta',\eta)\delta^3(\vec k+\vec p).
\end{align}
Then the $d^3p$ integrals convert the $\delta^3(\vec p_1 + \vec p_2 + \vec p_3)$ into $\delta^3(\vec k_1 + \vec k_2 + \vec k_3)$. Since the three-point function always enforces momentum conservation, we define a barred correlator to factor out the delta function. The bulk evolution factorizes into a kinematic component and a dynamical component. The external mode functions $f_{k}(\eta)$ describe the free propagation of the fields. The integral, however, acts as a cumulative weighting function. It sums the evolution history, effectively encoding how the vacuum correlations evolve with the mode functions.
\begin{align}
    \overline{\ev{\phi_{k_1}\phi_{k_2}\phi_{k_3}}}(\eta)
    &=2g\Im\big\{\overbrace{f_{k_1}(\eta)f_{k_2}(\eta)f_{k_3}(\eta)}^{\mathclap{\text{oscillatory boundary term}}}\underbrace{\int_{-\infty}^{\eta}D\eta' f^*_{k_1}(\eta')f^*_{k_2}(\eta')f^*_{k_3}(\eta')}_{\mathclap{\text{cumulative interaction weight}}}\big\},
\end{align}
where we defined $D\eta=d\eta a(\eta)^4$. We now extend this formalism to include the transition surface. The three-point function can naturally be decomposed into two parts, based on the epoch over which the mode functions are integrated, and each contribution can be evaluated separately. 
\begin{align}
\overline{\ev{\phi_{k_1}\phi_{k_2}\phi_{k_3}}}(\eta\ge\eta_{\rm inf})
    &=2g\Im{\int_{-\infty}^{\eta}D\eta' W_{k_1}(\eta, \eta')W_{k_2}(\eta,\eta')W_{k_3}(\eta,\eta')}\nt
    &=2g\Im{\int_{-\infty}^{\eta_{\text{inf}}}D\eta' W_{k_1}(\eta, \eta')W_{k_2}(\eta,\eta')W_{k_3}(\eta,\eta')}\nt
    &\qquad+2g\Im{\int_{\eta_{\text{inf}}}^{\eta}D\eta' W_{k_1}(\eta, \eta')W_{k_2}(\eta,\eta')W_{k_3}(\eta,\eta')}\nt
    &=\overline{\ev{\phi_{k_1}\phi_{k_2}\phi_{k_3}}}_{\text{inf}}(\eta\geq \eta_{\rm inf})+\overline{\ev{\phi_{k_1}\phi_{k_2}\phi_{k_3}}}_{\text{rad}}(\eta\ge\eta_{\rm inf}).
\end{align}
We can expand the imaginary part to make the connection to the in-in contour explicit:
\begin{align}
    \overline{\ev{\phi_{k_1}\phi_{k_2}\phi_{k_3}}}(\eta)&=2g\Im{\int_{-\infty}^{\eta}D\eta' W_{k_1}(\eta, \eta')W_{k_2}(\eta,\eta')W_{k_3}(\eta,\eta')}\nt
    &=-ig\int_{-\infty}^{\eta}D\eta' W_{k_1}(\eta, \eta')W_{k_2}(\eta,\eta')W_{k_3}(\eta,\eta')+ig\int_{-\infty}^{\eta}D\eta' W_{k_1}^*(\eta, \eta')W^*_{k_2}(\eta,\eta')W^*_{k_3}(\eta,\eta')\nt
    &=-ig\int_{-\infty}^{\eta}D\eta' W_{k_1}(\eta, \eta')W_{k_2}(\eta,\eta')W_{k_3}(\eta,\eta')+ig\int_{-\infty}^{\eta}D\eta' W_{k_1}(\eta', \eta)W_{k_2}(\eta',\eta)W_{k_3}(\eta',\eta).\label{bispectra-feyn}
\end{align}
The first term corresponds to the forward time-evolution (time-ordered), and the second term corresponds to the backward evolution (anti-time-ordered). The splitting of the integration limit at $\eta_{\text{inf}}$ corresponds to segmenting these contours, as illustrated in Fig.~\ref{fig:in-in-contour}. The Feynman diagrammatic representation of \eqref{bispectra-feyn} is given in Fig.~\ref{fig:feynman_diag}.
Let us focus on the inflation part first:
\begin{align} \label{infpart}
&\overline{\ev{\phi_{k_1}\phi_{k_2}\phi_{k_3}}}_{\text{inf}}(\eta\ge\eta_{\rm inf})\nt
    &\qquad=2g\Im{\int_{-\infty}^{\eta_{\text{inf}}}D\eta' W_{k_1}(\eta, \eta')W_{k_2}(\eta,\eta')W_{k_3}(\eta,\eta')}\nt
    &\qquad=2g\Im{\int_{-\infty}^{\eta_{\text{inf}}}d\eta' a(\eta')^4\frac{1}{a(\eta)^3a(\eta')^3}\frac{e^{-i(k_1+k_2+k_3)(\eta-\eta')}}{8k_1k_2k_3}}\nt
&\qquad=2g\frac{H^2_{\text{inf}}\eta_{\text{inf}}^6}{8k_1k_2k_3(2\eta_{\text{inf}}-\eta)^3}\Im{e^{-i(k_1+k_2+k_3)\eta}(\gamma+\frac{i\pi}{2}+\ln{[-(k_1+k_2+k_3)\eta_{\text{inf}}]})+\order{\eta_{\text{inf}}}}\nt
    &\qquad=\frac{\pi}{8}g\frac{H^2_{\text{inf}}\eta_{\text{inf}}^6}{k_1k_2k_3(2\eta_{\text{inf}}-\eta)^3}\cos{K\eta}-g\frac{H^2_{\text{inf}}\eta_{\text{inf}}^6}{4k_1k_2k_3(2\eta_{\text{inf}}-\eta)^3}(\ln{\abs{K\eta_{\inf}}}+\gamma)\sin{K\eta},
\end{align}
where we defined $K=k_1+k_2+k_3$ and used the following integral in the limit $\eta_{\text{inf}} \to 0$ at leading order:
\begin{equation}
    \int_{-\infty(1-i\epsilon)}^{\eta_{\text{inf}}}\frac{e^{iK\eta}}{\eta}d\eta =\int_{-\infty}^{iK\eta}\frac{e^t}{t}dt-\int_{-\infty}^{-\epsilon\infty-I\infty}\frac{e^t}{t}dt\approx\,\gamma+\frac{i\pi}{2}+\log(|K\eta_{\text{inf}}|).
\end{equation}
Note the second line in the equation (\ref{infpart}), inflationary contribution to the correlation function at present time involves mixing of mode functions of inflation as well as post-inflation phase through the Feynmann propagator $W_k(\eta,\eta')$. Nevertheless, let us stress the fact that for all practical purpose $\eta_{\rm inf}$ is very small and taking the limit $\eta_{\text{inf}} \to 0$ is justified for modes relevant to CMB scale.  Under this consideration, the radiation part of the bispectrum is found to be:
\begin{align}
    \overline{\ev{\phi_{k_1}\phi_{k_2}\phi_{k_3}}}_{\text{rad}}(\eta\ge\eta_{\rm inf})
    &=2g\Im{\frac{e^{-iK\eta}}{8a(\eta)^3k_1k_2k_3}\int_{\eta_{\text{inf}}}^{\eta}d\eta' a(\eta')e^{iK\eta'}}\nt
    &=2g\Im{\frac{e^{-iK\eta}}{8H_{\rm inf}a(\eta)^3k_1k_2k_3}\frac{-i K \eta _{\text{inf}}+e^{i K \left(\eta -\eta _{\text{inf}}\right)} \left(2 i K \eta _{\text{inf}}-i \eta  K+1\right)-1}{K^2\eta^2_{\text{inf}}}}.\nt
    \intertext{Taking the limit $\eta_{\text{inf}}\to0 ,$} 
     \overline{\ev{\phi_{k_1}\phi_{k_2}\phi_{k_3}}}_{\text{rad}}(\eta\ge\eta_{\rm inf})&=g\frac{H^2_{\text{inf}}\eta_{\text{inf}}^4}{4k_1k_2k_3K^2(\eta-2\eta_{\text{inf}})^3}\Im{(1-iK\eta)-e^{-iK\eta}}\nt
    &=g\frac{H^2_{\text{inf}}\eta_{\text{inf}}^4}{4k_1k_2k_3K^2(\eta-2\eta_{\text{inf}})^3}(-K\eta+\sin{K\eta}),
\end{align}
Collecting all the terms, we can express bispectrum at any time $\eta$ after the inflation as:
\begin{align}
    \overline{\ev{\phi_{k_1}\phi_{k_2}\phi_{k_3}}}(\eta>\eta_{\text{inf}})
    &=-\frac{1}{H_{\inf}a(\eta)^3}\frac{g}{k_1k_2k_3}\left(\frac{\pi}{8}\cos{K\eta}-\frac{(\ln{\abs{K\eta_{\rm inf}}+\gamma)}\sin{K\eta}}{4}+\frac{(K\eta-\sin{K\eta})}{4K^2\eta_{\inf}^2}\right)\label{cubic-bispectra}
\end{align}
To elucidate the physical implications of this result, we analyze the correlator's behavior in specific momentum configurations. We first note that on the transition hypersurface $\eta=\eta_{\rm inf}$, the momentum dependence of the growing mode matches the conformal bootstrap prediction for a scalar field of scaling dimension one~\cite{pajer2017conformal}. 
In this sense, the dominant mode at the transition is effectively conformal, taking the form:
\begin{align}
    \overline{\ev{\phi_{k_1}\phi_{k_2}\phi_{k_3}}}(\eta_{\text{inf}}) &=\frac{\pi}{8}g\frac{H^2_{\text{inf}}\eta_{\text{inf}}^3}{k_1k_2k_3}=-\frac{1}{H_{\text{inf}}a(\eta_{\text{inf}})^3}\frac{g}{k_1k_2k_3}\frac{\pi}{8}.
\end{align}
Given that the mode $k_{\rm rad}$ enters the horizon at some instant $\eta$ during radiation era ($k_{\rm rad}\eta=1-2k_{\rm rad}/k_{\rm inf}$), we can parametrizeq the bispectra as:
\begin{align}
    &\overline{\ev{\phi_{k_1}\phi_{k_2}\phi_{k_3}}}\nt
    &=\frac{-1}{H_{\inf}a(\eta)^3}\frac{g}{k_1k_2k_3}\left[\frac{\pi}{8}\cos{\left(\frac{K}{k_{\rm rad}}-\frac{2K}{k_{\rm inf}}\right)}-\frac{(\ln{\abs{K/k_{\rm inf}}+\gamma)}\sin{\left(\frac{K}{k_{\rm rad}}-\frac{2K}{k_{\rm inf}}\right)}}{4}+\frac{\frac{K}{k_{\rm rad}}-\frac{2K}{k_{\rm inf}}-\sin{\left(\frac{K}{k_{\rm rad}}-\frac{2K}{k_{\rm inf}}\right)}}{4K^2/k_{\inf}^2}\right]\label{cubic-bispectra}
\end{align}
\begin{figure}[t]
    \centering
    \begin{minipage}[t]{0.48\textwidth}
        \centering
        \includegraphics[width=\linewidth]{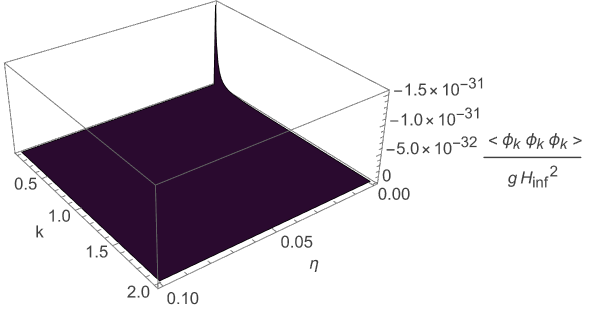}
        \captionsetup{style=centered_sub}\caption*{(a) Spatiotemporal evolution of the bispectrum}
    \end{minipage}
    \hfill
    \begin{minipage}[t]{0.48\textwidth}
        \centering
        \includegraphics[width=\linewidth]{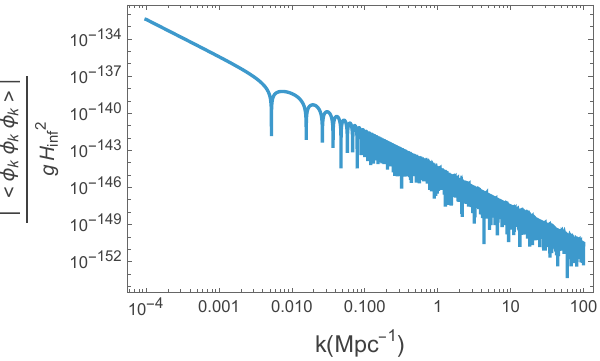}
        \captionsetup{style=centered_sub}\caption*{(b) Bispectra on $\eta=\eta_{\text{CMB}}$ hypersurface}
    \end{minipage}
    \caption{The bispectrum of the conformally coupled scalar field in the equilateral limit. Panel (a) illustrates the dependence on conformal time $\eta$ and wavenumber $k$ in the long-wavelength regime. The plot depicts that IR divergence mainly persists near the instant of transition. Panel (b) extends the analysis to smaller scales at the time of decoupling ($\eta_{\text{CMB}}$), showing the amplitude decay.}
    \label{fig:bispectrum_sidebyside}
\end{figure}
\paragraph{\bf The Equilateral limit:}
In the equilateral configuration ($k_1=k_2=k_3=k$), the equality of the momentum scale simplifies the expression. The bispectrum evolves as:
\begin{align}\label{eq:26}
\overline{\ev{\phi_{k}\phi_{k}\phi_{k}}}(\eta>\eta_{\text{inf}})&=g\frac{H^2_{\text{inf}}\eta_{\text{inf}}^6}{k^3(2\eta_{\text{inf}}-\eta)^3}\left(\frac{\pi}{8}\cos{3k\eta}-\frac{(\ln{\abs{3k\eta_{\rm inf}}}+\gamma)\sin{3k\eta}}{4}+\frac{3k\eta-\sin{3k\eta}}{36k^2\eta_{\inf}^2}\right).
\end{align}
The dominant contribution at late times ($\eta \to \infty$) comes from the modes satisfying $\frac{k}{k_{\rm inf}}\to 0$:
\begin{align}\label{eq:bispectrum_dimensionless_eq}
    k^3 a(\eta)^3 \overline{\langle\phi_{k}^3\rangle} &\approx g \frac{3k\eta-\sin(3k\eta)}{36(k/k_{\text{inf}})^2} .
\end{align}
Physically, this occurs because modes with smaller $k\eta_{\mathrm{inf}}$ exit the horizon much earlier during inflation. Thus, these extremely superhorizon modes are more correlated and as a result sensitive to the sharpness of the transition at $\eta = \eta_{\mathrm{inf}}$, resulting in the amplification of comoving dimensionless bispectra ($k^3 a(\eta)^3 \overline{\langle\phi_{k}^3\rangle}$) as seen in the Fig.~\ref{fig:bispectrum_sidebyside} and \ref{fig:radiation-bispectra}.

\paragraph{\bf The Squeezed limit:}
A distinct behavior emerges in the squeezed limit ($k_3 \ll k_1 \approx k_2 \approx k$), which probes the coupling between soft and hard modes. The bispectrum takes the form:
\begin{align}
    \overline{\ev{\phi_{k}\phi_{k}\phi_{k}}}(\eta>\eta_{\text{inf}})&=
    g\frac{H^2_{\text{inf}}\eta_{\text{inf}}^6}{k^2k_3(2\eta_{\text{inf}}-\eta)^3}\left(\frac{\pi}{8}\cos{2k\eta}-\frac{(\ln{\abs{2k\eta_{\rm inf}}}+\gamma)\sin{2k\eta}}{4}+\frac{2k\eta-\sin{2k\eta}}{16k^2\eta_{\inf}^2}\right).
\end{align}
The critical feature here is the $1/k_3$ pole in the pre-factor. When considering the dimensionless bispectrum (scaled by $\sim k^3$), this translates to an enhancement proportional to the ratio of scales $k/k_3$. This divergence implies that the small-scale modes are strongly modulated by the long-wavelength mode. 

\begin{figure}[t]
    \centering
    \begin{minipage}[t]{0.48\textwidth}
        \centering
        \includegraphics[width=\linewidth]{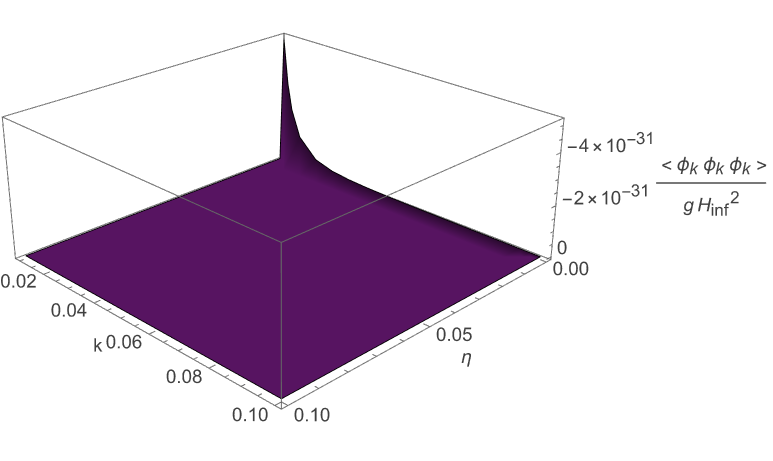}
        \captionsetup{style=centered_sub}\caption*{(a) Spatiotemporal evolution of the bispectrum}
    \end{minipage}
    \hfill
    \begin{minipage}[t]{0.48\textwidth}
        \centering
        \includegraphics[width=\linewidth]{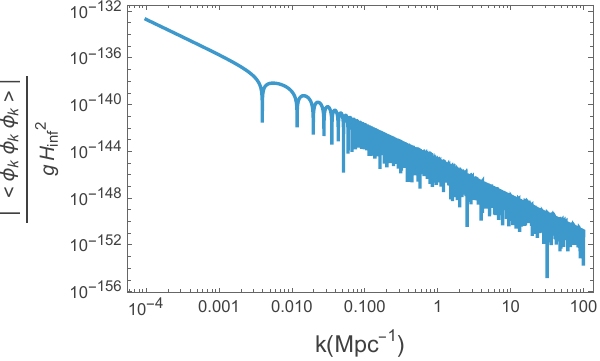}
        \captionsetup{style=centered_sub}\caption*{(b) Bispectra on $\eta=\eta_{\text{CMB}}$ hypersurface}
    \end{minipage}
    \caption{The bispectrum of the conformally coupled scalar field in the folded limit. Panel (a) illustrates the dependence on conformal time $\eta$ and wavenumber $k$ in the long-wavelength regime. Panel (b) extends the analysis to smaller scales at the time of decoupling ($\eta_{\text{CMB}}$), showing the amplitude decay.}
    \label{fig:bispectrum_conf_coupled}
\end{figure}

\paragraph{\bf The Folded limit:} Finally, for the folded limit ($k_1 = 2k, k_2 = k_3 = k$), the bispectrum reads:
\begin{align}
    \overline{\ev{\phi_{k_1}\phi_{k_2}\phi_{k_3}}}(\eta>\eta_{\text{inf}})&=g\frac{H^2_{\text{inf}}\eta_{\text{inf}}^6}{2k^3(2\eta_{\text{inf}}-\eta)^3}\bigg[\frac{\pi}{8}\cos{4k\eta}-\frac{(\ln{\abs{4k\eta_{\rm inf}}}+\gamma)\sin{4k\eta}}{4}+\frac{(4k\eta-\sin{4k\eta})}{64k^2\eta^2_{\rm inf}}\bigg] .
\end{align}
In this configuration, the three momenta are collinear in phase space. Despite this distinct geometry, the qualitative behavior closely mirrors the equilateral limit, as illustrated in Fig.~\ref{fig:bispectrum_conf_coupled}.

\begin{figure}[t]
    \centering
    \includegraphics[width=0.5\linewidth]{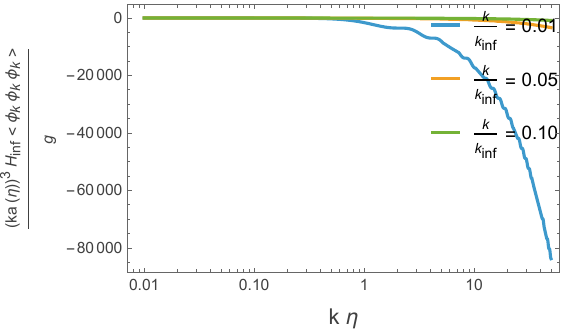}
    \caption{The dimensionless bispectrum shape function, scaled by $(ka)^3$, is plotted against $k\eta$ in the equilateral limit. By scaling out the phase-space suppression in comoving correlator, we isolate the non-trivial post-inflationary contributions.}
    \label{fig:radiation-bispectra}
\end{figure}

A few remarks are in order: First, we observe that the amplitude of the bispectrum generally scales as $a(\eta)^{-3}$. During inflation, where the scale factor behaves as $a(\eta) \propto |\eta|^{-1}$, this manifests as a suppression proportional to $|\eta|^3$. Conversely, during the radiation-dominated era where $a(\eta) \propto \eta$, the suppression follows a $1/\eta^3$ decay. The transition at $\eta_{\rm inf}$ therefore marks an inflection point in the suppression rate of the correlation functions.

However, one must be cautious when matching these regimes. Propagating the bispectrum from the end of inflation to end of reheating by the use of only linear transfer functions,\cite{babich2004primordial,fergusson2007primordial,duivenvoorden2020cmb}
\begin{equation}
    \overline{\langle \phi_{\mathbf{k}_1} \phi_{\mathbf{k}_2} \phi_{\mathbf{k}_3} \rangle}_{\eta > \eta_{\text{reh}}} \approx T(k_1)T(k_2)T(k_3) \, \overline{\langle \phi_{\mathbf{k}_1} \phi_{\mathbf{k}_2} \phi_{\mathbf{k}_3} \rangle}_{\eta_{\text{inf}}} \,,
\end{equation}
is not applicable. In the typical CMB correlation calculation, one assumes all the physical quantities become classical after the end of reheating. However, two cases arise where such assumption may not be completely valid. Since a non-zero bispectrum arises exclusively from non-linear (cubic) interaction terms in the Lagrangian, it is not a conserved quantity that merely rescales with the mode functions. The linear transfer function approach captures only the propagation of the initial Gaussian modes, failing to account for the continuous sourcing of new non-Gaussianity by quantum mechanical interference effective during the reheating and subsequently radiation eras. Consequently, the full bulk evolution must be treated as the sum of the linearly propagated primordial signal and the additive contribution from these subsequent quantum mechanical effect. Our present analysis seems to suggest therefore that assumption of post-inflationary processing of classical to quantum transition and consequent evolution of correlation function through transfer function may not capture the complete physical picture of the early universe.  

The sensitivity of these interactive contributions to the specific background evolution is fundamentally determined by the net scaling of the interaction vertex in the comoving frame. By considering the comoving correlators, $a^n \langle \phi^n \rangle$, one effectively strips away the kinematic redshift of the external fields to reveal the dynamical influence of the expanding spacetime. For a contact interaction of the form $\mathcal{L}_{\text{int}} \supset -\frac{g}{n!}\phi^n$, the rescaled vertex carries an effective weighting factor of $a^{4-n}$, arising from the competition between the metric determinant ($\sqrt{-g} \propto a^4$) and the field rescaling ($\phi \sim a^{-1}$). This leads to an illuminating observation: for purely four-point contact interaction ($n=4$), the factor $a(\eta)^4$ cancels precisely with the $a(\eta)^{-4}$ scaling of the fields associated with Wyle symmetry. Assuming a massless, conformally coupled scalar field, the mode functions behave essentially as plane waves ($f_k \sim e^{-ik\eta}$), and the bulk time integral reduces to that of a standard interacting field theory in Minkowski space. This enforces an effective energy conservation, without generating the era-dependent signatures characteristic of expanding backgrounds. As this computation offers no distinct cosmological signatures relative to the flat-space vacuum amplitude, we detail the derivation for the four point correlation function in Appendix~\ref{app:contact_interaction} for completeness. In contrast, the cubic interaction ($n=3$) does not admit such a cancellation; the residual factor of $a(\eta)$ in the vertex ensures that the bispectrum remains uniquely sensitive to the expansion history. This allows the three-point function to serve as a diagnostic of the specific cosmological epoch, such as radiation domination or reheating, which the contact trispectrum fails to resolve.

\section{Finite Duration of Reheating and Background Evolution}\label{section 2}
It is important to note that instantaneous reheating is an idealization. In physically realizable scenarios, reheating occurs over a finite duration, during which the inflaton gradually decays into other fields \cite{kofman1994reheating}. The timescale and dynamics of this process are commonly parametrized by the reheating temperature $T_{\rm reh}$ \cite{cook2015reheating}. Building on the instantaneous reheating case, we now consider a finite-duration reheating phase and investigate how its duration and corresponding reheating temperature affect the primordial power spectrum and the resulting cosmological correlators. 

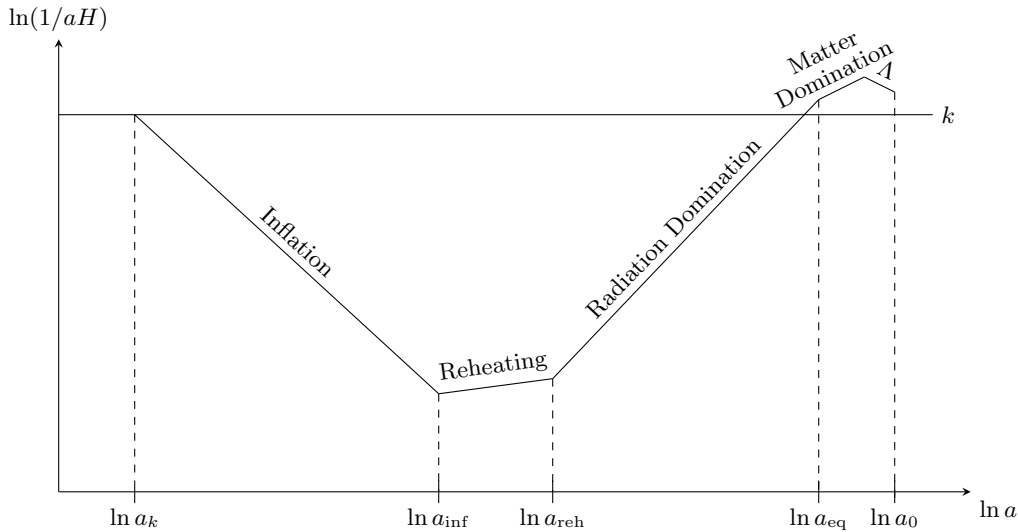
\begin{figure}[t]
			\centering
			\begin{tikzpicture}[scale=1.0,>=stealth]
				
				\draw[->] (0,0) -- (12,0) node[below right] {$\ln a$};
				\draw[->] (0,0) -- (0,6) node[above] {$\ln (1/aH)$};
				
				\draw (0,5) -- (11.5,5) node[right] {$k$};
				
				\draw[dashed] (1,5) -- (1,0.2);             
				\draw (1,5) -- (5,1.3) node[midway,sloped,above] {Inflation};
				\node[below] at (1,-0.1) {$\ln a_k$};
				\node[below] at (5,-0.1) {$\ln a_{\rm inf}$};
				
				\draw (5,1.3) -- (6.5,1.5) node[midway,above,sloped] {Reheating};
				\draw[dashed] (5,1.3) -- (5,0);    
				\draw[dashed] (6.5,1.5) -- (6.5,0);    
				\node[below] at (6.5,-0.1) {$\ln a_{\rm reh}$};
				
				\draw (6.5,1.5) -- (10,5.2) node[midway,sloped,above] {Radiation Domination};
				\draw[dashed] (10,5.2) -- (10,0);    
				\node[below] at (10,-0.1) {$\ln a_{\rm eq}$};
				
                \draw (10,5.2) -- (10.6,5.5)
                    node[midway,sloped,above] {\shortstack{Matter\\Domination}};

				\draw (10.6,5.5) -- (11,5.3) node[midway,sloped,above] {$\Lambda$};
				\draw[dashed] (11,5.3) -- (11,-0.1) node[below] {$\ln a_0$};

				\draw (1,0.15) -- (1,-0.15);
				\draw (5,0.15) -- (5,-0.15);
				\draw (6.5,0.15) -- (6.5,-0.15);
				\draw (10,0.15) -- (10,-0.15);
				\draw (11,0.15) -- (11,-0.15);
				
			\end{tikzpicture}
            \caption{Schematic evolution of the comoving Hubble horizon $(aH)^{-1}$ as a function of the scale factor $a$ in a logarithmic scale. The horizontal line represents a fixed comoving mode $k$. The mode exits the horizon during inflation at $a_k$ and subsequently re-enters during the radiation or matter-dominated epochs. The distinct slopes illustrate the changing expansion dynamics through the inflationary, reheating, radiation-dominated, and matter-dominated eras.}
            \label{fig:comoving-radius}
		\end{figure}
To illustrate the essential physics, we consider the simplest setup: a test scalar field $\phi$ evolving in the background of an inflaton $\varphi$ that slowly rolls toward its true minimum and subsequently undergoes oscillations about that minimum. During these oscillations, the inflaton decays into Standard Model particles as well as potential dark matter candidates, thereby reheating the universe and setting the stage for the standard hot Big Bang evolution. We consider the general Lagrangian describing the inflaton $\varphi$ and the spectator field $\phi$, assuming conformal coupling. All calculations are performed in the Jordan frame for the scalar field $\phi$.

\begin{equation}
    \mathcal{L}=-\sqrt{-g}\Bigg[\frac{1}{2}\partial_{\mu}\varphi\partial^{\mu}\varphi+V(\varphi)+\frac{1}{2}\partial_{\mu}\phi\partial^{\mu}\phi+\frac{1}{2}m_{\phi}\phi^2+\frac{1}{2}\xi \mathcal{R}\phi^2+\frac{g}{3!}\phi^3\Bigg],
\end{equation}
where $V(\varphi)$ depends on the choice of inflationary model. The equation of motion of background inflaton field is:
\begin{equation}
    \varphi''+2\mathcal{H}\varphi'+a^2\pdv{V(\varphi)}{\varphi}=0 ,
\end{equation}
and Friedman Equation is given as:
\begin{equation}
    \mathcal{H}^2= a^2 H^2 =\frac{8\pi G}{3}a(\eta)^2\rho=\frac{8\pi G}{3}\left[\frac{1}{2}(\varphi')^2+a^2V(\varphi)\right] .
\end{equation}
The EOM for scalar field $X=a(\eta)f_k(\eta)$ is given as:
\begin{align}
	\Box\phi-a^2\xi\mathcal{R}\phi =0 \implies 
	X''_{k}+\left[k^2-\frac{a''}{a}(1-6\xi)\right]X_{k}&=0.
\end{align}
where we used $\mathcal{R}=\nicefrac{6a''}{a^3}$. After inflation ends $\eta \ge\eta_{\mathrm{inf}}$, the inflaton undergoes coherent oscillations and gradually transfers its energy to relativistic degrees of freedom. During this phase, the universe is not in thermal equilibrium and its composition evolves continuously as the co-moving Hubble radius is growing as shown in Fig.~\ref{fig:comoving-radius}. 

There are two primary ways to define the effective equation of state (EoS) during this era. The first approach derives the EoS by time-averaging the field dynamics over an oscillation period, effectively treating the inflaton as a coarse-grain  fluid that decays instantaneously at a later time setting the end of inflation \cite{mishra2024cosmic,cembranos2016cosmological}. However, to capture the integrated expansion history of the evolving fluid mixture, we adopt the phenomenological approach ~\cite{dai2014reheating,munoz2015equation,german2024inflationary,cook2015reheating}. In this framework, we approximate the reheating era by a constant effective equation of state $w$, defined such that the total energy density scales as
\begin{equation}\label{energy-density}
    \rho_{\mathrm{tot}} = \rho_{\mathrm{inf}} a^{-3(1+w)} \,.
\end{equation}
This effective parameter represents a coarse-grained description of the combined inflaton--radiation rather than the instantaneous properties of the inflaton condensate. At the end of reheating at $\eta = \eta_{\rm reh}$, the universe becomes radiation dominated with the condition $\rho_{\rm tot} = \rho_{\rm rad}$, and the radiation component evolves as $\rho_{\rm rad} \propto a^{-4}$. Utilizing the standard Hubble equation 
$\mathcal{H}^2=(\nicefrac{a'}{a})^2  =\frac{8\pi G}{3}a^2\rho$,
%
we express the scale factor in the following form:

\begin{align}\label{scale-factor}
	a(\eta)=
	\begin{cases}
		-\frac{1}{H_{\rm inf}\eta},	&\text{when } \eta<\eta_{\text{inf}}\\
		-\frac{1}{H_{\rm inf}\eta_{\text{inf}}}\left(\frac{1+3w}{2\abs{\eta_{\text{inf}}}}\right)^{\frac{2}{1+3w}}\left(\eta-3\mu\eta_{\text{inf}}\right)^{\frac{2}{1+3w}}, &\text{when } \eta_{\text{inf}}<\eta<\eta_{\text{reh}}\\
        -\frac{1}{H_{\rm inf} \eta _{\text{inf}} }\left[\frac{1+3w}{2|\eta _{\text{inf}}|}\right]^{\frac{2}{1+3w}} \left[\eta _{\text{reh}}-3\mu\eta _{\text{inf}}\right]^{\frac{2}{3 w+1}} \left[1+\frac{2(\eta_{\text{reh}}-\eta)/(1+3w)}{\left[3\mu\eta _{\text{inf}}-\eta _{\text{reh}}\right]}\right]&\text{when } \eta>\eta_{\text{reh}} .
	\end{cases}
\end{align}
where we defined
\begin{equation}
    \mu=\frac{1+w}{1+3w},
\end{equation}
The scale factors stated above carries some unphysical cosmological singularity, which corresponds to the formal divergence of $a''/a$ term in the Mukhanov-Sasaki equation. Specifically, the mode equation for the reheating phase carries a singularity at $\eta_{s, \text{reh}} = 3\mu\eta_{\text{inf}} $, while the radiation-dominated era is associated with a pole at $\eta_{s, \text{RD}}$. Upon close inspection of these coordinate singularities relative to the physical boundaries of each era reveals that they do not impact the physical evolution of perturbations. In the case of reheating, the requirement $w > -1/3$ fixes the coefficient $3\mu > 1$; given that $\eta_{\text{inf}} < 0$, this ensures $\eta_{s, \text{reh}} < \eta_{\text{inf}}$, placing the singularity in the unphysical "pre-reheating" past. For the subsequent radiation-dominated era, the singularity is located at:
\begin{align}
    \eta_{s, \text{RD}}
= \eta_{\rm reh}-|\eta_{\inf}|-\frac{1}{2}|(\eta_{\inf}-\eta_{\rm reh})(1+3w)|
\end{align}
This enforces the inequality $\eta_{s, \text{RD}} < \eta_{\text{reh}}$ for $w>-1/3$. Since both singularities are strictly localized outside the time window of their respective eras, they possess no physical significance and cannot induce genuine divergences in the power spectrum. Therefore, any sharp features or localized power enhancements observed near $\eta \approx \eta_{\text{inf}}$ or $\eta \approx \eta_{\text{reh}}$ are not driven by these singularities, but are instead artifacts of the instantaneous transition approximation used to model the transition between equations of state.

\begin{figure}[h]
    \centering
    \begin{minipage}{0.48\linewidth}
        \centering
        \includegraphics[width=\linewidth]{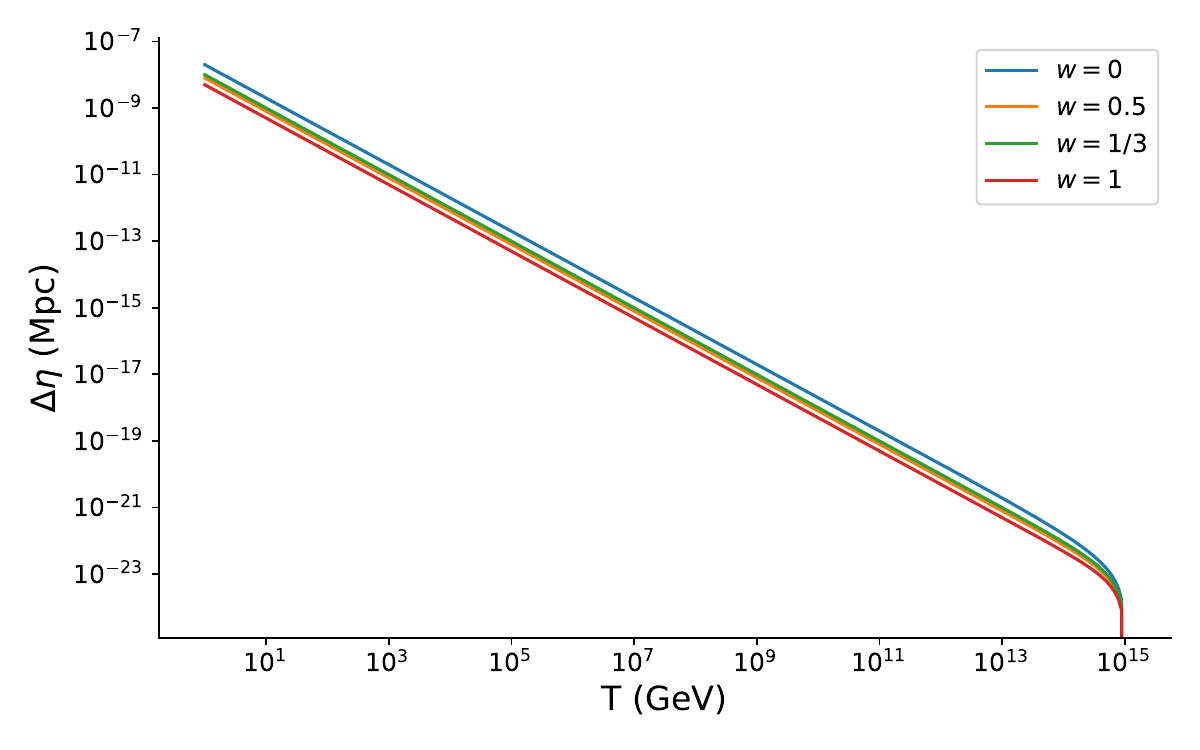}
        \caption{Constraints on the duration of the reheating phase consistent with observational bounds.}
        \label{fig:reheating_duration}
    \end{minipage}\hfill
    \begin{minipage}{0.48\linewidth}
        \centering
        \includegraphics[width=\linewidth]{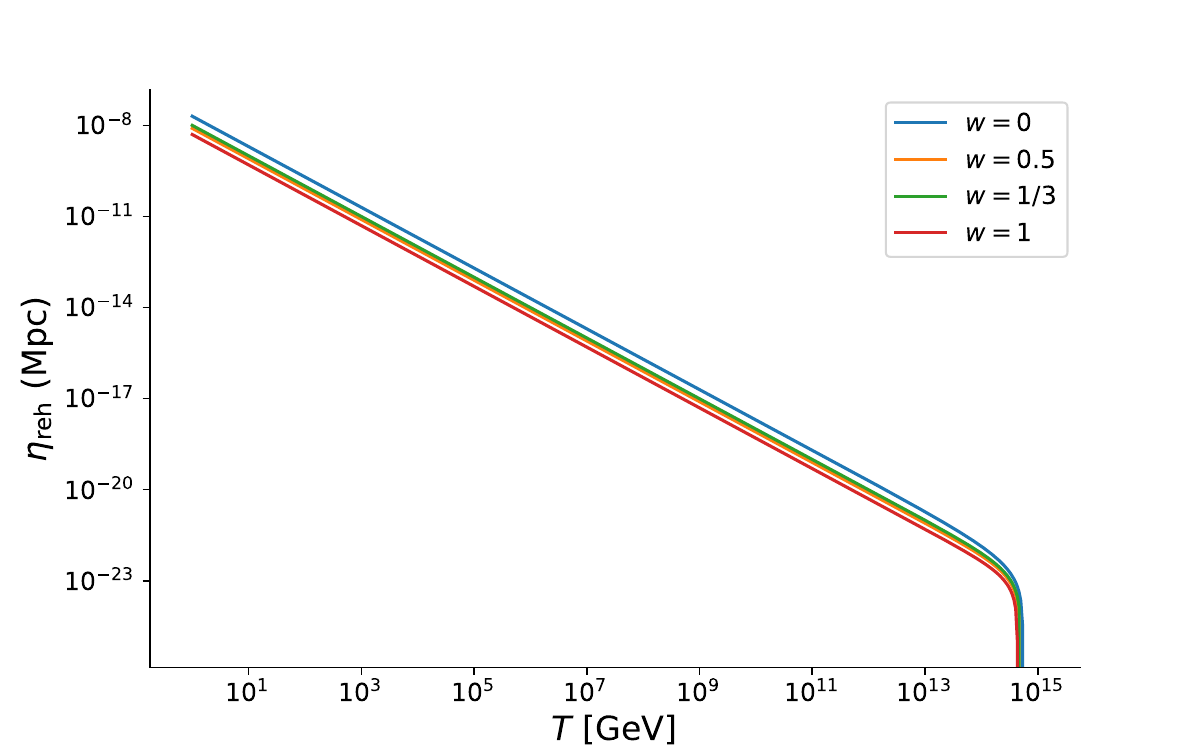}
        \caption{The conformal time at the end of reheating, $\eta_{\rm reh}$, as a function of the reheating temperature $T_{\rm reh}$.}
        \label{fig:eta_reheating}
    \end{minipage}
\end{figure}

Since both the scale factor and the Hubble parameter are continuous at the transition at $\eta=\eta_{\text{inf}}$, the mode functions and their first derivatives remain continuous as well. As a result, the Bogoliubov coefficients do not mix positive and negative frequency components, and there is no particle production for the conformally coupled scalar. This metric reduces to the standard radiation dominated form for $w=\nicefrac{1}{3}$ as discussed earlier. To contrast this smooth mode evolution with the sharp change in the background geometry, we now evaluate the Ricci scalar in the inflationary and reheating era:
\begin{align}
	\mathcal{R} =\frac{6a''}{a^3} 
	=\begin{cases}
		12H_{\inf}^2 ,	&\text{when } \eta<\eta_{\text{inf}}\\
		\frac{12H_{\inf}^2\eta_{\inf}^2(1-3w)}{(1+3w)^2}\left(\frac{1+3w}{2\abs{\eta_{\text{inf}}}}\right)^{-\frac{4}{1+3w}}\left(\eta-3\eta_{\text{inf}}\mu\right)^{-6\mu}, &\text{when } \eta>\eta_{\text{inf}} \\
        0 ,	&\text{when } \eta >\eta_{\text{reh}}
	\end{cases}
\end{align}
Note that the discontinuity in Ricci scalar at the instant of transition will manifest itself as the violation of adiabaticity condition $(\tfrac{\omega_k'}{\omega_k^2}\ll 1)$ and hence lead to particle production for non-conformally coupled scalar field.\cite{PhysRevD.101.083516} Since we are considering a sudden transition to reheating era, the temperature during this epoch provides a natural clock with which to trace the cosmic evolution. We will exploit this fact to parameterize the power spectrum and bispectrum as functions of the reheating temperature, \(T_{\rm reh}\), in the later sections of this work. We therefore begin with a brief review of the relevant ingredients. During this era, the energy density evolves according to Eq.~\eqref{energy-density}:
\begin{align}
    \rho_{\rm tot} (\eta)=\rho_{\text{inf}}\left[\frac{a(\eta)}{a_{\text{inf}}}\right]^{-3(1+w)}=\rho_{\text{inf}}e^{-3(1+w)\ln{\frac{a(\eta)}{a_{\text{inf}}}}}=\rho_{\text{inf}}e^{-3(1+w)N(\eta)},
\end{align}
which could be used to define number of e-folds reheating lasted \cite{garcia2021inflaton,dai2014reheating,afzal2023nanograv}, utilizing the condition $\rho_{\rm tot} (\eta_{\rm reh})=\rho_{\rm rad} (\eta_{\rm reh}) =\frac{\pi^2}{30}g_{*,re} T^4_{\rm reh}$, as
\begin{align}\label{e-folding_reheating}
    N_{\text{reh}}=N(\eta_{\rm reh})=\ln\left(\frac{a_{\rm reh}}{a_{\text{inf}}}\right)=\frac{1}{3(1+w)}\ln{\frac{\rho_{\text{inf}}}{\rho_{\text{rad}}}}=\frac{1}{3(1+w)}\ln{\frac{3H^2_{\rm inf}M^2_{pl}}{\frac{\pi^2}{30}g_{*,re} T^4_{\rm reh}}},
\end{align}
where $g_{*,re}$ is the effective number of relativistic degrees of freedom upon thermalisation. This can be inverted to parametrize the duration of reheating in terms of reheating temperature $(T_{\rm reh})$ at the end of this phase \cite{german2023model,dai2014reheating}:
\begin{equation}
    T^4_{\rm reh} =\frac{90 H^2 M_{\rm pl}^2}{\pi^2 g_{\star,re}}e^{-3(1+w)N_{\rm reh}} .
\end{equation}
This parametrizes the conformal time corresponding to the end of reheating as following:
\begin{align}
    \eta_{\text{reh}}&=\eta_{\inf}+\frac{2\eta_{\inf}}{1+3w}\left[1-\left(\frac{T_{\rm reh}^{\rm (ins)}}{T_{\text{reh}}}\right)^{\frac{2}{3\mu}}\right]\label{etareh} .
\end{align}
Given this, the comoving mode that re-enters the horizon at the end of reheating can likewise be expressed in terms of the same temperature scale using the horizon crossing condition, $k_{\rm reh}=\mathcal{H}(\eta_{\rm reh})$. For clarity we keep the present-day scale factor $a_0$ explicit in the following expressions, although it will be implicit in the final numerical result. Using \eqref{e-folding_reheating} and Eqs.~\eqref{etareh} to eliminate the explicit conformal time dependence in favor of the expansion history, we get:
\begin{align}
    \frac{k_{\rm reh}}{a_0} =\frac{a_{\rm reh}H_{\rm reh}}{a_0}
    &=\sqrt{\frac{\pi^2 g_{*,re}}{90}} \frac{T^2_{\rm reh}}{M_{\rm pl}} \frac{a_{\rm reh}}{a_0}=\left(\frac{43}{11 g_{\rm s,re}}\right)^{1/3}\frac{T_0}{T_{\rm reh}}\sqrt{\frac{\pi^2 g_{*,re}}{90}}\frac{T^2_{\rm reh}}{M_{\rm pl}} \approx1.67\times 10^{7}\left(\frac{T_{\rm reh}}{\rm GeV}\right)\text{Mpc}^{-1} ,
\end{align}
where we used \cite{dai2014reheating}
\begin{align}
    \frac{a_{\rm re}}{a_0}
    &=\left(\frac{43}{11 g_{\rm s,re}}\right)^{1/3}\frac{T_0}{T_{\rm reh}},
\end{align}
and $g_{*,re}=g_{s,re}=106.7$, representing the total number of relativistic degrees of freedom contributing to the energy and entropy densities after thermalization, respectively. The present-day CMB photon temperature is taken as $T_0=9.453\times 10^{-32}M_{\rm pl}$. For unit consistency, we employ the conversion $1{\rm GeV}=1.38\times 10^{38}{\rm Mpc}^{-1}$, where $T_{\text{reh}}^{\text{(ins)}}$ denotes the temperature for the limit of instantaneous reheating. This result is insensitive to the particular choice of metric used to model the reheating epoch. Likewise, one can determine the comoving mode that last crossed the horizon at the end of inflation using $k_{\rm inf}=\mathcal{H}_{\rm inf} = a_{\rm inf} H_{\rm inf}$, and \eqref{e-folding_reheating}:
\begin{align*}
   \frac{k_{\rm inf}}{a_0}
    &=\frac{a_{\rm inf}}{a_{\rm re}}\frac{a_{\rm re}}{a_0}H_{\rm inf}
    =3.66\times 10^{25}\left(\frac{43}{11 g_{\rm s,re}}\right)^{1/3}\left(\frac{T_{\rm reh}}{T_{\rm reh}^{\rm (ins)}}\right)^{\frac{4}{3(1+w)}}\left(\frac{H_{\rm inf}}{T_{\rm reh}}\right)\text{Mpc}^{-1}.
\end{align*}
There is an inflection point in the definition: for $w>1/3$, $k_{\rm inf}$ decreases with temperature, while for $w<1/3$ it increases with temperature. At $w=1/3$, $k_{\rm inf}$ becomes independent of reheating temperature. The lower bound $(\tfrac{k}{k_{\rm inf}}>\frac{k_{\rm reh}}{k_{\rm inf}})$ on the modes that enter during reheating can be given as:
\begin{align*}
      \frac{k_{\rm reh}}{k_{\rm inf}}&=\sqrt{\frac{\pi^2 g_{*,re}}{90}}\frac{T^2_{\rm reh}}{H_{\rm inf}M_{\rm pl}}\left(\frac{T_{\rm reh}}{T_{\rm reh}^{\rm (ins)}}\right)^{-\frac{4}{3(1+w)}}=\left(\frac{T_{\rm reh}^{\rm (ins)}}{T_{\rm reh}}\right)^{-\frac{2}{3\mu}}=\left(\frac{T_{\rm reh}}{T_{\rm reh}^{\rm (ins)}}\right)^{\frac{2}{3\mu}}\le 1.
\end{align*}
For finite duration of reheating, we can utilise the following bound
\begin{align}
    A_s &=\frac{H_{\rm inf}}{8\pi^2\epsilon M_{\rm pl}^2}=\frac{2H_{\rm inf}^2}{\pi^2 r M_{\rm pl}^2}\implies H_{\rm inf} = \pi M_{\rm pl}\sqrt{\frac{r A_s}{2}}\lesssim 10^{-5}M_{\rm pl}.
\end{align}
Using this, the rough estimate on $k_{\rm inf}$ can be given as:
\begin{align}
    \frac{k_{\rm inf}}{a_{0}} &\sim1.22\times 10^{25}\left(\frac{T_{\rm reh}}{T_{\rm reh}^{\rm (ins)}} \right)^{\frac{4}{3(1+w)}}\frac{10^{-5}\times 2.435\times 10^{18}\text{GeV}}{T_{\rm reh}}\text{Mpc}^{-1}\nt
    &\sim 10^{38}T_{\rm reh}^{\rm (ins)^{-\frac{1}{3(1+w)}}}T_{\rm reh}^{\frac{1-3w}{3(1+w)}}\text{GeV }\text{Mpc}^{-1}\label{kinf}.
\end{align}
For the limiting case of instantaneous reheating, Eq. \eqref{kinf} yields a conformal time at the end of inflation of $a_0|\eta_{\rm inf}|\approx 10^{-22}-10^{-23}\text{Mpc}$. From this point onward the factor of $a_0$ will again be kept implicit. The horizon-crossing condition $k_{\rm inf}|\eta_{\rm inf}|=1$ then allows this result to be generalized to arbitrary reheating temperatures. This relation allows us to parametrize the reheating duration $\Delta\eta$ in terms of the reheating temperature $T_{\rm reh}$, as illustrated in Fig.~\ref{fig:eta_reheating}. This interval characterizes the period during which the field evolution is subjected to the reheating equation of state $w$:
\begin{align}
    \Delta\eta&=-\frac{2}{1+3w}\left(\frac{T_{\rm reh}^{\rm (ins)}}{T_{\rm reh}}\right)^{\frac{4}{3(1+w)}}\left(\frac{11 g_{\rm s,re}}{43}\right)^{1/3}\left(\frac{T_{\rm reh}}{T_0H_{\rm inf}}\right)\left[1-\left(\frac{T_{\rm reh}^{\rm (ins)}}{T_{\text{reh}}}\right)^{\frac{2}{3\mu}}\right].
\end{align}

\begin{figure}[h]
    \includegraphics[width=1\linewidth]{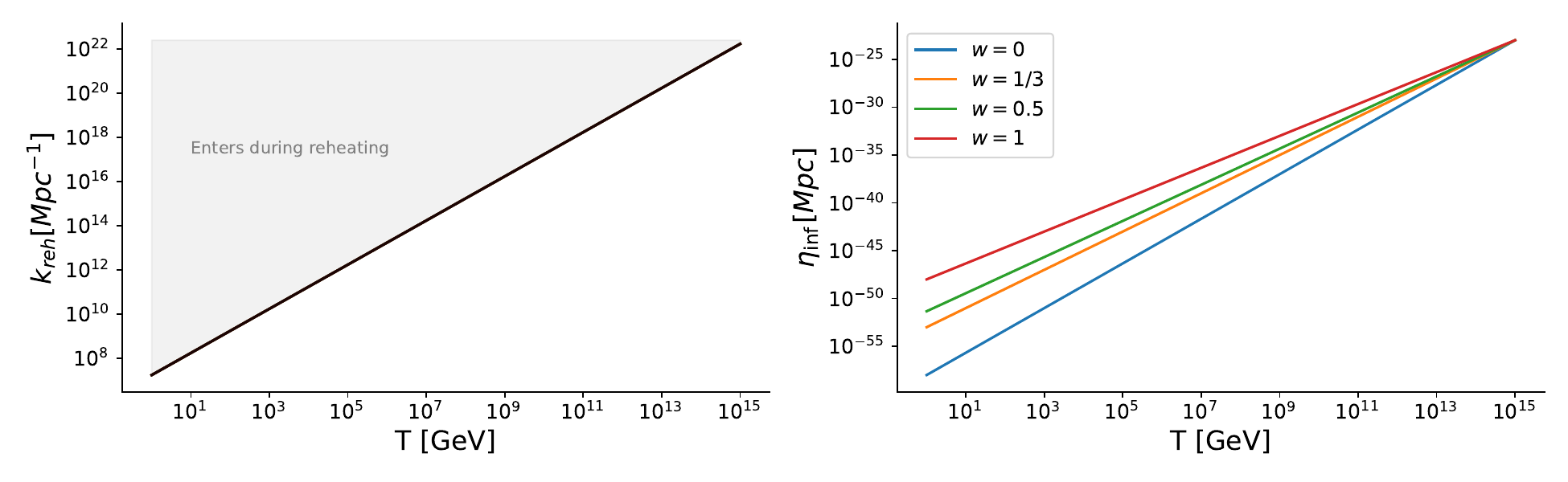}
    \caption{Dependence of physical scales on the reheating temperature $T_{\rm reh}$. (Left) The mode re-entry scale; modes re-entering during reheating ($k > k_{\rm reh}$) correspond to short wavelengths negligible for CMB physics. (Right) The conformal time at the end of inflation, $\eta_{\rm inf}$, as a function of $T_{\rm reh}$ for different equations of state.}
    \label{fig:k_eta_temp} 
\end{figure}

\begin{figure}[h]
    \centering
    \includegraphics[width=0.5\linewidth]{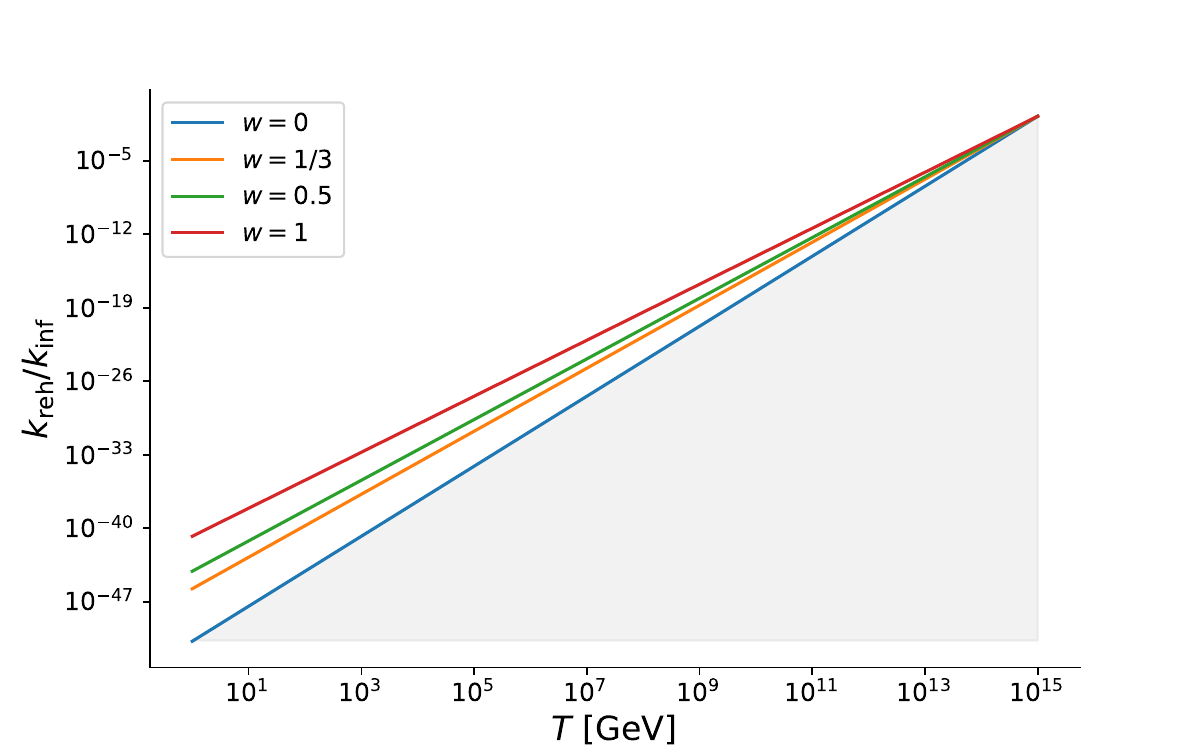}
        \caption{The threshold for the scale ratio $k/k_{\rm inf}$. Modes exceeding this value re-enter the horizon during the reheating epoch.}
    \label{fig:ratio_scale}
\end{figure}
\subsection{Field Dynamics}
\begin{figure}[t]
    \centering
    \begin{minipage}{0.48\linewidth}
        \centering
        \includegraphics[width=\linewidth]{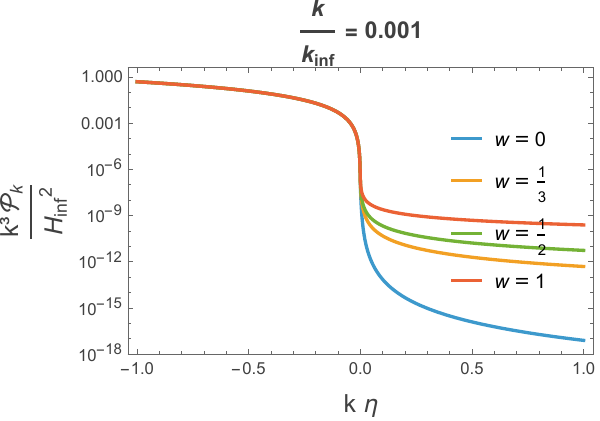}
        \label{fig:w1}
    \end{minipage}\hfill
    \begin{minipage}{0.48\linewidth}
        \centering
        \includegraphics[width=\linewidth]{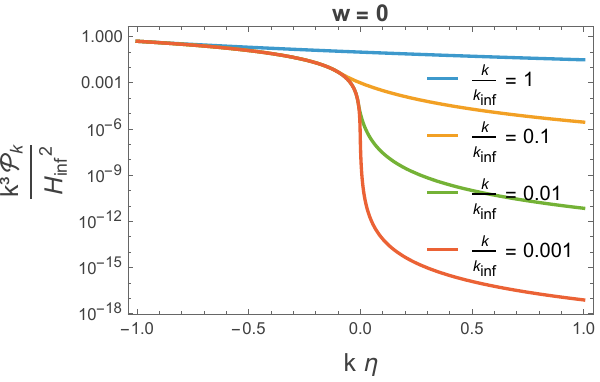}
        \label{fig:w0}
    \end{minipage}
    \captionsetup{style=centered_sub}
    \caption{Power spectrum for conformally coupled scalar during reheating for fixed $\frac{k}{k_{\rm inf}}$ but varying $w$ is plotted in (a) and for fixed $w$ but varying $\frac{k}{k_{\rm inf}}$ is plotted in (b).}
    \label{fig:inf-to-reh-powerspectra}
\end{figure}

We have now specified the metric and scale-factor evolution throughout this era, and we expressed the background in terms of the reheating temperature and the effective equation of state. We also identified the comoving mode that exited the horizon at the end of inflation and the mode that re-entered at the end of reheating, so the relevant horizon scales for this epoch are now fixed. With this background structure in place, we proceed to study the dynamics of spectator scalar field, described by the action:
\begin{equation}
    S=-\int d^4x\sqrt{|g|}\left[\frac{1}{2}\partial_\mu \phi\partial^\mu\phi+\frac{1}{2}m_{\phi}\phi^2+\frac{1}{2}\xi \mathcal{R}\phi^2+\frac{g}{3!} \phi^3 \right].
\end{equation}
The free field EoM in the $m_{\phi}/H_{\rm inf}\ll1$ limit, for comoving field can be given as:
\begin{align}
        X''_{k}+\left[k^2-\frac{a''}{a}(1-6\xi)\right]X_{k}&=0\tag{with $X_k=a(\eta)f_k$}
\end{align}
The solution in each era for $\xi=\tfrac{1}{6}$ is trivial and therefore we will collectively state them here:
\begin{align}
	X_{\vec{k}}=
	\begin{cases}
		\frac{1}{\sqrt{2k}}e^{-ik\eta}&\text{when } \eta<\eta_{\text{inf}}\\
		\frac{1}{\sqrt{2k}}e^{-ik\eta} &\text{when } \eta_{\text{inf}}<\eta<\eta_{\text{reh}}\\
        \frac{1}{\sqrt{2k}}e^{-ik\eta} &\text{when } \eta>\eta_{\text{reh}}
	\end{cases}
\end{align}
Let us now look at how the Power spectrum changes during each phases of their evolution.

\begin{align}
    \mathcal{P}_k =  \frac{X_k X_k^*}{a(\eta)^2}
    =\frac{1}{a(\eta)^2}\frac{1}{2k}
\end{align}
We can express the power spectrum as piecewise defined function in following manner.
\begin{equation}
    \mathcal{P}_k=\frac{1}{2k}\frac{1}{a(\eta)^2}=
\begin{cases}
       \frac{H^2\eta^2}{2k} \qquad&\text{for }\eta<\eta_{\text{inf}}	\\
\frac{H^2_{\text{inf}}\eta_{\text{inf}}^2}{2k}\left(\frac{1+3w}{2\abs{\eta_{\text{inf}}}}\right)^{-\frac{4}{1+3w}}\left(\eta-3\eta_{\text{inf}}\mu\right)^{-\frac{4}{1+3w}}\qquad&\text{for }\eta_{\text{inf}}<\eta\le \eta_{\rm reh}\\
       \frac{H_{\rm inf}^2\eta _{\text{inf}}^2}{2k}\left[\frac{1+3w}{2|\eta _{\text{inf}}|}\right]^{-\frac{4}{1+3w}} \left[\eta _{\text{reh}}+3|\eta _{\text{inf}}|\mu\right]^{-\frac{4}{3 w+1}} \left[1+\frac{2(\eta_{\text{reh}}-\eta)/(1+3w)}{\left[3\mu\eta _{\text{inf}}-\eta _{\text{reh}}\right]}\right]^{-2}\qquad&\text{for }\eta> \eta_{\rm reh}
\end{cases}
\end{equation}
To highlight the sensitivity of the power spectrum to the reheating equation of state $w$, we evaluate the piecewise evolution during the reheating phase in terms of $k_{\rm inf}$ and subsquently plot in Fig.~\ref{fig:inf-to-reh-powerspectra}.
\begin{equation}
k^3\mathcal{P}_k(\eta) = \frac{1}{2}\left(\frac{k}{a(\eta)}\right)^2 
\propto \begin{cases} 
(k\eta)^2 \qquad&\text{for }k\eta<k\eta_{\text{inf}} \\
\left[ k\eta - \frac{3\mu k}{k_{\text{inf}}} \right]^{-\frac{4}{1+3w}} \qquad&\text{for }k\eta>k\eta_{\text{inf}}
\end{cases}
\end{equation}
The suppression of the power spectrum is inherently mode-dependent and determined by the horizon exit time relative to the reheating transition. Modes that exit the horizon well before the transition $k/k_{\inf}\to0$ experience significant suppression, with the power spectrum vanishing as $\Delta_{\phi}\to0$. In contrast, modes deep inside the horizon effectively track the vacuum solution as expected. For an arbitrary reheating equation of state $w$, the dimensionless spectrum scales as:
\begin{align}
    \Delta_{\phi}
    \approx\frac{H^2_{\text{inf}}k^2/k_{\text{inf}}^2}{4\pi^2}\left[\frac{2}{3(1+w)}\right]^{\frac{4}{1+3w}}.
\end{align}
While the spectral shape retains the characteristic $k^2$ behavior, the amplitude is modulated by the background expansion history, explicitly encoding the reheating equation of state $w$. Moreover, this behavior is a direct consequence of the finite duration of inflation. Had inflation persisted indefinitely $\eta\to0$, the expansion-induced suppression $(k\eta)^2$ would have eventually suppressed all modes, driving the entire spectrum to zero. By terminating inflation at a finite time, the reheating transition effectively captures the instantaneous vacuum state of the high-$k$ modes, preserving the spectral profile that would have otherwise decayed away. In the following discussions, we look into the the evolution of bispectrum during different phases namely reheating and radiation separately to better understand their impact.  

\subsection{Bispectrum Dynamics during Reheating}
\begin{figure}[t]
    \begin{minipage}{0.48\linewidth}
        \centering
        \includegraphics[width=\linewidth]{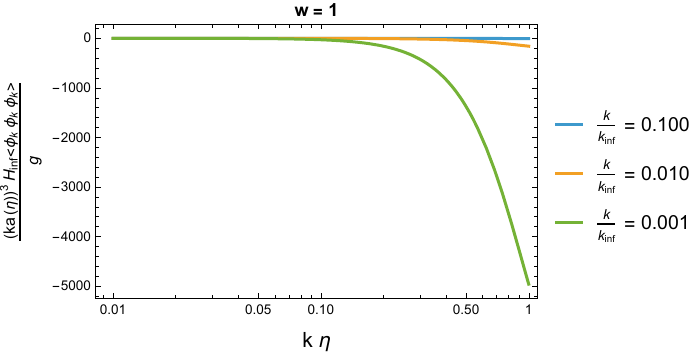}
        \captionsetup{style=centered_sub}
        \captionsetup{style=centered_sub}\caption*{(a) $w=1$}
    \end{minipage}
    \hfill
    \begin{minipage}{0.48\linewidth}
        \centering
        \includegraphics[width=\linewidth]{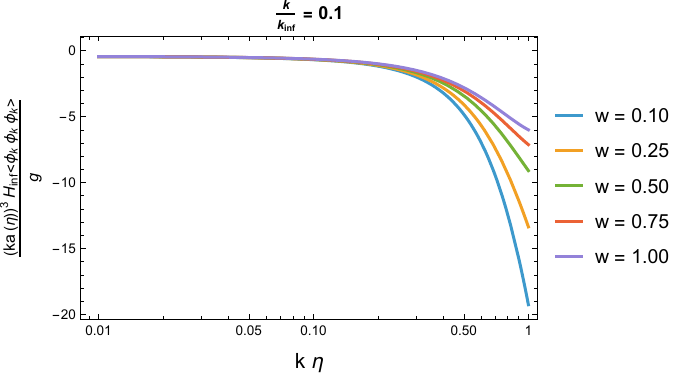}
        \captionsetup{style=centered_sub}
        \caption*{(b) $k/k_{\rm inf}=0.1$}
    \end{minipage}
   \caption{The dimensionless bispectrum $(ka(\eta))^3\langle\phi_k\phi_k\phi_k\rangle$ in the equilateral limit. (a) In the deep superhorizon regime ($k/k_{\rm inf}=0.01$), the evolution is dominated by a steep monotonic decay (the ``falling" envelope), which obscures the oscillatory features dependent on the reheating EoS. (b) For fixed EoS $w$, the modes closer to the horizon ($k/k_{\rm inf}=0.1$), the monotonic drift is less pronounced while the modes far away from horizon experience maximum drift.}
    \label{fig:bispectrum-combined}
\end{figure}

Introducing a reheating phase prior to radiation era means that we need to consider the bispectrum at the end of inflation and at the end of reheating before we proceed to study the bispectrum for superhorizon modes during radiation dominated era. Similar to earlier cases, we split the bulk evolution into respective eras depending on their stage of evolution. We first consider the inflationary part of conformally coupled massless scalar field.
\begin{align}  
\ev{\phi_{k_1}\phi_{k_2}\phi_{k_3}}_{\text{inf}}
    &=2g\Im{\int_{-\infty}^{\eta_{\text{inf}}}D\eta' W_{k_1}(\eta, \eta')W_{k_2}(\eta,\eta')W_{k_3}(\eta,\eta')}\nt
    &=2g\Im{\int_{-\infty}^{\eta_{\text{inf}}}d\eta' a(\eta')^4\frac{1}{a(\eta)^3a(\eta')^3}\frac{e^{-i(k_1+k_2+k_3)(\eta-\eta')}}{8k_1k_2k_3}}\nt
    &=\frac{-2g}{8k_1k_2k_3H_{\rm inf}a(\eta)^3}\Im\bigg\{e^{-i(k_1+k_2+k_3)\eta}(\gamma+\frac{i\pi}{2}+\ln{[-(k_1+k_2+k_3)\eta_{\text{inf}}]})
    +\order{\eta_{\rm inf}}\bigg\}\nt
    &=-\frac{\pi}{8}\frac{g}{k_1k_2k_3H_{\rm inf}a(\eta)^3}\cos{K\eta}+\frac{g}{4k_1k_2k_3H_{\rm inf}a(\eta)^3}\ln{\abs{K\eta_{\inf}}}\sin{K\eta}
    +\frac{g}{4k_1k_2k_3H_{\rm inf}a(\eta)^3}\gamma\sin{K\eta}\label{conformally-coupled-during-inflation}
\end{align}
where we defined $K=k_1+k_2+k_3$. 
The reheating part of the bispectrum is found to be:
\begin{align}
\overline{\ev{\phi_{k_1}\phi_{k_2}\phi_{k_3}}}_{\text{reh}}  
    &=2g\Im{\frac{e^{-iK\eta}}{8a(\eta)^3k_1k_2k_3}\int_{\eta_{\text{inf}}}^{\eta}d\eta' a(\eta')e^{iK\eta'}}\nt
    &=2ga_{\rm inf}\left(\frac{1+3w}{2\abs{\eta_{\text{inf}}}}\right)^{\frac{2}{1+3w}}\Im{\frac{e^{-iK\eta}}{8a(\eta)^3k_1k_2k_3}\int_{\eta_{\text{inf}}}^{\eta}d\eta' \left(\eta'-3\mu\eta_{\text{inf}}\right)^{\frac{2}{1+3w}}e^{iK\eta'}}.
\end{align}
The reheating contribution can be given as:
\begin{align}\label{conformally-coupled-during-reheating}
\overline{\ev{\phi_{k_1}\phi_{k_2}\phi_{k_3}}}_{\text{reh}}=\frac{-2ga_{\rm inf}K^{-3\mu}}{8a(\eta)^3k_1k_2k_3}\left(\frac{1+3w}{2\abs{\eta_{\text{inf}}}}\right)^{\frac{2}{1+3w}} \Im{i e^{\frac{i(\pi+3 K\eta_{\rm inf}(w+1))}{3 w+1}-iK\eta} \Gamma\left(3\mu, -i K\eta + 3\mu i K\eta_{\text{inf}}, \frac{2i K\eta_{\text{inf}}}{1+3w}\right)}.
\end{align}
where $\Gamma(a,b,c)$ is the generalized incomplete gamma function. Hence, the full bispectrum can be given as:
\begin{align} 
\overline{\ev{\phi_{k_1}\phi_{k_2}\phi_{k_3}}} &=\frac{g}{ k_1k_2k_3H_{\rm inf}a(\eta)^3}\bigg(-\frac{\pi}{8}\cos{K\eta}+\frac{\ln{\abs{K\eta_{\inf}}}\sin{K\eta}+\gamma\sin{K\eta}}{4}+\frac{2}{8K\eta_{\rm inf}}\left(\frac{1+3w}{2K\abs{\eta_{\text{inf}}}}\right)^{\frac{2}{1+3w}}\nt
&\qquad\times\bigg[\cos\left(\frac{(\pi+3 K\eta_{\rm inf}(w+1))}{3 w+1}-K\eta\right)\Re{ \Gamma\left(3\mu, -i K\eta + 3\mu i K\eta_{\text{inf}}, \frac{2i K\eta_{\text{inf}}}{1+3w}\right) }\nt
&\qquad-\sin\left(\frac{(\pi+3 K\eta_{\rm inf}(w+1))}{3 w+1}-K\eta\right)\Im{ \Gamma\left(3\mu, -i K\eta + 3\mu i K\eta_{\text{inf}}, \frac{2i K\eta_{\text{inf}}}{1+3w}\right)}\bigg]\bigg).
\end{align}
To visualize how the reheating equation of state (EoS) affects the bispectrum, we study the evolution of the dimensionless comoving bispectrum as a function of $k\eta$ in Fig.~\ref{fig:bispectrum-combined}. Qualitatively, this resembles Fig.~\ref{fig:radiation-bispectra} and we find that the reheating contribution to bispectrum at early times (superhorizon scale) are negligible. On the contrary, the sensitivity of these modes to the reheating EoS is larger at late time. This reflects in the amplification of dimensionless comoving bispectra for modes with smaller $k/k_{\mathrm{inf}}$ as seen in panel (a). While the dependence of subhorizon profile for different EoS is shown in panel (b), we find that the qualitative feature is essentially identical for all modes. While the modes deep inside the horizon during inflation are heavily suppressed, the maximum correlation exists between modes that are superhorizon at the end of inflation. The same modes evolve to encode sensitivity to reheating EoS and generates isocurvature which depend on reheating history.

\subsection{Evolution in the Radiation Era}
\begin{figure}[t]
    \centering
    \begin{minipage}[t]{0.48\linewidth}
        \centering
        \includegraphics[width=\linewidth]{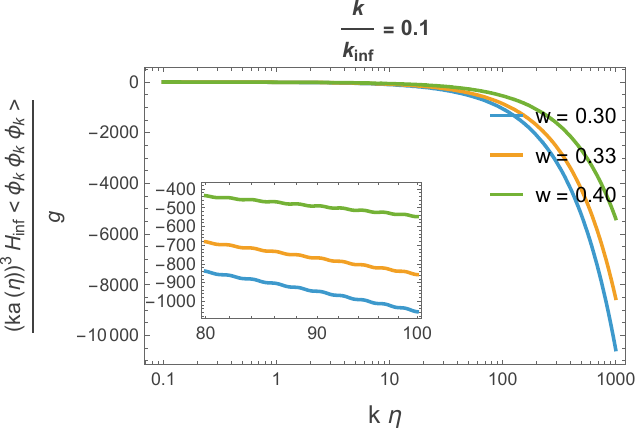}
        \captionsetup{style=centered_sub}\caption*{(a) Post-reheating evolution ($\eta > \eta_{\text{reh}}$).}
        \label{fig:with_reheating}
    \end{minipage}
    \hfill
    \begin{minipage}[t]{0.48\linewidth}
        \centering
        \includegraphics[width=\linewidth]{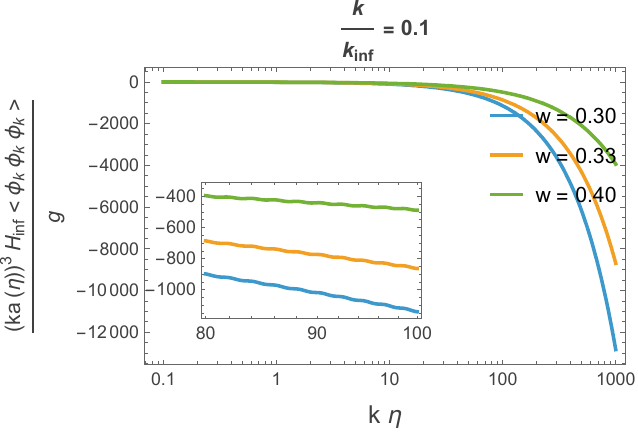}
        \captionsetup{style=centered_sub}\caption*{(b) Evolution during the reheating phase.}
        \label{fig:without_reheating}
    \end{minipage}
    
    \caption{Evolution of the bispectrum for a fixed reheating temperature of $T_{\text{reh}} = 10^{13}\,\text{GeV}$. Panel (a) shows the signal settling into the radiation-dominated era ($w=1/3$) with the transition marked by $k\eta=k\eta_{\rm reh}$ where $\eta_{\rm reh}(T_{\rm reh})$ is given in Eq.~\eqref{etareh}, while Panel (b) details the behavior during the reheating epoch if the transition to radiation didn't take place.}
    \label{fig:bispec_evolution}
\end{figure}

\begin{figure}[t]
    \centering
    \begin{minipage}{0.48\linewidth}
        \centering
        \includegraphics[width=\linewidth]{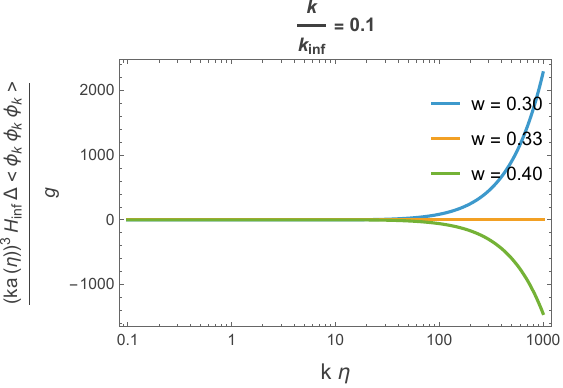}
        \captionsetup{style=centered_sub}\caption*{(a) Temporal evolution of the $\ev{\phi_k^3}_{T_{\rm reh}=10^{13}GeV}-\ev{\phi_k^3}_{T_{\rm reh}=10^{15}GeV}$.}
    \end{minipage}
    \hfill
    \begin{minipage}{0.48\linewidth}
        \centering
        \includegraphics[width=\linewidth]{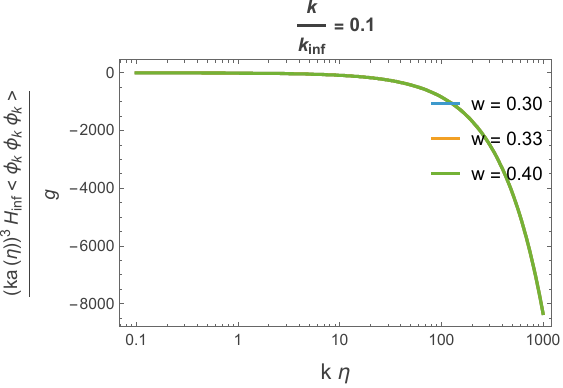}
        \captionsetup{style=centered_sub}\caption*{(b) The high-temperature limit ($T_{\text{re}} = 10^{15}$ GeV).}
    \end{minipage}

    \caption{Comparison of the bispectrum generated during a finite reheating phase against the standard instantaneous transition prediction. Panel (a) illustrates the growth of the discrepancy after the end of reheating. Panel (b) demonstrates the consistency check: as the reheating temperature approaches the scale of inflation ($T_{\text{re}} \approx 10^{15}$ GeV), the finite-duration effects vanish and the instantaneous result is recovered.}
    \label{fig:bispec-combined}
\end{figure}

The bispectrum during the reheating era depends sensitively on the equation of state of the reheating phase. The same technique can be used to study bispectrum during the radiation-dominated era. We can split the bispectrum into three contributions, depending on which era is used to evaluate the time integral and focus on them one by one. Since we already calculated the inflation part and reheating part in \eqref{conformally-coupled-during-inflation} and \eqref{conformally-coupled-during-reheating}. We just need to replace the scale factor in those results. Therefore, in this part we solely focus on radiation part:
\begin{align}\label{conformal:radiation_part}
    \overline{\ev{\phi_{k_1}\phi_{k_2}\phi_{k_3}}}_{\text{rad}}
    &=2g\Im{\int_{\eta_{\text{reh}}}^{\eta}D\eta' W_{k_1}(\eta, \eta')W_{k_2}(\eta,\eta')W_{k_3}(\eta,\eta')}\nt
    &=2g\Im{\frac{e^{-iK\eta}}{a(\eta)^3}\int_{\eta_{\text{reh}}}^{\eta}d\eta' a(\eta')^4\frac{1}{a(\eta')^3}\frac{e^{i(k_1+k_2+k_3)\eta'}}{8k_1k_2k_3}}\nt
    &=\frac{2ga_{\text{reh}}[-2\sin(K(\eta_{\rm reh}-\eta))+K\{(1+3w)\eta_{\rm reh}-3(1+w)\eta_{\rm inf}\}\cos{K(\eta_{\rm reh}-\eta)}]}{2a(\eta)^3k_1k_2k_3K^2(1+3w)\left(\eta _{\text{reh}}-3 \mu  \eta _{\inf }\right)}\nt
    &\qquad-\frac{2ga_{\text{reh}}K[(-1+3w)\eta_{\rm reh}+2\eta-3(1+w)\eta_{\inf}]}{8a(\eta)^3k_1k_2k_3K^2(1+3w)\left(\eta _{\text{reh}}-3 \mu  \eta _{\inf }\right)}.
\end{align}
We have explicitly retained the overall momentum scale $K$ in this final term to make it manifest that, in the limit $k_i \eta \to 0$, these contributions vanish relative to the leading-order pole. The full bispectra can be given as:
\begin{align}
    \overline{\ev{\phi_{k_1}\phi_{k_2}\phi_{k_3}}}&=-\frac{\pi}{8}g\frac{1}{k_{1}k_{2}k_{3}H_{\rm inf}a(\eta)^3}\cos{K\eta}+g\frac{1}{4k_{1}k_{2}k_{3}H_{\rm inf}a(\eta)^3}\ln{\abs{K\eta_{\inf}}}\sin{K\eta}\nt
    &\qquad+g\frac{1}{4k_{1}k_{2}k_{3}H_{\rm inf}a(\eta)^3}\gamma\sin{K\eta}-\frac{2g}{8a(\eta)^3H_{\rm inf}k_{1}k_{2}k_{3}K|\eta_{\rm inf}|} \left(\frac{1+3w}{2K\abs{\eta_{\text{inf}}}}\right)^{\frac{2}{1+3w}}\nt
    &\qquad\times\bigg[\cos\left(\frac{(\pi+3 K\eta_{\rm inf}(w+1))}{3 w+1}-K\eta\right)\Re{ \Gamma\left(3\mu,-i K\eta_{\rm reh}+3\mu iK\eta_{\rm inf},\frac{2i K\eta_{\rm inf}}{3 w+1}\right)}\nt
    &\qquad-\sin\left(\frac{(\pi+3 K\eta_{\rm inf}(w+1))}{3 w+1}-K\eta\right)\Im{ \Gamma\left(3\mu,-i K\eta_{\rm reh}+3\mu iK\eta_{\rm inf},\frac{2i K\eta_{\rm inf}}{3 w+1}\right)}\bigg]\nt
    &\qquad+\frac{2ga_{\text{reh}}[-2\sin(K(\eta_{\rm reh}-\eta)) + K \{(1+3w) \eta_{\rm reh} - 3(1+w)\eta_{\rm inf}\}\cos{K(\eta_{\rm reh}-\eta)}]}{2a(\eta)^3k_1k_2k_3K^2(1+3w)\left(\eta _{\text{reh}}-3 \mu\eta _{\inf }\right)}\nt
    &\qquad-\frac{2ga_{\text{reh}}K[(-1+3w)\eta_{\rm reh}+2\eta-3(1+w)\eta_{\inf}]}{8a(\eta)^3k_1k_2k_3K^2(1+3w)\left(\eta_{\text{reh}}-3 \mu\eta_{\inf }\right)}.
\end{align}
It takes the following form for long wavelength modes in the equilateral limit:

\begin{align}
    \overline{\ev{\phi_{k}\phi_{k}\phi_{k}}} &=-\frac{\pi}{8}g\frac{1}{k^3H_{\rm inf}a(\eta)^3}+g\frac{1}{4k^3H_{\rm inf}a(\eta)^3}\ln{\abs{3k\eta_{\inf}}}3k\eta+g\frac{1}{4k^3H_{\rm inf}a(\eta)^3}3\gamma k\eta\nt
    &\qquad-\frac{2ga_{\rm inf}(3k)^{-3\mu}}{8a(\eta)^3k^3}\left(\frac{1+3w}{2\abs{\eta_{\text{inf}}}}\right)^{\frac{2}{1+3w}}\bigg[\cos\left(\frac{\pi}{3 w+1}-3k\eta\right)\Re{\frac{\left(-i3k\eta_{\rm reh}+9\mu ik\eta_{\rm inf}\right)^{3\mu}}{3\mu}}\nt
    &\qquad-\sin\left(\frac{\pi}{3 w+1}-3k\eta\right)\Im{\frac{\left(-3ik\eta_{\rm reh}+9\mu ik\eta_{\rm inf}\right)^{3\mu}}{3\mu}}
    \bigg]\nt
    &\qquad-\frac{2ga_{\text{reh}}[2(k\eta_{\rm reh}-k\eta)-\{(1+3w)k\eta_{\rm reh}-3(1+w)k\eta_{\rm inf}\}]}{24a(\eta)^3k^5(1+3w)\left(\eta _{\text{reh}}-3 \mu  \eta _{\inf }\right)}\nt
    &\qquad-\frac{2ga_{\text{reh}}[(-1+3w)k\eta_{\rm reh}-2k\eta+3(1+w)k\eta_{\inf}]}{24a(\eta)^3k^5(1+3w)\left(\eta _{\text{reh}}-3 \mu  \eta _{\inf }\right)}.
\end{align}
Since the result depends over $\eta_{\rm reh}$ and $\eta_{\rm inf}$, the bispectrum is parametrized by $w$ and $T_{\rm reh}$ through \eqref{kinf} and \eqref{etareh}. At the leading order in the limit $k\eta\to0$, this reduces to
\begin{align}
    \overline{\ev{\phi_{k}\phi_{k}\phi_{k}}}=-\frac{\pi}{8}g\frac{1}{k^3H_{\rm inf}a(\eta)^3}.
\end{align}
We find that the leading behavior on superhorizon scale is identical to that of inflation however, the introduction of reheating era fundamentally changes the time evolution of bispectrum. Assuming the interaction channel remains active during reheating and persists post-thermalization, the resulting spectrum differs from the sudden transition case. The evolution of the bispectra during reheating dominates the subsequent radiation era evolution (see Fig.~\ref{fig:bispec_evolution}). This deviation from instantaneous reheating is quantified in Fig.~\ref{fig:bispec-combined}. 

\section{BEYOND CONFORMAL COUPLING}\label{section 3}

\subsection{Field Dynamics}
We now extend our analysis to a non–minimally coupled scalar field. Such coupling is assumed to be inevitable in the low energy effective theory for the scalar field when coupled with gravity. Non-minimal gravitational coupling has been extensively explored in the context of inflation \cite{faraoni1996nonminimal,tsujikawa2000power,komatsu1999complete,lucchin1986inflation,spokoiny1984inflation,futamase1989chaotic,shokri2021nonminimal,capozziello1994nother,nozari2008non,sarkar2022non,gomes2017inflation}, reheating \cite{bassett1998geometric,tsujikawa1999resonant,tsujikawa1999preheating,ema2017violent,dimopoulos2018non,figueroa2023lattice,opferkuch2019ricci,bettoni2022hubble,laverda2024ricci,figueroa2025geometric,laverda2024rise}, DM \cite{markkanen2017dark,fairbairn2019despicable,kainulainen2023tachyonic,Markkanen:2017edu,Kainulainen:2022lzp,Lebedev:2022vwf,Kolb:2023ydq,Ema:2018ucl,PhysRevD.108.123509,cembranos2024vector,kolb2021completely,capanelli2024gravitational,capanelli2024runaway}, and dark energy \cite{setare2010non,sami2012cosmological,kase2020scalar,ye2025hints}. The action for such a system is given by:
\begin{equation}
S_\phi = -\int d^4x\,\sqrt{-g}\,\Big[ \frac12g^{\mu\nu}\partial_\mu\phi\,\partial_\nu\phi +\frac{1}{2}m_{\phi}\phi^2+ \frac12\xi \mathcal{R}\phi^2+\frac{g}{3!} \phi^3 \Big].
\end{equation}
We introduce the canonically normalised mode function as before
\begin{equation}
    X_k(\eta)=a(\eta)f_k(\eta),
\end{equation}
such that the EoM in momentum space can be written as:
\begin{equation}
 X_k'' + \Big[k^2 +(\xi-\tfrac16)a(\eta)^2\mathcal{R}(\eta) \Big]X_k +a^2m^2_{\phi}f_k= 0.
\end{equation}
During inflation, with $a(\eta)= -\frac{1}{H_{\rm inf}\eta}$, $m_{\phi}/H_{\rm inf}\ll1$ and $\mathcal{R}= 12H_{\rm inf}^2$, the EoM reduces to:
\begin{equation}
    X''_{k}+\underbrace{\left[k^{2}-\frac{2}{\eta^{2}}(1-6\xi)\right]}_{\omega_k^{2}}X_{k}=0
\end{equation}
which is a Bessel-type equation for each mode $k$ and 
the solution would be \cite{bowman2012introduction},
\begin{equation}
    X_{k}^{\rm inf}(\eta)=C_1\sqrt{\abs{\eta}}J_{\nu_1}(k\abs{\eta})+C_2\sqrt{\abs{\eta}}Y_{\nu_1}(k\abs{\eta});\qquad\text{where }\nu_1=\frac{1}{2}\sqrt{9-48\xi}\label{inflation_mode_function}
\end{equation}
Following \eqref{scale-factor}, the mode function during reheating is found by solving the following:
\begin{align}
X''_{k}+\Bigg[k^2-\frac{2(1-3w)}{(1+3w)^2}\frac{1}{\left[\eta-\frac{3\eta_{\text{inf}}(1+w)}{1+3w}\right]^2}(1-6\xi)\Bigg]X_{k}&=0
\end{align}
We can note that for $\xi=\nicefrac{1}{6}$, we again have the same set of solution we found earlier. Let us define $$\mu=\frac{1+w}{1+3w}\qquad\text{and}\qquad\alpha^2=\frac{2(1-6\xi)(1-3w)}{(1+3w)^2}$$
and, consider the following:
\begin{align}
	X''_{k}(\eta)+\Bigg[k^2-\frac{\alpha^2}{\left[\eta-3\mu\eta_{\text{inf}}\right]^2}\Bigg]X_{k}(\eta)&=0 .
\end{align}
The solution to above can be given as:
\begin{equation}
    X_{k}^{\rm reh}(\eta)=C_{3}\sqrt{\tilde{\eta}}J_{\nu_2}(k\tilde{\eta})+C_{4}\sqrt{\tilde{\eta}}Y_{\nu_2}(k\tilde{\eta});\qquad\text{where }\nu_2
=\frac{\sqrt{3[3(1-w)^2+16\xi(3w-1)]}}{|2(1+3w)|}\label{reheating_mode_function}
\end{equation}
It is worth noting that for $w > -1/3$, the parameter satisfies $\mu > 1/3$ and for $w=0$ have $\nu_2=\nu_1$. Consequently, the shifted conformal time variable, $\tilde{\eta}=\eta-3\mu\eta_{\text{inf}}$ remains strictly positive throughout the domain of interest. The connection between $\tilde{\eta}$ and the physical evolution of the scales is made manifest through the comoving Hubble radius, which behaves as:
\begin{equation}
    \frac{1}{\cal H}=\frac{(1+3w)}{2}\underbrace{\left(\eta-\eta_{\rm inf}+\frac{2|\eta_{\rm inf}|}{1+3w}\right)}_{>0}
\end{equation}
The sign of the prefactor determines whether modes enter or exit the horizon during this era:
\begin{equation}
\frac{d}{d\eta} \left( \frac{1}{\mathcal{H}} \right) =
    \begin{cases} 
        < 0 & \text{for } w < -\frac{1}{3} \quad (\text{modes exit}) \\[6pt]
        > 0 & \text{for } w > -\frac{1}{3} \quad (\text{modes enter})
    \end{cases}
\end{equation}
Therefore the boundary condition that we impose assumes a bound on EoS during their respective era. With this caveat in mind, we impose the boundary condition assuming $w=-1$ during inflation and $w>-1/3$ during reheating:
\begin{alignat}{2}
    X_{k}^{\rm inf}(\eta) &\xrightarrow{|\eta|\to\infty} \frac{e^{-ik\eta}}{\sqrt{2k}} &&\qquad  (\text{for } \eta < \eta_{\text{inf}}) \label{adiabatic_inf} \\
    X_{k}^{\rm reh}(\eta) &\xrightarrow{\eta\to\infty} \frac{e^{-ik\eta}}{\sqrt{2k}}  &&\qquad  (\text{for } \eta > \eta_{\text{inf}}) \label{adiabatic_reh}
\end{alignat}
In the limit $k|\eta| \gg 1$, the mode solution \eqref{inflation_mode_function} becomes
\begin{equation}\label{inf_lim}
X_k^{\rm inf}(\eta) \approx \frac{1}{\sqrt{2\pi k}}
\bigg[(C_1 - i C_2) e^{-i (k\eta + \pi/4 + \pi\nu_1/2)}
+ \underbrace{(C_1 + i C_2)}_{=0} e^{i (k\eta + \pi/4 + \pi\nu_1/2)}\bigg].
\end{equation}
Comparing \eqref{inf_lim} with \eqref{adiabatic_inf}, we obtain

\begin{equation}
C_1 = \frac{\sqrt{\pi}}{2} e^{i(\pi/4 + \pi\nu_1/2)},
\qquad
C_2 = iC_1=i \frac{\sqrt{\pi}}{2} e^{i(\pi/4 + \pi\nu_1/2)}.
\end{equation}
This fixes the unknown coefficients for mode function during inflation and thus the resulting adiabatic vacuum solution during inflation can be given as:
\begin{equation}
    a(\eta)f^{\text{inf}}_k(\eta)=X_{k}^{\text{inf}}=\frac{\sqrt{-\pi\eta}}{2}e^{i(\nicefrac{\pi}{4}+\nicefrac{\pi\nu_1}{2})}H_{\nu_1}^{(1)}(-k\eta).
\end{equation}
This solution has infrared divergence in the long–wavelength limit whenever $\nu$ is real. In particular, $\nu>0$ for $\xi<3/16$, but we only have $\nu>1/2$ (a stronger IR divergence) when $\xi<1/6$. In physical terms, this means the effective time–dependent frequency
\begin{equation}
    \omega_k^{2}(\eta)=k^{2}-\frac{2}{\eta^{2}}(1-6\xi)
\end{equation}
becomes negative in the $k\to 0$ limit, i.e. $\omega_k^{2}<0$ for sufficiently long wavelengths\cite{fairbairn2019despicable}. This regime is sometimes referred to as ``tachyonic'' in the loose sense that the curved background amplifies the intrinsic flat-space infrared behaviour. 
This behaviour is made precise by the late-time asymptotics of the mode function. In the super-horizon limit $k\to0$, the Hankel function asymptotics give
\begin{align}\label{long-approx}
X_{k}^{\inf}\sim -i\sqrt{\frac{|\eta|}{4\pi}}e^{\frac{i\pi}{4}}\left[e^{i\frac{\pi\nu_1}{2}}\Gamma(\nu_1)\left(\frac{-k\eta}{2}\right)^{-\nu_1}+e^{-i\frac{\pi\nu_1}{2}}\Gamma(-\nu_1)\left(\frac{-k\eta}{2}\right)^{\nu_1}\right],
\end{align}
leading to:
\begin{equation}
f_{k}^{\rm inf} \approx -iH_{\rm inf}\sqrt{\frac{1}{4\pi k^3}}e^{\frac{i\pi}{4}}\left[e^{i\frac{\pi\nu_1}{2}}\Gamma(\nu_1)\left(\frac{-k\eta}{2}\right)^{\frac{3}{2}-\nu_1}+e^{-i\frac{\pi\nu_1}{2}}\Gamma(-\nu_1)\left(\frac{-k\eta}{2}\right)^{\frac{3}{2}+\nu_1}\right],
\end{equation}
so the field decomposes into two branches in the long wavelength limit $k\to 0$,
\begin{equation}\label{boundary:decomp}
\phi(\eta,\vec{x})\xrightarrow{k \to 0}\eta^{\Delta_-^{\rm inf}}\mathcal{O}^{\rm inf}+\eta^{\Delta_+^{\rm inf}}\widetilde{\mathcal{O}}^{\rm inf},
\end{equation}
where the scaling dimensions are
\begin{equation}
    \Delta_\pm^{\rm inf} = \frac{3}{2} \pm \nu_1.
\end{equation}
For $\xi=\frac{1}{6}$, the growing part has $\Delta_{-}=1$ while decaying part has $\Delta_{+}=2$ and the mode function reduces to $X_k^{\rm inf} =\frac{1}{\sqrt{2k}}(1-ik\eta)\approx\frac{1}{\sqrt{2k}}$. Comparing this to the asymptotic expansion in \eqref{long-approx}, the relative infrared scaling between the non-minimally coupled and conformally coupled fields is given by:
\begin{equation}\label{tachyonic-scaling-both}
\lim_{k\to0}\frac{|X_k^{\rm inf\,\xi\ne\frac{1}{6}}|}{|X_k^{\rm inf\,\xi=\frac{1}{6}}|} \sim \mathcal{A}_k k^{-(\nu_1 - 1/2)} + \mathcal{B}_k k^{\nu_1 + 1/2},
\end{equation}
where $\mathcal{A}_k$ and $\mathcal{B}_k$ are $k-$independent coefficients corresponding to growing and decaying branch. The first term in \eqref{tachyonic-scaling-both} corresponds to the growing mode which undergoes tachyonic enhancement for $\nu_1>\frac{1}{2}$
enhancing the long-wavelength modes relative to their conformal modes. For $0<\nu_1<1/2$ the situation reverses: the curvature term suppresses the IR divergence, effectively ``stiffening'' the field rather than destabilizing it.
The second term represents the decaying mode, which remains subdominant as $k\to0$. \footnote{A related interpretation arises in the dS/CFT correspondence, where the de Sitter bulk is conjectured to be dual to a three-dimensional Euclidean conformal field theory~\cite{odintsov2001quantum,strominger2001ds,mcinnes2002exploring,Dey:2024zjx}. In that context, the late-time boundary limit $\eta\to0$ during inflation coincides with the infrared limit $k\to0$, since dilatations act as exact isometries of the background and fix the scaling of bulk fields. This equivalence ceases to hold during reheating, where dilatations are no longer an isometry, and the infrared behavior must instead be determined by matching onto reheating mode functions.}

To complete the description of the field's evolution, we fix the remaining coefficients of reheating mode function by imposing the Bunch Davies boundary condition in the limit $k\gg \mathcal{H}$ during reheating:
\begin{align}
	X_{k}^{\rm reh}(\eta)&\approx\sqrt{\frac{2}{\pi k}}\left[C_3\cos(k\tilde{\eta}-\nicefrac{\pi}{4}-\nicefrac{\pi\nu_2}{2})+C_4\sin(k\tilde{\eta}-\nicefrac{\pi}{4}-\nicefrac{\pi\nu_2}{2})\right]\nt
	&= \frac{1}{\sqrt{2\pi k}}\bigg[(C_3+iC_4)e^{-i(k\tilde{\eta}-\nicefrac{\pi}{4}-\nicefrac{\pi\nu_2}{2})}+\underbrace{(C_3-iC_4)}_{=0}e^{i(k\tilde{\eta}-\nicefrac{\pi}{4}-\nicefrac{\pi\nu_2}{2})}\bigg].
\end{align}
Comparing it to \eqref{adiabatic_reh}, we obtain:
\begin{equation}
    C_3=\frac{\sqrt{\pi}}{2}e^{-3\mu\,ik\eta_{\rm inf}}e^{-i(\nicefrac{\pi}{4}+\nicefrac{\pi\nu_2}{2})}\qquad\&\qquad C_4=-iC_3=-i\frac{\sqrt{\pi}}{2}e^{-3\mu\,ik\eta_{\rm inf}}e^{-i(\nicefrac{\pi}{4}+\nicefrac{\pi\nu_2}{2})}.
\end{equation}
Hence, the adiabatic vacuum solution during reheating is also Hankel with order $\nu_2$:
\begin{equation}
    X_{k}^{\rm reh}(\eta)=\frac{\sqrt{\pi\tilde{\eta}}}{2}e^{-i\left[\nicefrac{\pi}{4}+\nicefrac{\pi\nu_2}{2}+3\mu\,k\eta_{\text{inf}}\right]}H_{\nu_2}^{(2)}(k\tilde{\eta}).
\end{equation}
These mode functions describe the particle-like solutions in the infinite past and infinite future, as well as the subhorizon modes at any intermediate time. Superhorizon modes, on the other hand, do not have a particle interpretation, so we do not treat them as particle solutions. One can similarly take the long wavelength limit as before:
\begin{align}
X_{k}^{\rm reh}\sim i\sqrt{\frac{\tilde\eta}{4\pi}}e^{-i(\frac{\pi}{4}+3\mu k\eta_{\inf})}\left[e^{-i\frac{\pi\nu_2}{2}}\Gamma(\nu_2)\left(\frac{k\tilde\eta}{2}\right)^{-\nu_2}+e^{i\frac{\pi\nu_2}{2}}\Gamma(-\nu_2)\left(\frac{k\tilde\eta}{2}\right)^{\nu_2}\right],
\end{align}
leading to
\begin{align}
    f_k^{\rm reh} &= iH_{\rm inf}|\eta_{\rm inf}|\left(\frac{2|\eta_{\rm inf}|}{1+3w}\right)^{\frac{2}{1+3w}}\sqrt{\frac{1}{4\pi k^3}}e^{-i(\frac{\pi}{4}+3\mu k\eta_{\inf})}\left[e^{-i\frac{\pi\nu_2}{2}}\Gamma(\nu_2)\left(\frac{k\tilde\eta}{2}\right)^{\frac{3}{2}-\nu_2}+e^{i\frac{\pi\nu_2}{2}}\Gamma(-\nu_2)\left(\frac{k\tilde\eta}{2}\right)^{\frac{3}{2}+\nu_2}\right].
\end{align}
This generally decomposes into:
\begin{equation}\label{boundary2:decomp}
\phi(\eta,\vec{x})\xrightarrow{k \to 0}\eta^{\Delta_-^{\rm reh}}\mathcal{O}^{\rm reh}+\eta^{\Delta_+^{\rm reh}}\widetilde{\mathcal{O}}^{\rm reh},
\end{equation}
where 
\begin{equation}
    \Delta_{\pm}^{\rm reh}=\frac{3}{2}\pm\nu_2.
\end{equation}
This leads to similar expression as \eqref{tachyonic-scaling-both}:
\begin{equation}
\lim_{k\to0}\frac{|X_k^{\rm reh\,\xi\ne\frac{1}{6}}|}{|X_k^{\rm reh\,\xi=\frac{1}{6}}|} \sim \mathcal{C}_k k^{-(\nu_2 - 1/2)} + \mathcal{D}_k k^{\nu_2 + 1/2},
\end{equation}
where $\mathcal{C}_k$ and $\mathcal{D}_k$ are $k-$independent coefficients corresponding to growing and decaying branch. The decomposition of field into growing and decaying mode in long wavelength limit as shown in \eqref{boundary:decomp} and \eqref{boundary2:decomp} are written in different basis. The tachyonic enhancement of these modes within their respective bases is encoded in $\Delta_{\pm}$. These two bases are related by a Bogoliubov transformation, the details of which are discussed in the following section. 

\subsection{Matching Conditions for Multi-Phase Transitions}
\subsubsection{From inflation to reheating transition}
\noindent
The crucial difference with conformally coupled scalar field is that the non-minimally coupled scalar field undergoes particle production and therefore on super-horizon scale Bogoliubov mixing becomes very relevant. Having made the choice of adiabatic vacuum during inflation and reheating for non-minimally coupled scalar field; the remaining step is simply to match these two vacua across the transition surface. The Bogoliubov coefficients then quantify how the sudden transition relates the vacuum states of the two cosmological phases. Our focus will be on the super-horizon modes at the end of inflation, which subsequently re-enter the growing co-moving Hubble horizon during reheating or the radiation-dominated era. The two basis defined during the respective era are connected by the following transformation:
$f^{\rm inf}_k(\eta_{\inf})=\alpha_kf^{\rm reh}_k(\eta_{\inf})+\beta_k f^{\rm reh*}_k(\eta_{\inf}).$
Depending on the values of the non-minimal coupling $\xi$ and the equation-of-state parameter $w$, the parameters $\nu_1$ and $\nu_2$ can be real or imaginary, which in turn affects both the super-horizon and the asymptotic behavior of the mode functions. In general, the mode function can be written as:
\begin{align}
    a(\eta)\hat{\phi}_k = \begin{cases}X^{\rm inf}\hat{a}_k + X^{\rm inf*}\hat{a}^\dagger_{-k},\qquad &\eta<\eta_{\rm inf},\\
     (\alpha_{k}X^{\rm reh}+\beta_{k}X^{\rm reh*})\hat{a}_k + (\alpha_{k}X^{\rm reh}+\beta_{k}X^{\rm reh*})^*\hat{a}^\dagger_{-k},\qquad &\eta\ge \eta_{\inf}.
    \end{cases}
\end{align}
Imposing continuity of field and its conjugate momenta, we get the following equations for determining $\alpha_k$ and $\beta_k$:
\begin{align}
    X^{\text{inf}}\bigg\lvert_{\eta=\eta_{\text{inf}}} = \alpha_{k}X^{\rm reh}+\beta_{k}X^{\rm reh*}\bigg\lvert_{\eta=\eta_{\text{inf}}}~~;~~
    X^{\text{inf}'}\bigg\lvert_{\eta=\eta_{\text{inf}}} = \alpha_{k}X^{\rm reh'}+\beta_{k}X^{\rm reh*'}\bigg\lvert_{\eta=\eta_{\text{inf}}}
\end{align}
Solving the above system of equation yields:\cite{de1993spectrum}
\begin{align}
	\alpha_k &= i(X^{\text{inf}'}_{k}X^{\rm reh*}_{k}-X^{\text{inf}}_{k}X^{\rm reh'*}_{k}),\nt
	\beta_k &= -i(X^{\text{inf}'}_{k}X^{\rm reh}_{k}-X^{\text{inf}}_{k}X^{\rm reh'}_{k}),
\end{align}
where $'$ denotes derivative with respect to conformal time. We used the following condition for normalization in comoving coordinates:
\begin{align}
	\begin{bmatrix}
		X^{\rm reh}_{k} & X^{\rm reh*}_{k}\\
		\partial_\eta X^{\rm reh}_{k} & \partial_\eta X^{\rm reh*}_{k}
	\end{bmatrix}=i ,
\end{align}
and by using the following identity;
$$\frac{\partial H_{\nu}^{(1,2)}(k\eta)}{\partial \eta }=\frac{1}{2} k \left[H_{\nu-1}^{(1,2)}(k \eta)-H_{\nu+1}^{(1,2)}(k\eta)\right], $$
the bogolibov coefficients for inflation to arbitrary EoS transition are then given as:

\begin{align}\label{bogoliubov_during_reheating}
    \alpha_k    &= \frac{\pi e^{\frac{1}{2} i \left(6 k \mu  \eta _{\text{inf}}+\pi  \left(\nu _1+\nu^* _2\right)\right)} (-k \eta _{\text{inf}})(3 \mu -1)\left[H_{\nu _1-1}^{(1)}\left(-k \eta _{\text{inf}}\right)-H_{\nu _1+1}^{(1)}(-k \eta _{\text{inf}})\right] H_{\nu _2^*}^{(1)}\left(k \eta _{\text{inf}} (1-3 \mu )\right)}{8 \sqrt{3\mu-1}}\nt
    &\qquad-\frac{\pi e^{\frac{1}{2} i \left(6 k \mu  \eta _{\text{inf}}+\pi  \left(\nu _1+\nu _2^*\right)\right)}k\eta _{\text{inf}} (3 \mu -1) H_{\nu _1}^{(1)}\left(-k \eta _{\text{inf}}\right)\left[H_{\nu _2^*-1}^{(1)}\left(k\eta _{\text{inf}}(1-3 \mu )\right)-H_{\nu _2^*+1}^{(1)}(k\eta _{\text{inf}}(1-3 \mu ))\right]}{8\sqrt{3\mu-1}}\nt
    &\qquad+3 \mu  \frac{\pi e^{\frac{1}{2} i \left(6 k \mu  \eta _{\text{inf}}+\pi  \left(\nu _1+\nu _2^*\right)\right)}H_{\nu _1}^{(1)}(-k\eta_{\inf})H_{\nu _2^*}^{(1)}\left(k \eta _{\text{inf}}(1-3 \mu )\right)}{8\sqrt{3\mu-1}},\nt
    \\
    \beta_k&=-i\frac{\pi e^{-\frac{1}{2} i \left(6 k \mu  \eta _{\text{inf}}+\pi  \left(\nu_2-\nu _1\right)\right)} k \eta _{\text{inf}}(3 \mu -1)\left[H_{\nu _1-1}^{(1)}\left(-k \eta      _{\text{inf}}\right)-H_{\nu _1+1}^{(1)}(-k \eta _{\text{inf}})\right] H_{\nu _2}^{(2)}\left(k     \eta _{\text{inf}} (1-3 \mu )\right)}{8 \sqrt{3\mu-1}}\nt
    &\qquad-i\frac{\pi e^{-\frac{1}{2} i \left(6 k \mu  \eta _{\text{inf}}+\pi  \left(\nu _2-\nu _1\right)\right)}k\eta _{\text{inf}} (3 \mu -1) H_{\nu _1}^{(1)}\left(-k \eta _{\text{inf}}\right)\left[H_{\nu _2-1}^{(2)}\left(k\eta _{\text{inf}}(1-3 \mu )\right)-H_{\nu _2+1}^{(2)}(k\eta _{\text{inf}}(1-3 \mu ))\right]}{8\sqrt{3\mu-1}}\nt
    &\qquad+3 \mu i\frac{\pi e^{-\frac{1}{2} i \left(6 k \mu  \eta _{\text{inf}}+\pi  \left(\nu _2-\nu _1\right)\right)} H_{\nu_1}^{(1)}(-k \eta _{\text{inf}})H_{\nu _2}^{(2)}(k \eta _{\text{inf}}(1-3 \mu ))}{8\sqrt{3\mu-1}}.
\end{align}
For $\xi=\frac{1}{6}$ or $\nu_1=\nu_2=\tfrac{1}{2}$, this reduces to $\alpha_k =1~,~    \beta_k =0$, in agreement with conformally coupled scenario.
\subsubsection{From inflation to reheating to radiation transition}
Building on the two-stage case, we now consider a single scalar field $\hat{\phi}(\eta)$ propagating through three successive stages of expansion, each defined by a distinct scale factor and its corresponding mode equation. In each phase, the mode equation admits two linearly independent solutions, allowing the quantized field to be expanded as
\begin{align}
    a(\eta)\hat{\phi}_k = \begin{cases}X^{\rm inf}(\eta)\hat{a}_k + X^{\rm inf*}(\eta)\hat{a}^\dagger_{-k},\qquad &\eta<\eta_{\rm inf},\\
     X^{\rm reh}(\eta)\hat{b}_k + X^{\rm reh*}(\eta)\hat{b}^\dagger_{-k},\qquad &\eta_{\rm inf}<\eta<\eta_{\rm reh},\\
    X^{\rm rad}(\eta)\hat{c}_k + X^{\rm rad*}(\eta)\hat{c}^\dagger_{-k},\qquad &\eta>\eta_{\rm reh}.
    \end{cases}
\end{align}
With only a single instantaneous transition, extracting the Bogoliubov coefficients is straightforward: one first imposes the Bunch Davies boundary condition to determine the mode function and then match the mode function and its first derivative across the transition surface to determine the Bogoliubov coefficients. Introducing a second transition needs to be studied with caution, since modes may not re-enter the horizon during the intermediate era. The natural boundary conditions are instead imposed at the onset of inflation and deep in the radiation era. For clarity, we briefly outline how all coefficients are determined from the junction conditions.

The mode functions must satisfy the canonical Wronskian normalization, which in general requires them to be complex. Under this consideration, the mode function is a linear combination of those linearly independent complex solutions with undetermined coefficients. We impose the adiabatic vacuum, i.e., the Bunch-Davies vacuum, as our initial condition to fix the physical state. Note that the vacuum is only unambiguously defined for subhorizon modes \cite{mukhanov2007introduction}; therefore, we define the vacua for all modes on the spacelike hypersurface at the infinite past during inflation, and at the infinite future during the radiation era, where all the modes are subhorizon. 

However, at a generic intermediate time, one cannot, in general, define a unique state that behaves as a vacuum for all modes. Instead, we focus on a state that serves as a vacuum only for subhorizon modes. At the inflation-to-reheating transition, this is defined by
\begin{equation}
b_{k}\ket{0(\eta_{\rm inf}^{+})}=0 
\qquad \text{for } k|\eta_{\rm inf}^+|\gg 1 ,
\end{equation}
where \(\eta_{\rm inf}^{+}\) denotes a time immediately after the end of inflation. This does not affect our analysis, since we are not concerned with the physics in this intermediate vacuum, but rather with using its mathematically well-defined basis to track the evolution of the initial state defined in the infinite past. Similarly for radiation dominated era, we have 
$$c_{k}\ket{0(\eta^+_{\rm reh})}=0\qquad \text{for}~~k|\eta_{\rm reh}^+|\gg1 .$$
This requires the mode functions to form an oscillatory basis given as
\begin{align}\label{mode_function}
	X_{\vec{k}}=
	\begin{cases}
		\frac{\sqrt{-\pi\eta}}{2}e^{i(\nicefrac{\pi}{4}+\nicefrac{\pi\nu_1}{2})}H_{\nu_1}^{(1)}(-k\eta) &\text{when } \eta<\eta_{\text{inf}}\\
		\frac{\sqrt{\pi\tilde{\eta}}}{2}e^{-i\left[\nicefrac{\pi}{4}+\nicefrac{\pi\nu_2}{2}+3\mu k\eta_{\text{inf}}\right]}H_{\nu_2}^{(2)}(k\tilde{\eta})&\text{when } \eta_{\text{inf}}<\eta<\eta_{\text{reh}}\\
        \frac{1}{\sqrt{2k}}e^{-ik\eta} &\text{when } \eta>\eta_{\text{reh}} ,
	\end{cases}
\end{align}
which satisfies the condition that in the limit $aH/k \to 0$, the solutions reduce to their flat-space counterparts \cite{stewart1993more,Parker:1971pt}. However, since the mode functions evolve through a curved spacetime, the vacuum state defined at the onset of reheating does not coincide with the initial Bunch-Davies vacuum. To account for this particle production, we relate the solutions in the reheating era to the operators of the asymptotic past using Bogoliubov transformations: 
\begin{align}
    a(\eta)\hat{\phi}&=X^{\rm reh}(\eta)\hat{b}_k+X^{\rm reh*}(\eta)\hat{b}^{\dagger}_{-k} \nt
    &=X^{\rm reh}(\eta)[\alpha^{(1)}_k\hat{a}_k+\beta^{(1)*}_k\hat{a}_{-k}^\dagger]+X^{\rm reh}(\eta)^*[\beta^{(1)}_k\hat{a}_k+\alpha ^{(1)*}_k\hat{a}_{-k}^\dagger]\nt
    &=[\alpha^{(1)}_k X^{\rm reh}+\beta^{(1)}_kX^{\rm reh*}]\hat{a}_k+[\alpha^{(1)}_k X^{\rm reh}+\beta^{(1)}_{k} X^{\rm reh*}]^{*}\hat{a}_{-k}^\dagger .
\end{align}
The same issue persists in the radiation-dominated era: even if we fix the vacuum in the asymptotic future, there is no physical requirement that this state coincides with the vacuum defined in the asymptotic past. To maintain consistency across these epochs, the radiation-era mode functions must therefore be related to the initial inflationary Bunch-Davies solution through a Bogoliubov transformation.
\begin{align}
    a(\eta)\hat{\phi}_k
    &= X^{\rm rad}(\eta)\hat{c}_k + X^{\rm rad*}(\eta)\hat{c}_{-k}^\dagger \nt
    &= X^{\rm rad}(\eta)\left(\alpha^{(2)}_k\hat{a}_k + \beta^{(2)*}_k \hat{a}_{-k}^\dagger\right)
       + X^{\rm rad*}(\eta)\left(\beta^{(2)}_k \hat{a}_k + \alpha^{(2)*}_k \hat{a}_{-k}^\dagger\right) \nt
    &= \left(\alpha^{(2)}_k X^{\rm rad} + \beta^{(2)}_k X^{\rm rad*}\right)\hat{a}_k
       + \left(\alpha^{(2)}_k X^{\rm rad} + \beta^{(2)}_k X^{\rm rad*}\right)^* \hat{a}_{-k}^\dagger.
\end{align}
Hence the quantum field can be written in a unified form across all three eras:
\begin{align}
    a(\eta)\hat{\phi}_k = \begin{cases} X^{\rm inf}(\eta)\hat{a}_k + X^{\rm inf*}(\eta)\hat{a}_{-k}^\dagger,
       & \eta < \eta_{\rm inf}, \\\\
     [\alpha^{(1)}_k X^{\rm reh}+\beta^{(1)}_kX^{\rm reh*}]\hat{a}_k+[\alpha^{(1)}_k X^{\rm reh}+\beta^{(1)}_kX^{\rm reh*}]^*\hat{a}^{\dagger}_{-k},
        &\eta_{\rm inf} < \eta < \eta_{\rm reh}, \\\\
    [\alpha^{(2)}_k X^{\rm rad} + \beta^{(2)}_k X^{\rm rad*}]\hat{a}_k
       + [\alpha^{(2)}_k X^{\rm rad} + \beta^{(2)}_k X^{\rm rad*}]^* \hat{a}_{-k}^\dagger,
        &\eta > \eta_{\rm reh}.
        \end{cases}
\end{align}
where $a_k$ is the annihilation operator mapping the vacuum to the null vector in Fock space when the modes are deep into the horizon during inflation:
\begin{equation}
   \hat  a_k \ket{0} = 0,
\end{equation}
For $\xi=\tfrac{1}{6}$ in \eqref{mode_function}, we have $\nu_1=\nu_2=\frac{1}{2}$, which can be simplified using
$$H_{\nicefrac{1}{2}}^{(2)}(k\eta)=i\sqrt{\frac{2}{\pi k\eta}}e^{-ik\eta}\qquad\text{and}\qquad \tilde\eta=\eta-3\mu \eta_{\rm inf}.$$
Thus, the solutions reduces to earlier case:
\begin{align}
	X_{\vec{k}}=
	\begin{cases}
		\frac{1}{\sqrt{2k}}e^{-ik\eta}&\text{when } \eta<\eta_{\text{inf}}\\
		\frac{1}{\sqrt{2k}}e^{-ik\eta} &\text{when } \eta_{\text{inf}}<\eta<\eta_{\text{reh}}\\
        \frac{1}{\sqrt{2k}}e^{-ik\eta} &\text{when } \eta>\eta_{\text{reh}} .
	\end{cases}
\end{align}
\subsection{Bogoliubov Coefficients and Particle Production}
\begin{figure}[h]
    \centering
    \begin{minipage}{0.48\linewidth}
        \centering
        \includegraphics[width=\linewidth]{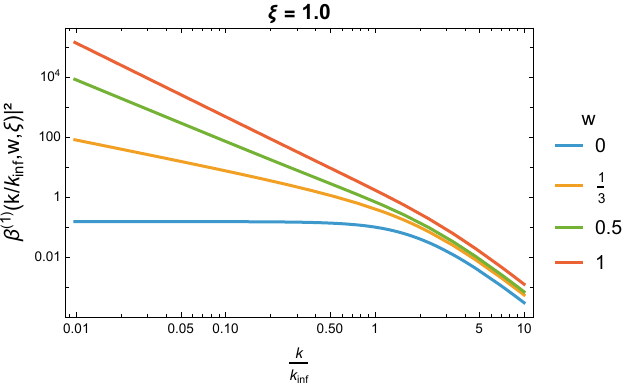}
        \captionsetup{style=centered_sub}\caption*{(a) at $\eta=\eta_{\text{inf}}$}
    \end{minipage}
    \hfill
    \begin{minipage}{0.48\linewidth}
        \centering
        \includegraphics[width=\linewidth]{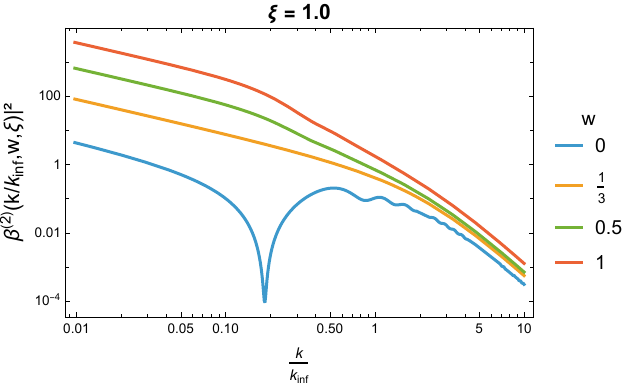}
        \captionsetup{style=centered_sub}\caption*{(b) at $\eta=\eta_{\text{reh}}$}
    \end{minipage}
    \begin{minipage}{0.48\linewidth}
        \centering
        \includegraphics[width=\linewidth]{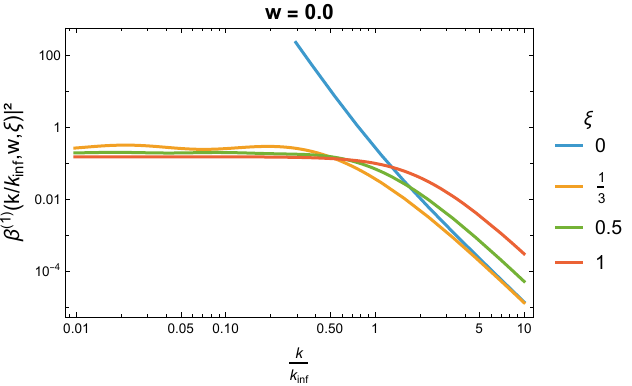}
        \captionsetup{style=centered_sub}\caption*{(c) at $\eta=\eta_{\text{inf}}$}
    \end{minipage}
    \hfill
    \begin{minipage}{0.48\linewidth}
        \centering
        \includegraphics[width=\linewidth]{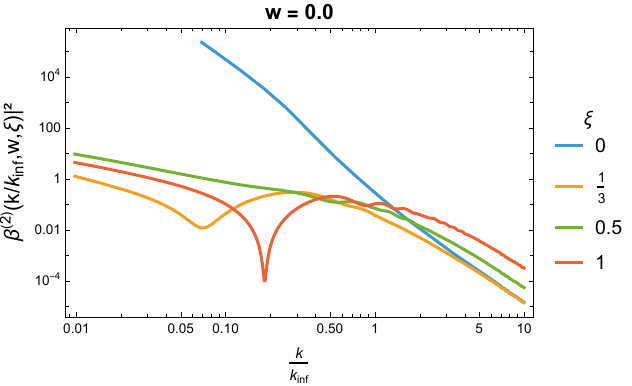}
        \captionsetup{style=centered_sub}\caption*{(d) at $\eta=\eta_{\text{reh}}$}
    \end{minipage}
    \caption{The plots show the spectrum of the Bogoliubov coefficient squared, $|\beta_k|^2$, representing particle production induced by non-adiabatic transitions in the presence of non-minimal coupling $\xi$. (a,c) The spectrum computed at the end of inflation ($\eta = \eta_{\mathrm{inf}}$) captures the particle creation resulting solely from the first transition (Inflation $\to$ Reheating). (b,d) The final spectrum at the end of the reheating epoch ($\eta = \eta_{\mathrm{reh}}$), which incorporates the additional contributions from the second transition (Reheating $\to$ Radiation). The distinct curves correspond to different values of the reheating equation of state $w$ for the respective $\xi$ in (a,b) and vice-versa in (c,d).}
    \label{fig:beta2}
\end{figure}

The objective of this section is to compute the corresponding Bogoliubov coefficients and express the mode functions in each regime. We can express the mode function in each era as:

\begin{align}
	a(\eta)f_{k} =\begin{cases}
		X^{\text{inf}}&\text{when }\eta<\eta_{\text{inf}}\\
		\alpha_{k}^{(1)}X^{\rm reh}+\beta_{k}^{(1)}X^{\rm reh*} &\text{when }\eta_{\text{inf}}<\eta<\eta_{\text{reh}}\\
        \alpha_{k}^{(2)}X^{\rm rad}+\beta_{k}^{(2)}X^{\rm rad*}&\text{when }\eta>\eta_{\text{reh}},
	\end{cases}
\end{align}
with the following conditions imposed over them:
\begin{align}
    X^{\text{inf}}\bigg\lvert_{\eta=\eta_{\text{inf}}} &= \alpha_{k}^{(1)}X^{\rm reh}+\beta_{k}^{(1)}X^{\rm reh*}\bigg\lvert_{\eta=\eta_{\text{inf}}}\nt
    X^{\text{inf}'}\bigg\lvert_{\eta=\eta_{\text{inf}}} &= \alpha_{k}^{(1)}X^{\rm reh'}+\beta_{k}^{(1)}X^{\rm reh*'}\bigg\lvert_{\eta=\eta_{\text{inf}}} .
\end{align}
Solving the above system of equation representing the continuity of field and its conjugate momenta, we find the Bogoliubov coefficient corresponding to the inflation-to-reheating transition:

\begin{align}\label{bogo1}
	\alpha_k^{(1)} &= i(X^{\text{inf}'}_{k}X^{\rm reh*}_{k}-X^{\text{inf}}_{k}X^{\rm reh'*}_{k}),\nt
	\beta_k^{(1)} &= -i(X^{\text{inf}'}_{k}X^{\rm reh}_{k}-X^{\text{inf}}_{k}X^{\rm reh'}_{k}) ,
\end{align}
where $'$ denotes derivative with respect to conformal time. We used the following condition for normalization in comoving coordinates for simplifying the expression further.
\begin{align}
	\begin{bmatrix}
		X^{\rm reh}_{k} & X^{\rm reh*}_{k}\\
		\partial_\eta X^{\rm reh}_{k} & \partial_\eta X^{\rm reh*}_{k}
	\end{bmatrix}=i .
\end{align}
The system of equations \eqref{bogo1} is same as two phase case from eq.~\eqref{bogoliubov_during_reheating}, therefore we briefly summarize the result as following:
\begin{align}
    \alpha_k^{(1)}    &= \frac{\pi e^{\frac{1}{2} i \left(6 k \mu  \eta _{\text{inf}}+\pi  \left(\nu _1+\nu^* _2\right)\right)} (-k \eta _{\text{inf}})(3 \mu -1)\left[H_{\nu _1-1}^{(1)}\left(-k \eta _{\text{inf}}\right)-H_{\nu _1+1}^{(1)}(-k \eta _{\text{inf}})\right] H_{\nu _2^*}^{(1)}\left(k \eta _{\text{inf}} (1-3 \mu )\right)}{8 \sqrt{3\mu-1}}\nt
    &\qquad-\frac{\pi e^{\frac{1}{2} i \left(6 k \mu  \eta _{\text{inf}}+\pi  \left(\nu _1+\nu _2^*\right)\right)}k\eta _{\text{inf}} (3 \mu -1) H_{\nu _1}^{(1)}\left(-k \eta _{\text{inf}}\right)\left[H_{\nu _2^*-1}^{(1)}\left(k\eta _{\text{inf}}(1-3 \mu )\right)-H_{\nu _2^*+1}^{(1)}(k\eta _{\text{inf}}(1-3 \mu ))\right]}{8\sqrt{3\mu-1}}\nt
    &\qquad+3 \mu  \frac{\pi e^{\frac{1}{2} i \left(6 k \mu  \eta _{\text{inf}}+\pi  \left(\nu _1+\nu _2^*\right)\right)}H_{\nu _1}^{(1)}(-k\eta_{\inf})H_{\nu _2^*}^{(1)}\left(k \eta _{\text{inf}}(1-3 \mu )\right)}{8\sqrt{3\mu-1}},\\
    \beta_k^{(1)}&=-i\frac{\pi e^{-\frac{1}{2} i \left(6 k \mu  \eta _{\text{inf}}+\pi  \left(\nu_2-\nu _1\right)\right)} k \eta _{\text{inf}}(3 \mu -1)\left[H_{\nu _1-1}^{(1)}\left(-k \eta      _{\text{inf}}\right)-H_{\nu _1+1}^{(1)}(-k \eta _{\text{inf}})\right] H_{\nu _2}^{(2)}\left(k     \eta _{\text{inf}} (1-3 \mu )\right)}{8 \sqrt{3\mu-1}}\nt
    &\qquad-i\frac{\pi e^{-\frac{1}{2} i \left(6 k \mu  \eta _{\text{inf}}+\pi  \left(\nu _2-\nu _1\right)\right)}k\eta _{\text{inf}} (3 \mu -1) H_{\nu _1}^{(1)}\left(-k \eta _{\text{inf}}\right)\left[H_{\nu _2-1}^{(2)}\left(k\eta _{\text{inf}}(1-3 \mu )\right)-H_{\nu _2+1}^{(2)}(k\eta _{\text{inf}}(1-3 \mu ))\right]}{8\sqrt{3\mu-1}}\nt
    &\qquad+3 \mu i\frac{\pi e^{-\frac{1}{2} i \left(6 k \mu  \eta _{\text{inf}}+\pi  \left(\nu _2-\nu _1\right)\right)} H_{\nu_1}^{(1)}(-k \eta _{\text{inf}})H_{\nu _2}^{(2)}(k \eta _{\text{inf}}(1-3 \mu ))}{8\sqrt{3\mu-1}}.
\end{align}
The bogoliubov coefficient is simplest for matter like reheating with $\xi=w=0$, we have: $\nu_1=\nu_2=\tfrac{3}{2}$ and hence:
\begin{align}
    \alpha_k^{(1)}&=\frac{k \eta _{\inf } \left[-9+4 k \eta _{\inf } \left(2 k \eta _{\inf }-3 i\right)\right]+3 i}{8 k^3 \eta _{\inf }^3},\nt
    \beta_k^{(1)}&=\frac{3(i+k\eta_{\rm inf})e^{-2ik\eta_{\rm inf}}}{8k^3\eta_{\rm inf}^3} .
\end{align}
We now apply the junction conditions at the second transition hypersurface. Imposing continuity of the mode functions and their first derivatives at conformal time $\eta=\eta_{\rm reh}$, we obtain the following system of equations:
\begin{align}
    \alpha_{k}^{(1)}X^{\rm reh}+\beta_{k}^{(1)}X^{\rm reh*}\bigg\lvert_{\eta=\eta_{\text{reh}}} &= \alpha_{k}^{(2)}X^{\rm rad}+\beta_{k}^{(2)}X^{\rm rad*}\bigg\lvert_{\eta=\eta_{\text{reh}}},\nt
    \alpha_{k}^{(1)}X^{\rm reh'}+\beta_{k}^{(1)}X^{\rm reh*'}\bigg\lvert_{\eta=\eta_{\text{reh}}} &= \alpha_{k}^{(2)}X^{\rm rad'}+\beta_{k}^{(2)}X^{\rm rad*'}\bigg\lvert_{\eta=\eta_{\text{reh}}} .
\end{align}
Inverting this system of equations allow us to express the Bogoliubov coefficients in the radiation-dominated era:
\begin{align}
	\alpha^{(2)}_k &= i\alpha^{(1)}_k(X^{\rm reh'}_{k}X^{\rm rad*}_{k}-X^{\rm reh}_{k}X^{\rm rad'*}_{k})+i\beta_{k}^{(1)}(X^{\rm reh'*}_{k}X^{\rm rad*}_{k}-X^{\rm reh*}_{k}X^{\rm rad'*}_{k})\nt\nt
	\beta_k ^{(2)}
    &= -i\alpha_k^{(1)}(X^{\rm reh'*}_{k}X^{\rm rad *}_{k}-X^{\rm reh*}_{k}X^{\rm rad'*}_{k})^*-i\beta_{k}^{(1)}(X^{\rm reh'}_{k}X^{\rm rad*}_{k}-X^{\rm reh}_{k}X^{\rm rad'*}_{k})^* .
\end{align}
Here we have used the standard Wronskian normalization condition for the mode functions in the radiation-dominated era for further simplification:
\begin{align}
	\begin{bmatrix}
		X^{\rm rad}_{k} & X^{\rm rad*}_{k}\\
		\partial_\eta X^{\rm rad}_{k} & \partial_\eta X^{\rm rad*}_{k}
	\end{bmatrix}=i .
\end{align}
To completely determine $\alpha_k^{(2)}$ and $\beta_k^{(2)}$, we only need to evaluate the following:
\begin{multline*}
    X^{\rm reh'*}_{k}X^{\rm rad*}_{k}-X^{\rm reh*}_{k}X^{\rm rad'*}_{k}\\ 
    = \frac{(1+i)\sqrt{\pi} e^{\frac{1}{2} i \left(6 k \mu  \eta _{\inf }+2 k \eta _{\text{reh}}+\pi  \nu^* _2\right)} \left[2 k \left(\eta _{\text{reh}}-3 \mu  \eta _{\inf }\right) H_{\nu _2^*-1}^{(1)}\left(k \left(\eta _{\text{reh}}-3 \mu  \eta _{\inf }\right)\right)\right]}{8\sqrt{k(\eta _{\text{reh}}-3 \mu  \eta _{\inf }})}\nt
    +\frac{(1+i)\sqrt{\pi} e^{\frac{1}{2} i \left(6 k \mu  \eta _{\inf }+2 k \eta _{\text{reh}}+\pi  \nu^*_2\right)} \left[\left(6 i k \mu  \eta _{\inf }-2 i k \eta _{\text{reh}}-2 \nu^* _2+1\right) H_{\nu^*_2}^{(1)}\left(k \left(\eta _{\text{reh}}-3 \mu  \eta _{\inf }\right)\right)\right]}{8\sqrt{k(\eta _{\text{reh}}-3 \mu  \eta _{\inf }})} ,
\end{multline*}
\begin{multline*}
    X^{\rm reh'}_{k}X^{\rm rad*}_{k}-X^{\rm reh}_{k}X^{\rm rad'*}_{k}\\
    =\frac{(1-i)\sqrt{\pi } e^{-\frac{1}{2} i \left(6 k \mu  \eta _{\inf }-2 k \eta _{\text{reh}}+\pi  \nu _2\right)} \left[2 k \left(\eta _{\text{reh}}-3 \mu  \eta _{\inf }\right) H_{\nu _2-1}^{(2)}\left(k \left(\eta _{\text{reh}}-3 \mu  \eta _{\inf }\right)\right)\right]}{8\sqrt{k(\eta _{\text{reh}}-3 \mu  \eta _{\inf }})}\nt
    +\frac{(1-i)\sqrt{\pi } e^{-\frac{1}{2} i \left(6 k \mu  \eta _{\inf }-2 k \eta _{\text{reh}}+\pi  \nu _2\right)} \left[\left(6ik \mu  \eta _{\inf }-2ik \eta _{\text{reh}}-2\nu _2+1\right) H_{\nu _2}^{(2)}\left(k \left(\eta _{\text{reh}}-3 \mu  \eta _{\inf }\right)\right)\right]}{8\sqrt{k(\eta _{\text{reh}}-3 \mu  \eta _{\inf }})} .
\end{multline*}
For $\nu_1=\nu_2=\tfrac{1}{2}$, the above reduces to:
\begin{align*}
    X^{\rm reh'*}_{k}X^{\rm rad*}_{k}-X^{\rm reh*}_{k}X^{\rm rad'*}_{k}&= 0 ,\\
    X^{\rm reh'}_{k}X^{\rm rad*}_{k}-X^{\rm reh}_{k}X^{\rm rad'*}_{k}&=-i .
\end{align*}
These values ensure that conformally coupled massless scalar do not undergo particle production. The purpose of the Bogoliubov coefficients $(\alpha_k^{(1,2)}, \beta_k^{(1,2)})$ is to relate the initial free theory vacuum to current free theory vacuum. This signifies the mixing of positive and negative frequency modes, resulting in particle production. It is important to note that $|\beta_k|^2$ interprets strictly as a particle number density only for subhorizon modes ($k \gg aH$). On superhorizon scales ($k \ll aH$), this quantity merely characterizes the mode mixing, as the particle concept is ill-defined in the non-adiabatic regime \cite{mukhanov2007introduction,giudice2005heavy}. The spectrum of occupied states at the end of inflation ($\eta_{\text{inf}}$) and at reheating ($\eta_{\text{reh}}$) for various equation of state and non-minimal coupling parameter $\xi$ are presented in Fig. \ref{fig:beta2} corresponding to the reheating temperature of $T_{\rm reh}=10^{14}\rm GeV$. 

\begin{figure}[h]
    \centering
    \begin{minipage}{0.48\linewidth}
        \centering
        \includegraphics[width=\linewidth]{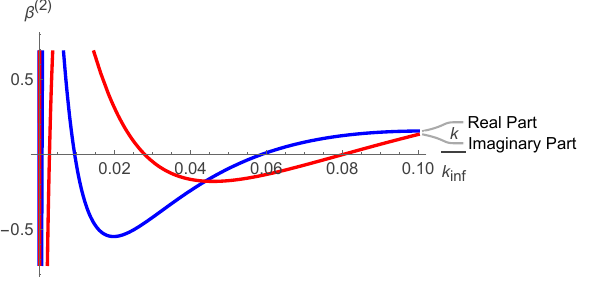}
        \captionsetup{style=centered_sub}\caption*{(a)For $\xi=\frac{1}{3}$, the maximal destructive interference takes place near $\tfrac{k}{k_{\rm inf}}\approx0.07$}
    \end{minipage}
    \hfill
    \begin{minipage}{0.48\linewidth}
        \centering
        \includegraphics[width=\linewidth]{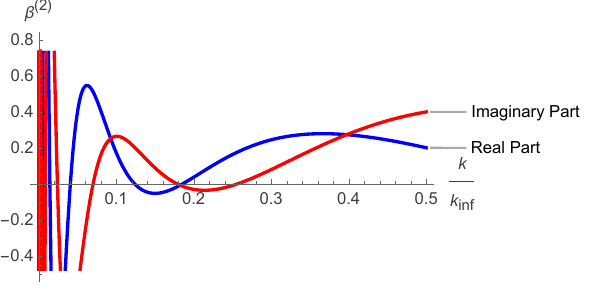}
        \captionsetup{style=centered_sub}\caption*{(b) For $\xi=1$, the maximal destructive interference takes place near $\tfrac{k}{k_{\rm inf}}\approx0.2$.}
    \end{minipage}
    \caption{The plot of real part and imaginary part of $\beta_k^{(2)}$ for $w=0$.}
    \label{fig:imreal}
\end{figure}

Some observation are in order: The dip observed in the spectrum of $|\beta_k^{(2)}|^2$ in Fig.~\ref{fig:beta2} for $\xi=1,\tfrac{1}{3}$ and $w=0$, is a general feature arising from the destructive interference of modes on superhorizon scale. By decomposing the Bogoliubov coefficient into its real and imaginary parts $|\beta_k|^2 = (\text{Re}\,\beta_k)^2 + (\text{Im}\,\beta_k)^2$, as shown in Fig.~\ref{fig:imreal}, we find that the spectral minimum corresponds to a point of maximal destructive interference.
This behavior is characteristic of the superhorizon limit $k\eta \to 0$ when the Hankel index $\nu$ is purely imaginary, which causes the modes to oscillate logarithmically rather than decay monotonically. For the non-minimal coupling $\xi=1$, this oscillatory regime persists for equations of state $w \lesssim 0.3$. 
We further check that in the limit of instantaneous reheating ($T_{\rm reh} = 10^{15}$ GeV), the Bogoliubov coefficient $\beta^{(2)}_{k}$ recovers the standard instantaneous result. 
\subsection{Two Point Function}
\begin{figure}[t]
    \centering
    \begin{minipage}[b]{0.48\linewidth}
        \centering
        \includegraphics[width=\linewidth]{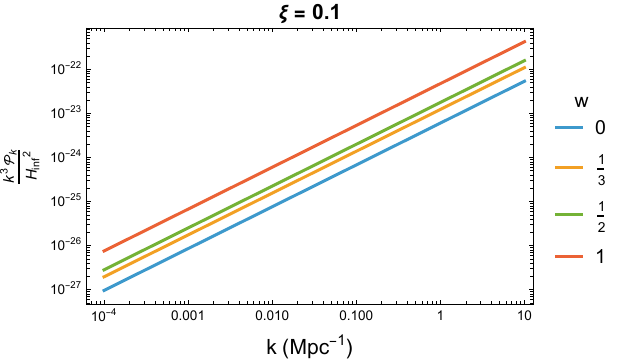}
        \captionsetup{style=centered_sub}\caption*{(a) Spectrum at $T_{\mathrm{reh}}=10^{13}\rm GeV$}
    \end{minipage}
    \hfill
    \begin{minipage}[b]{0.48\linewidth}
        \centering
        \includegraphics[width=\linewidth]{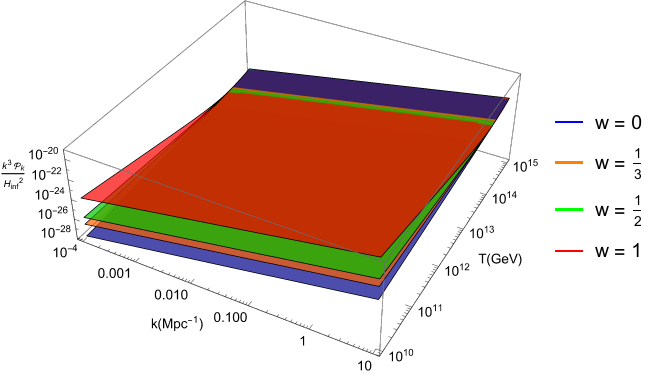}
        \captionsetup{style=centered_sub}\caption*{(b) Spectrum for varying $T_{\mathrm{reh}}$}
    \end{minipage}
    
    \caption{The primordial power spectra of a non-minimally coupled scalar field. (a) The spectra evaluated at the end of reheating for a fixed temperature $T_{\mathrm{reh}}=10^{13}$ GeV. The spectral behavior is governed by the scaling index $\nu$, where lower values of the equation of state $w$ lead to a pronounced enhancement on superhorizon scales. (b) The 3D evolution of the power spectrum as a function of the reheating temperature, illustrating how the superhorizon amplitude is sensitive to the duration of the reheating phase.}
    \label{fig:combined_spectra}
\end{figure}

\begin{figure}[h]
    \centering
    \begin{minipage}{0.48\linewidth}
        \centering
        \includegraphics[width=\linewidth]{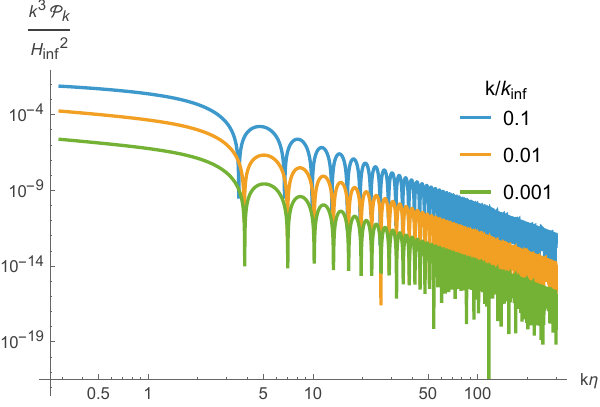}
        \captionsetup{style=centered_sub}\caption*{(a) $w=0$}
        \label{fig:ps_w0_xi01}
    \end{minipage}
    \hfill
    \begin{minipage}{0.48\linewidth}
        \centering
        \includegraphics[width=\linewidth]{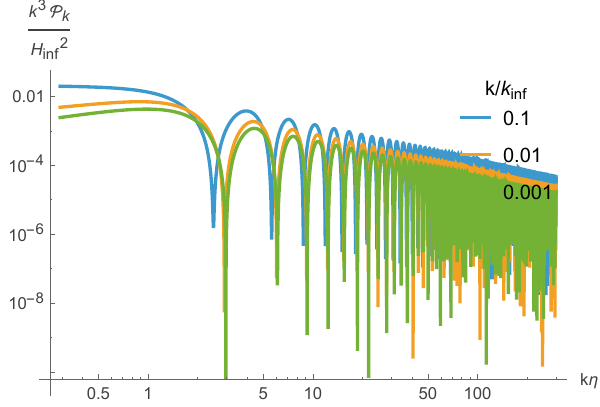}
        \captionsetup{style=centered_sub}\caption*{(b) $w=1$}
        \label{fig:ps_w1_xi01}
    \end{minipage}
    \caption{Power spectrum for modes entering the horizon during reheating phase with $\xi=0.1$.}
    \label{non-minimal1}
\end{figure}

\begin{figure}[h]
    \centering
    \begin{minipage}{0.48\linewidth}
        \centering
        \includegraphics[width=\linewidth]{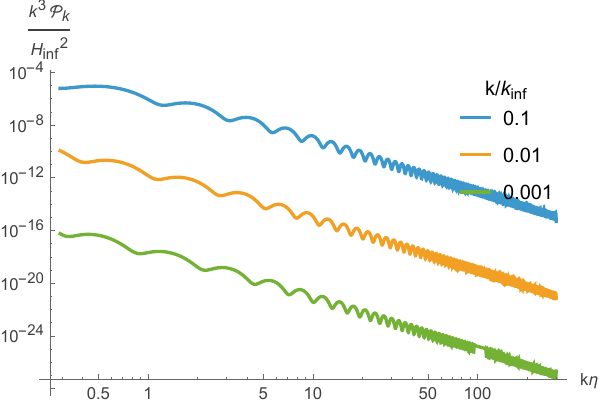}
        \captionsetup{style=centered_sub}\caption*{(a) $w=0$}
        \label{fig:ps_w0_xi1}
    \end{minipage}
    \hfill
    \begin{minipage}{0.48\linewidth}
        \centering
        \includegraphics[width=\linewidth]{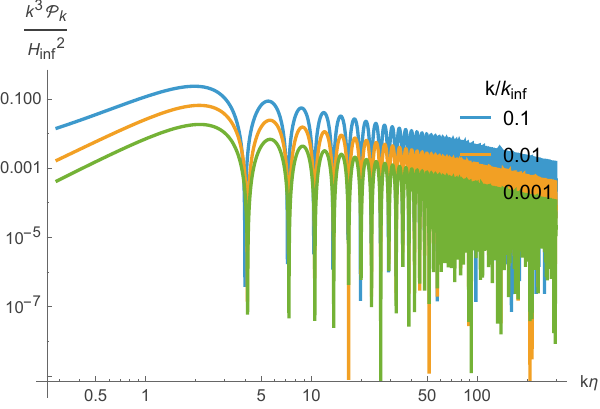}
        \captionsetup{style=centered_sub}\caption*{(b) $w=1$}
        \label{fig:ps_w1_xi1}
    \end{minipage}
    \caption{Power spectrum for modes entering the horizon during reheating phase with $\xi=1$.}
    \label{non-minimal2}
\end{figure}

During inflation, the power spectrum for a non-minimally coupled scalar is determined by the vacuum expectation value of the mode functions $X_k^{\rm inf}$:
\begin{align}
    \mathcal{P}^{\rm inf}_k = f_k f_k^*= \frac{X_k^{\inf} X_k^{\inf*}}{a(\eta)^2}
    =\frac{\pi}{4}H^2(-\eta)^3\abs{H_{\nu_1}^{(1)}(-k\eta)}^2.
\end{align}
Transitioning into the reheating era, the spectrum incorporates the mixing of positive and negative frequency modes induced by the change in the background expansion. The resulting power spectrum is given by:
\begin{align}\label{Powerspectrum-nonminimal}
 k^3  \mathcal{P}^{\rm reh}_k&=\frac{k^3}{a(\eta)^2}\left(
		\alpha_{k}^{(1)}X^{\rm reh}+\beta_{k}^{(1)}X^{\rm reh*} \right)\left(
		\alpha_{k}^{(1)*}X^{\rm reh*}+\beta_{k}^{(1)*}X^{\rm reh} \right)\nt
    &=\left(\frac{k}{a(\eta)}\right)^2\left[\left(\abs{\alpha_k^{(1)}}^2+\abs{\beta_k^{(1)}}^2\right)\frac{\pi k\tilde{\eta}}{4}\abs{H_{\nu_2}^{(2)}(k\tilde{\eta})}^2+\frac{\pi k\tilde{\eta}}{2}\Re{\alpha_k^{(1)*}\beta_k^{(1)}e^{i\left(\frac{\pi}{2}+\pi\nu_2+6k\mu\eta_{\text{inf}}\right)}(H_{\nu_2}^{(1)}(k\tilde{\eta}))^2}\right].
\end{align}
On superhorizon scale for $\xi<\frac{3}{16}$:
\begin{align}\label{power-spec-longwave}
 k^3  \mathcal{P}^{\rm reh}_k
    &\approx
    \left(\frac{k}{a(\eta)}\right)^2\frac{2^{2\nu_2-1}\Gamma(\nu_2)^2}{2\pi}\left(k\tilde{\eta}\right)^{1-2\nu_2}\left[\abs{\alpha_k^{(1)}}^2+\abs{\beta_k^{(1)}}^2+2\Re{\alpha_k^{(1)*}\beta_k^{(1)}e^{i\left(\frac{\pi}{2}+\pi\nu_2\right)}}\right].
\end{align}
The dimensionless power spectrum scales with the physical momentum $(k/a)^2$, where the redshift of the momentum is dictated by the reheating equation of state $w$. For $\xi>\frac{3}{16}$ and $w<\frac{1}{3}$ where $\nu_1$ and $\nu_2$ are both imaginary, the power spectrum suffers from Boltzman-like suppression. The last term in Eq.~\eqref{Powerspectrum-nonminimal} or \eqref{power-spec-longwave} encodes phase term originating from the interference effect due to the transition to reheating era.
The primordial power spectra exhibit both superhorizon and subhorizon features determined by the spectral tilt, parameterized by the non-minimal coupling $\xi$, and the background equation of state $w$. These dependencies are illustrated in Figs.~\ref{non-minimal1} and \ref{non-minimal2}. The reheating scale (temperature) primarily dictates the duration of the reheating epoch, thereby altering the infrared extent to which modes undergo superhorizon evolution, as shown in Fig.~\ref{fig:combined_spectra}.

During radiation dominated era:
\begin{align}
\mathcal{P}^{\rm rad}_k &=\frac{1}{a(\eta)^2}\left[\left(
		\alpha_{k}^{(2)}X^{\text{rad}}+\beta_{k}^{(2)}X^{\text{rad}*} \right)\left(
		\alpha_{k}^{(2)*}X^{\text{rad}*}+\beta_{k}^{(2)*}X^{\text{rad}} \right)\right]\nt
    &=\frac{1}{a(\eta)^2}\frac{1}{2k}\left[\abs{\alpha_k^{(2)}}^2+\abs{\beta_k^{(2)}}^2+2\Re{\alpha_k^{(2)*}\beta_k^{(2)}e^{2ik\eta}}\right].
\end{align}
The physics of reheating is encoded in the Bogoliubov coefficients, $\alpha_k^{(2)}$ and $\beta_k^{(2)}$, which carry the information of the reheating history into the radiation era. The specific equation of state during reheating and the duration of that phase determine the final amplitude and the oscillatory profile of the power spectrum. These results could be parametrized by replacing $\eta_{\inf}=-\frac{1}{k_{\inf}}$ and $\eta_{\rm reh}$ from \eqref{kinf} and \eqref{etareh} which allows us to parametrize the correlator during radiation dominated era in terms of $(w,\xi,T_{\rm reh})$. 

\begin{figure}[h]
    \centering
    \begin{minipage}{0.45\textwidth}
        \centering
        \includegraphics[width=\linewidth]{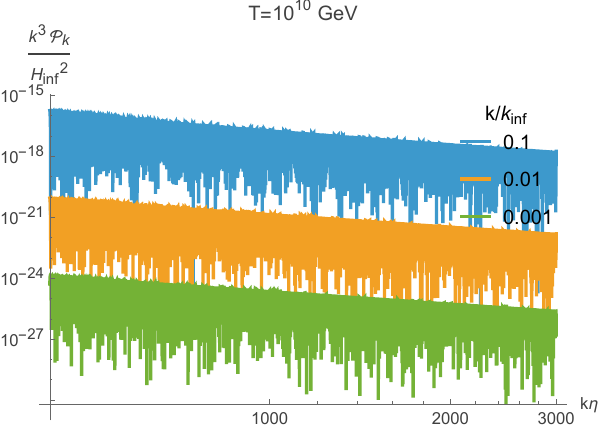}
        \caption{Dimensionless power spectrum evolution in the radiation-dominated era for $w=0$, $\xi=1$, and $T_{\rm reh} = 10^{10}$~GeV.}
        \label{fig:rad_xi1}
    \end{minipage}
    \hfill
    \begin{minipage}{0.45\textwidth}
        \centering
        \includegraphics[width=\linewidth]{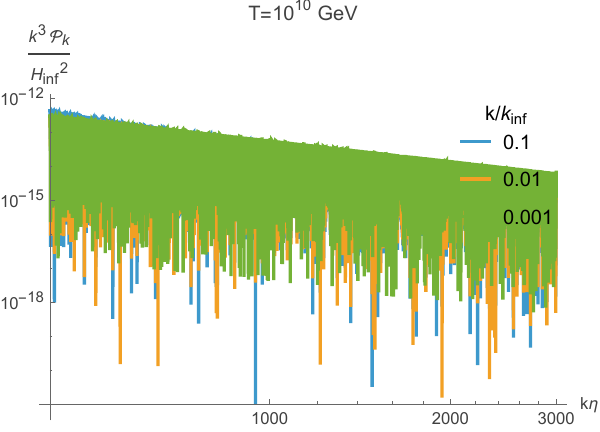}
        \caption{Oscillatory behavior of the power spectrum in the radiation regime for $\xi=0.1$ and $w=0$.}
        \label{fig:rad_xi0.1}
    \end{minipage}
\end{figure}
\begin{figure}[t]
    \centering
    \includegraphics[width=1\linewidth]{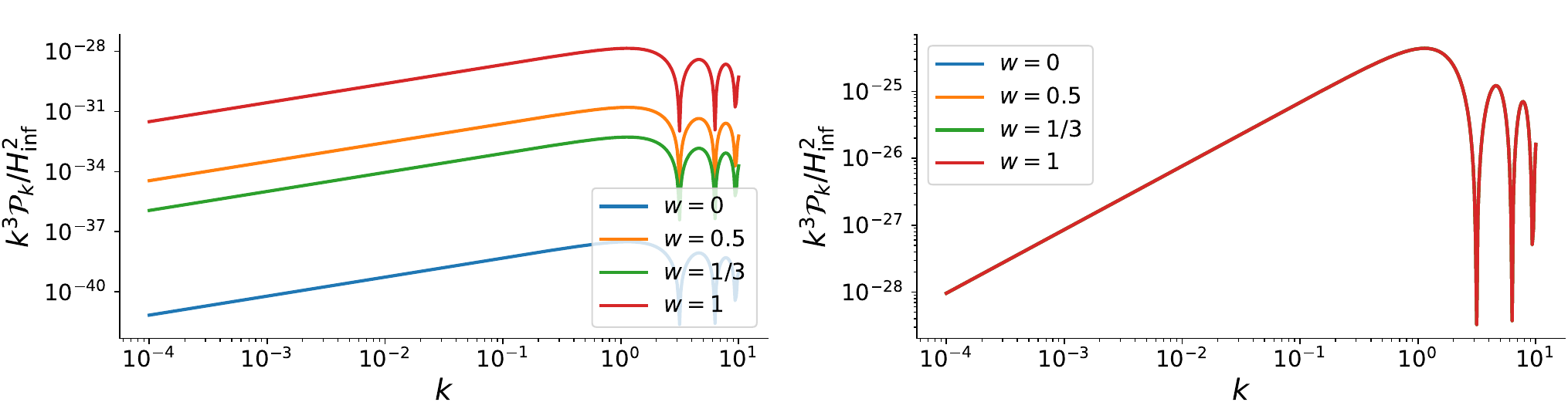}
    \caption{We have used \eqref{etareh} to parametrize the power spectrum as a fuction of reheating temperature. \textbf{Left:} $T = 10^{10}\,\mathrm{GeV}$. \textbf{Right:} $T = 10^{15}\,\mathrm{GeV}$. The power spectrum is evaluated at $\eta = \eta_{\rm reh} + 1\,\mathrm{Mpc}$, chosen arbitrarily to probe the spectrum immediately after the transition. For $T = 10^{15}\,\mathrm{GeV}$, $\eta_{\rm reh} = \eta_{\rm inf}$. All results are shown for $\xi = 0.1$ satisfying isocurvature constraint, and reheating induces a significant enhancement on superhorizon scales.}
    \label{fig:rad-power_spectrum}
\end{figure}

To place these results in a broader context, it is instructive to consider the case of an instantaneous reheating transition, corresponding to $T_{\text{reh}} = 10^{15}$~GeV. By maintaining $\xi=0.1$, we remain in a regime consistent with the previously discussed cases while isolating the effects of the reheating duration. As seen in Fig. \ref{fig:rad-power_spectrum}, the characteristic lack of rapid oscillations for super-horizon modes persists; however, the instantaneous transition precludes the equation-of-state dependent tachyonic enhancement observed in the finite-duration cases. This confirms that the final normalization of the spectrum, which is subsequently inherited by the radiation era in Fig.~\ref{fig:rad_xi1} and Fig.~\ref{fig:rad_xi0.1}, is uniquely sensitive to the specific history of the reheating phase.

\subsubsection{Isocurvature constraints}\label{sec:isocurv}
In this section, we apply the formalism developed above in the context of multifield inflation. We consider a two-field system comprising the inflaton $\varphi \equiv \phi_1$, which dominates the background energy density and drives the expansion, and a spectator field $\phi \equiv \phi_2$. A distinctive feature of such multifield scenarios is the generation of isocurvature perturbations: unlike single-field inflation where non-adiabatic modes are either absent or strictly decaying, the presence of a second field allows for persistent entropy perturbations sourced by the gravitational amplification of vacuum fluctuations.

To characterize these perturbations, we adopt the standard decomposition of field fluctuations into adiabatic and entropy components. The adiabatic perturbation $\delta\sigma$ represents fluctuations along the background trajectory, while the entropy perturbation $\delta s$ describes fluctuations orthogonal to it. The gauge-invariant \textit{intrinsic entropy perturbation}, $\mathcal{S}$, is defined as the relative fluctuation in the field configuration normalized by the background evolution dynamics:
\begin{equation}
    \mathcal{S} \equiv \frac{H}{\dot{\sigma}}\delta s\,.
\end{equation}
Here, $\dot{\sigma}^2 = \dot{\phi}_1^2 + \dot{\phi}_2^2$ represents the kinetic energy of the background trajectory. We consider a scenario where the second field $\phi$ remains subdominant during inflation, characterized by a vanishing vacuum expectation value $\phi \approx 0$. In this limit, the background trajectory aligns with the inflaton direction such that $\dot{\sigma} \simeq \dot{\varphi}$, and the orthogonal entropy perturbation simplifies directly to the fluctuation of the second field, $\delta s \simeq \delta\phi\,$. Because the test field remains near its minimum $\phi\approx0$, the field-space trajectory is nearly straight, making the turning rate $\dot\theta$˙ negligibly small. As a result, the conversion of isocurvature perturbations into curvature perturbations is highly suppressed during inflation. The intrinsic entropy perturbation is linearly proportional to the spectator field fluctuation:
\begin{equation}\label{multi-entropy}
    \mathcal{S} \simeq \frac{H}{\dot{\varphi}}\delta\phi\,.
\end{equation}
Since $\mathcal{S}$ is directly proportional to $\delta\phi$, the behavior of the entropy mode is determined by the quantum dynamics of the spectator field. The correlators computed in the preceding sections therefore constitute the fundamental stochastic quantities from which observable isocurvature statistics are derived; the transfer from field-space correlators to observable spectra depends on the inflationary trajectory and reheating history, and we treat this as a model-dependent step separate from the present calculation. 

The spectator field fluctuations $\delta\phi$, amplified by the spacetime curvature, sources a non-zero entropy power spectrum $\mathcal{P}_{\mathcal{S}}(k) \propto \mathcal{P}_{\phi}(k)$, and the bispectrum sourced by the interactions of these gravitationally produced dark matter particles is related in analogous fashion via \eqref{multi-entropy} as $\overline{\langle \mathcal{S}_{k_1}\mathcal{S}_{k_2}\mathcal{S}_{k_3}\rangle} \propto \overline{\langle\delta\phi_{k_1}\delta\phi_{k_2}\delta\phi_{k_3}\rangle}$. While the overall amplitudes of these spectra are sensitive to the equation-of-state history and are rescaled by the factor $(H/\dot{\varphi})^2$ evaluated at the reheating junction, the spectral shapes are not. The tilt, scale-dependence, and momentum structure of both the power spectrum and bispectrum are entirely determined by the inflationary dynamics and are directly inherited from $\mathcal{P}_\phi(k)$ and $\langle\delta\phi_{k_1}\delta\phi_{k_2}\delta\phi_{k_3}\rangle$ respectively. 

While the intrinsic entropy perturbation $\mathcal{S}$ outlined above provides an intuitive measure of multifield dynamics, recent literature suggests that this definition can yield predictions that deviate significantly from late-time observables, potentially by several orders of magnitude~\cite{cicoli2023choice}. A more rigorous connection to cosmological observations requires evaluating the \textit{relative} isocurvature perturbation of cold dark matter (CDM), following Ref.~\cite{Liddle:1999pr}:
\begin{equation}
    S = \frac{\delta\rho_{\phi}}{\rho_{\phi}}-\frac{3}{4}\frac{\delta\rho_{\gamma}}{\rho_{\gamma}}\,.
\end{equation}
In the isothermal limit where $\delta\rho_{\gamma} \to 0$, this reduces 
to $S = \delta\rho_{\phi}/\rho_{\phi}$. We model the CDM as an ultralight spectator field $(\frac{m}{H_{\rm inf}}\ll1)$of non-relativistic particles with a vanishing vacuum expectation value, $\bar{\phi}=0$, such that the energy density is dominated by the quadratic mass term, $\rho_{\phi}\approx \tfrac{1}{2}m^2\delta\phi^2$. The isocurvature perturbation is therefore sourced by the quadratic field operator:
\begin{equation}
    S_k = \frac{\delta\phi^2_k}{\langle\delta\phi^2\rangle}\,\tag{for $k\ne0$}
\end{equation}
The power spectrum of $S$ is then determined by the four-point function of $\delta\phi$, and qualitatively it follows the convolution of the field power spectrum:
\begin{equation}
    \overline{\langle S_{k}S_{-k}\rangle} \propto 
    \int \frac{d^3q}{(2\pi)^3}\overline{\langle\delta\phi_{q}\delta\phi_{-q}\rangle}\, \overline{\langle\delta\phi_{k-q}\delta\phi_{-(k-q)}\rangle\,}.
\end{equation}
The upper bound of the amplitude of isocurvature power spectrum at CMB scale $\mathcal{P}_{\mathcal{S}}(k_{\ast})\lesssim 8.3\times 10^{-11}$\cite{aghanim2020planck}. The second-order isocurvature power spectrum is evaluated by using the following expression as \cite{Kolb:2023ydq, Liddle:1999pr} 
\begin{equation}\label{isocurvpow}
\mathcal{P}_{\mathcal{S}}(k)= \frac{1}{\rho_{\phi}^2}\frac{k^3}{2\pi^2}\int d^3\vec{x}\langle\delta\rho_{\phi}(\vec{x})\delta\rho_{\phi}(0)\rangle e^{-i \vec{k}.\vec{x}}.
\end{equation}
 Recent numerical and semi-analytical studies have shown that for $w=0$, $\xi \gtrsim 0.02$, while for $w > 1/2$, the constraint shifts to $\xi \lesssim 4.0$~\cite{chakraborty2025probing}. 

\subsection{Squeezed limit: Three point function}
For a massless scalar field with non-minimal coupling to curvature, the linear equations of motion admit exact solutions in terms of Hankel functions. While the mode functions are known explicitly, a direct evaluation of the three-point function remains non-trivial as it involves time integrals over triple products of Hankel functions with arbitrary momenta. For generic configurations, these integrals do not admit closed-form expressions.

To retain analytic control and isolate the asymptotic behavior, we evaluate the bispectrum in the \textbf{squeezed limit}, where $k_1 \approx k_2 =k$ such that $\tfrac{k_3}{k}\to0$. This hierarchy allows for a systematic separation of scales: the long mode $k_3$ exits the horizon significantly earlier relative to $k$ and can be replaced by its asymptotic super-horizon expansion, while the short modes remain sub-horizon during the dominant period of interaction~\cite{Chen:2009zp}. The long-wavelength behavior is governed by:
\begin{align}
    H_{\nu}^{(1)}(x) &\approx -\frac{i}{\pi}\left[\Gamma(\nu)\left(\frac{x}{2}\right)^{-\nu}+e^{-i\pi\nu}\Gamma(-\nu)\left(\frac{x}{2}\right)^{\nu}\right],
\end{align}
which allows the time integral to be carried out analytically. Note that since $H_{\nu}^{(1)}(x)$ is a complex function, we take into account both real and imaginary part upto leading order in $x$. Dropping the second term in the limit $x\to0$ amounts to setting the series expansion of real part to zero, which breaks the normalization of mode function. Under this approximation, the bispectrum at the end of inflation is governed by the following in-in integral:
\begin{align}
    \overline{\ev{\phi_{k_1}\phi_{k_2}\phi_{k_3}}} (\eta_{\rm inf}) &= 2g \, \text{Im} \left\{ f_{k_1}(\eta_{\inf})f_{k_2}(\eta_{\inf})f_{k_3}(\eta_{\inf}) \int_{-\infty}^{\eta_{\text{inf}}} d\eta' \, a(\eta')^4 f_{k_1}^*(\eta')f_{k_2}^*(\eta')f_{k_3}^*(\eta') \right\} \nt
    &= 2g \, \text{Im} \left\{ f_{k}^2 f_{k_3} \left[\mathcal{N}(\nu_1^*, k, k_3)\int_{-\infty}^{k\eta_{\text{inf}}} dx \, [H_{\nu_1^*}^{(2)}(-x)]^2 (-x)^{-\nu_1^* + 1/2}+(\nu_1\leftrightarrow-\nu_1) \right]\right\},
\end{align}
where $k \approx k_1 \approx k_2$ and the prefactor $\mathcal{N}$ contains the squeezed-limit kinematics. Defining the dimensionless integral ,
\begin{align}
\mathcal{I}_1(\nu) &= \int_{-\infty}^{k\eta_{\text{inf}}} dx [H_{\nu}^{(2)}(-x)]^2 (-x)^{-\nu + 1/2};\qquad&
\mathcal{\tilde I}_1(\nu) &=\int_{-\infty}^{k\eta_{\text{inf}}} dx [H_{\nu}^{(2)}(-x)]^2 (-x)^{\nu + 1/2}
=e^{2i\pi\nu}\mathcal{I}_1(-\nu). 
\end{align}
We can express the bispectra as:
\begin{align}
    \overline{\ev{\phi_{k}\phi_{k}\phi_{k_3}}}(\eta_{\rm inf})
    &= 2g \, \text{Im} \bigg\{
    \left[ \frac{\sqrt{-\pi \eta_{\inf}}}{2 a(\eta_{\inf})} e^{i(\frac{\pi}{4} + \frac{\pi\nu_1}{2})} H_{\nu_1}^{(1)}(-k\eta_{\inf}) \right]^2 \left[ \frac{\sqrt{-\pi \eta_{\inf}}}{2 a(\eta_{\inf})} e^{i(\frac{\pi}{4} + \frac{\pi\nu_1}{2})} H_{\nu_1}^{(1)}(-k_3\eta_{\inf}) \right]\nt
    &\qquad\times[\mathcal{N}(\nu_1^*, k, k_3)  \mathcal{I}_1 (\nu_1^*)+\mathcal{N}(-\nu_1^*, k, k_3)\mathcal{I}_1 (-\nu_1^*) ]\bigg\}.
\end{align}
where,
\begin{align}
    \mathcal{N}(\nu,k,k_3) &
    =i 2^{\nu} \frac{\sqrt{\pi}\Gamma(\nu)}{8H_{\rm inf}k_3^{3/2}} \left(\frac{k_3}{k}\right)^{\tfrac{3}{2}-\nu} e^{-3i(\frac{\pi}{4} + \frac{\pi\nu}{2})}.
\end{align}
Finally, taking the super-horizon limit $(k_3\eta\to0)$, the expression reduces to:
\begin{align}
    \overline{\ev{\phi_{k}\phi_{k}\phi_{k_3}}}(\eta_{\rm inf})
    &=g\frac{\pi (k\cdot k_{\rm inf})^{-3/2}}{32 H_{\rm inf} a(\eta_{\rm inf})^3} \, \text{Im} \Bigg\{ [H_{\nu_1}^{(1)}(-k\eta_{\rm inf})]^2\bigg[ 2^{\nu_1+\nu_1^*} |\Gamma(\nu_1)|^2 \left(\frac{k_3}{k}\right)^{-\nu_1-\nu_1^*} (k/k_{\rm inf})^{-\nu_1} \mathcal{I}_1(\nu_1^*) \nt
    &\quad \quad + 2^{\nu_1-\nu_1^*}\Gamma(\nu_1)\Gamma(-\nu_1^*) e^{3i\pi\nu_1^*} (k/k_{\rm inf})^{-\nu_1}k_3^{\nu_1-\nu_1^*} \mathcal{I}_1(-\nu_1^*) \nt
    &\qquad+ 2^{-\nu_1+\nu_1^*}\Gamma(\nu_1^*)\Gamma(-\nu_1) e^{-i\pi\nu_1} (k/k_{\rm inf})^{\nu_1} k_3^{\nu_1-\nu_1^*}\mathcal{I}_1(\nu_1^*) \nt
    &\quad \quad + 2^{-\nu_1-\nu_1^*} |\Gamma(-\nu_1)|^2 e^{2i\pi\nu_1^*} \left(\frac{k_3}{k}\right)^{\nu_1+\nu_1^*} (k/k_{\rm inf})^{\nu_1} \mathcal{I}_1(-\nu_1^*) \bigg] \Bigg\} .
\end{align}
This result captures the leading-order squeezed bispectrum, providing a primordial template that is sensitive to the asymptotic behavior induced by the non-minimal coupling. Although the bispectrum is defined through the imaginary part of the time integral, extracting it explicitly at this stage would merely amount to expanding the expression into its complex conjugates without offering further simplification. Since the phase information is already manifest, we retain this compact representation for the primordial signal. To extend this result beyond the $\eta=\eta_{\rm inf}$ hypersurface, we take the $k_3/k\to0$ limit inside the in-in integral. Subsequently, based on the era where the in-in contour lies, we decompose the bispectrum into three distinct temporal regions. The contribution originating from the inflationary era is evaluated as:
\begin{align}
    \overline{\ev{\phi_{k}\phi_{k}\phi_{k_3}}}_{\text{inf}}(\eta) =2g \, \text{Im} \bigg\{ f_{k}(\eta)^2 f_{k_3}(\eta)[\mathcal{N}(\nu_1^*, k, k_3)  \mathcal{I}_1 (\nu_1^*)+\mathcal{N}(-\nu_1^*, k, k_3)\mathcal{I}_1 (-\nu_1^*) ]\}.
\end{align}
Once the explicit $k$-dependence is factored out, $\mathcal{I}_1$ reduces to a definite integral that remains finite for generic (non-integer) values of $\nu_1$. While this integral admits a formal representation in terms of generalized hypergeometric functions. The integral is ill defined for integral values of $\nu_1$, therefore we focus on non-integer values of $\nu_1$, where it is well-defined and stable. The reheating contribution is evaluated by expressing the mode function via the Bogoliubov coefficients:
\begin{align}
  \alpha_{k_3}^{(1)*}X_{k_3}^{\text{reh}*} + \beta_{k_3}^{(1)*}X_{k_3}^{\text{reh}}
&= \frac{i}{2} \sqrt{\frac{\tilde{\eta}}{\pi}}\bigg[\beta_{k_3}^{(1)*}e^{-i(\frac{\pi}{4}+\frac{\pi\nu_2}{2})}\Gamma(\nu_2)\left(\frac{k_3\tilde{\eta}}{2}\right)^{-\nu_2}+\beta_{k_3}^{(1)*}e^{-i(\frac{\pi}{4}-\frac{\pi\nu_2}{2})}\Gamma(-\nu_2)\left(\frac{k_3\tilde{\eta}}{2}\right)^{\nu_2}\nt
&\qquad\qquad\quad-\alpha_{k_3}^{(1)*}e^{i(\frac{\pi}{4}+\frac{\pi\nu_2^*}{2})}\Gamma(\nu_2^*)\left(\frac{k_3\tilde{\eta}}{2}\right)^{-\nu_2^*}-\alpha_{k_3}^{(1)*}e^{i(\frac{\pi}{4}-\frac{\pi\nu_2^*}{2})}\Gamma(-\nu_2^*)\left(\frac{k_3\tilde{\eta}}{2}\right)^{\nu_2^*}\bigg].
\end{align}
The resulting integral for the reheating phase is given by:
\begin{align}\label{reheating-part-nonminimal}
&\overline{\langle\phi_{k}\phi_{k}\phi_{k_3}\rangle}_{\text{reh}}\nt
&=
2g\,\text{Im}\Bigg\{\frac{f_k^2 f_{k_3}}{k\sqrt{k}}\frac{i\sqrt{\pi}}{8}\Big[(\alpha_k^{(1)*})^2
e^{i(\frac{\pi}{2}+\pi\nu_2+6\mu k\eta_{\inf})}\Big(W^\beta_{+\nu_2}e^{i2\pi\nu_2\delta_I}\mathcal I_3(\nu_2)
+W^\beta_{-\nu_2}e^{i2\pi\nu_2(\delta_I-1)}\mathcal I_3(-\nu_2)\nt
&\qquad-W^\alpha_{-\nu_2}e^{i2\pi\nu_2(\delta_I-1)}\mathcal I_3(-\nu_2)-W^\alpha_{+\nu_2}e^{i2\pi\nu_2\delta_I}\mathcal I_3(\nu_2)\Big)+2\alpha_k^{(1)*}\beta_k^{(1)*}\Big(W^\beta_{+\nu_2}\mathcal I_2(\nu_2)+W^\beta_{-\nu_2}\mathcal I_2(-\nu_2)\nt
&\qquad-W^\alpha_{-\nu_2}\mathcal I_2((-1)^{\delta_I}\nu_2)-W^\alpha_{+\nu_2}\mathcal I_2((-1)^{1-\delta_I}\nu_2)\Big)+(\beta_k^{(1)*})^2 e^{-i(\frac{\pi}{2}+\pi\nu_2+6\mu k\eta_{\inf})}\Big(W^\beta_{+\nu_2}\mathcal I_3^*((-1)^{\delta_I}\nu_2)\nt
&\qquad+W^\beta_{-\nu_2}e^{i2\pi\nu_2}\mathcal I_3^*((-1)^{1-\delta_I}\nu_2)-W^\alpha_{-\nu_2}e^{i2\pi\nu_2}\mathcal I_3^*((-1)^{1-\delta_I}\nu_2)-W^\alpha_{+\nu_2}\mathcal I_3^*((-1)^{\delta_I}\nu_2)\Big)\Big]\Bigg\},
\end{align}
where we defined:
\begin{equation}
\begin{array}{c c}
\displaystyle 
\delta_I =
\begin{cases}
1 & \nu_2 \in i\mathbb{R} \\
0 & \nu_2 \in \mathbb{R}
\end{cases}
&\qquad\qquad;\qquad\qquad
\displaystyle
\begin{aligned}
W^{\alpha}_{\pm\nu_2} &\equiv
\alpha_{k_3}^{(1)*}\,e^{i(\frac{\pi}{4}\pm\frac{\pi\nu_2}{2})}
\,2^{\pm\nu_2}\Gamma(\pm\nu_2)
\left(\frac{k_3}{k}\right)^{\mp\nu_2}\\[4pt]
W^{\beta}_{\pm\nu_2} &\equiv
\beta_{k_3}^{(1)*}\,e^{-i(\frac{\pi}{4}\pm\frac{\pi\nu_2}{2})}
\,2^{\pm\nu_2}\Gamma(\pm\nu_2)
\left(\frac{k_3}{k}\right)^{\mp\nu_2}
\end{aligned}
\end{array}
\end{equation}
Finally, the contribution originating from the radiation-dominated era ($\eta \ge \eta_{\rm reh}$) is evaluated. In this phase, the interaction remains active and is integrated from the reheating surface to the time of observation:
\begin{align}
    &\overline{\ev{\phi_{k}\phi_{k}\phi_{k_3}}}_{\text{rad}}(\eta\ge\eta_{\rm reh})
    =2g\Im{f_{k}(\eta)f_{k}(\eta)f_{k_3}(\eta)\int_{\eta_{\rm reh}}^{\eta}D\eta' f_{k}^*(\eta')f_{k}^*(\eta') f_{k_3}^*(\eta')}\nt
    &=\frac{g}{a(\eta)^3\,2k^2k_3}\Im\bigg\{(\alpha_{k}^{(2)}e^{-ik\eta}+\beta_{k}^{(2)}e^{ik\eta})\,^2(\alpha_{k_3}^{(2)}e^{-ik_3\eta}+\beta_{k_3}^{(2)}e^{ik_3\eta})\bigg[(\alpha_{k}^{(2)*})^2\alpha_{k_3}^{(2)*}\mathcal{I}_4(2k+k_3)+(\alpha_{k}^{(2)*})^2\beta_{k_3}^{(2)*}\mathcal{I}_4(2k-k_3)\nt
    &\qquad+2\alpha_k^{(2)*}\beta_{k}^{(2)*}(\beta_{k_3}^{(2)*}\mathcal{I}_4(-k_3)+\alpha_{k_3}^{(2)*}\mathcal{I}_4(k_3))+(\beta_k^{(2)*})^2(\alpha_{k_3}^{(2)}\mathcal{I}_4(k_3-2k)+\beta_{k_3}^{(2)*}\mathcal{I}_4(-2k-k_3))\bigg].
\end{align}
The resulting expression for the full bispectrum is given by:
\begin{align}
    &\overline{\ev{\phi_{k}\phi_{k}\phi_{k_3}}}(\eta\ge\eta_{\rm reh})=\overline{\ev{\phi_{k}\phi_{k}\phi_{k_3}}}_{\text{inf}}+\overline{\ev{\phi_{k}\phi_{k}\phi_{k_3}}}_{\text{reh}}+\overline{\ev{\phi_{k}\phi_{k}\phi_{k_3}}}_{\text{rad}}\nt
    &=\frac{g}{a(\eta)^3k\sqrt{2k_3}}\Im\bigg\{(\alpha_{k}^{(2)}e^{-ik\eta}+\beta_{k}^{(2)}e^{ik\eta})^2(\alpha_{k_3}^{(2)}e^{-ik_3\eta}+\beta_{k_3}^{(2)}e^{ik_3\eta})\bigg[\frac{i\sqrt{\pi} e^{-3i(\frac{\pi}{4} + \frac{\pi\nu_1^*}{2})}}{8 H_{\text{inf}}k^{3/2}}\bigg[\Gamma(\nu_1^*) \left(\frac{k_3}{2k}\right)^{-\nu_1^*} \mathcal{I}_1(\nu_1^*)\nt
    &\qquad+\Gamma(-\nu_1^*) \left(\frac{k_3}{2k}\right)^{\nu_1^*}e^{3i\pi\nu_1^*}\mathcal{I}_1(-\nu_1^*)\bigg]+\frac{1}{k\sqrt{k}}\frac{i\sqrt{\pi}}{8}\Big[(\alpha_k^{(1)*})^2 e^{i(\frac{\pi}{2}+\pi\nu_2+6\mu k\eta_{\inf})}\Big(W^\beta_{+\nu_2}e^{i2\pi\nu_2\delta_I}\mathcal I_3(\nu_2)\nt
&\qquad+W^\beta_{-\nu_2}e^{i2\pi\nu_2(\delta_I-1)}\mathcal I_3(-\nu_2)-W^\alpha_{-\nu_2}e^{i2\pi\nu_2(\delta_I-1)}\mathcal I_3(-\nu_2)-W^\alpha_{+\nu_2}e^{i2\pi\nu_2\delta_I}\mathcal I_3(\nu_2)\Big)+2\alpha_k^{(1)*}\beta_k^{(1)*}\Big(W^\beta_{+\nu_2}\mathcal I_2(\nu_2)\nt
&\qquad+W^\beta_{-\nu_2}\mathcal I_2(-\nu_2)-W^\alpha_{-\nu_2}\mathcal I_2((-1)^{\delta_I}\nu_2)-W^\alpha_{+\nu_2}\mathcal I_2((-1)^{1-\delta_I}\nu_2)\Big)+(\beta_k^{(1)*})^2 e^{-i(\frac{\pi}{2}+\pi\nu_2+6\mu k\eta_{\inf})}\Big(W^\beta_{+\nu_2}\mathcal I_3^*((-1)^{\delta_I}\nu_2)\nt
&\qquad+W^\beta_{-\nu_2}e^{i2\pi\nu_2}\mathcal I_3^*((-1)^{1-\delta_I}\nu_2)-W^\alpha_{-\nu_2}e^{i2\pi\nu_2}\mathcal I_3^*((-1)^{1-\delta_I}\nu_2)-W^\alpha_{+\nu_2}\mathcal I_3^*((-1)^{\delta_I}\nu_2)\Big)\Big]\nt
&\qquad+\frac{1}{2k\sqrt{2k_3}}\bigg((\alpha_{k}^{(2)*})^2\alpha_{k_3}^{(2)*}\mathcal{I}_4(2k+k_3)+(\alpha_{k}^{(2)*})^2\beta_{k_3}^{(2)*}\mathcal{I}_4(2k-k_3)+2\alpha_k^{(2)*}\beta_{k}^{(2)*}(\beta_{k_3}^{(2)*}\mathcal{I}_4(-k_3)+\alpha_{k_3}^{(2)*}\mathcal{I}_4(k_3))\nt
&\qquad+(\beta_k^{(2)*})^2(\alpha_{k_3}^{(2)}\mathcal{I}_4(k_3-2k)+\beta_{k_3}^{(2)*}\mathcal{I}_4(-2k-k_3))\bigg)\bigg]\bigg\}.
\end{align}

The quantities $\mathcal{I}_{1,2,3}$ arise from time integrals involving products of Hankel mode functions and can be evaluated analytically once the background parameters are fixed; we therefore write them as explicit coefficients. The remaining contribution, denoted by $\mathcal{I}_4(K)$, corresponds to an oscillatory time integral whose momentum dependence is kept explicit. Different arguments of $\mathcal{I}_4(K)$ simply reflect different momentum combination appearing due to bogoliubov mixing. The term $\mathcal{I}_4(K)$ describes radiation-era contribution to the integral and, is given by
\begin{align}\label{I4}
\mathcal{I}_4(K)&=a_{\text{reh}}\frac{i e^{i K \eta _{\text{reh}}} \left(K (3 w+1) \left(\eta _{\text{reh}}-3 \mu  \eta _{\inf }\right)+2 i\right)}{K^2 (3 w+1) \left(\eta _{\text{reh}}-3 \mu  \eta _{\inf }\right)}\nt
    &\qquad-a_{\text{reh}}\frac{i e^{i K\eta} \left(-3 K \mu  (3 w+1) \eta _{\inf }+K (3 w-1) \eta _{\text{reh}}+2 K \eta+2 i\right)}{K^2 (3 w+1) \left(\eta _{\text{reh}}-3 \mu  \eta _{\inf }\right)},
\end{align}
with
\begin{equation}
    a_{\rm reh}=\frac{1}{H_{\rm inf}|\eta_{\text{inf}}|}\left(\frac{1+3w}{2\abs{\eta_{\text{inf}}}}\right)^{\frac{2}{1+3w}}\left(\eta_{\rm reh}+|3\eta_{\text{inf}}|\mu\right)^{\frac{2}{1+3w}}.
\end{equation}
On the superhorizon scale during reheating, we have: 
\begin{align}\label{I2}
\mathcal{I}_3&\approx\frac{1}{H_{\rm inf}(k/k_{\rm inf})}
\left(\frac{1+3w}{2k/k_{\rm inf}}\right)^{\frac{2}{1+3w}}
\Bigg[\frac{2}{3}(3w+1)(k\eta_{\rm reh})^{-\nu_2+\frac{5}{2}+\frac{2}{1+3w}}
\nonumber\Bigg(\frac{4^{\nu_2}\csc^2(\pi\nu_2)(k\eta_{\rm reh})^{-2\nu_2}}{\Gamma(1-\nu_2)^2(2\nu_2+6\nu_2 w-5w-3)}\nonumber \\[4pt]
&\qquad-\frac{3\times4^{-\nu_2}(\cot(\pi\nu_2)-i)^2(k\eta_{\rm reh})^{2\nu_2}}
{\Gamma(\nu_2+1)^2(2\nu_2+6\nu_2 w+15w+9)}-\frac{6(\cot(\pi\nu_2)-i)\csc(\pi\nu_2)}{\Gamma(1-\nu_2)\Gamma(\nu_2+1)(2\nu_2+6\nu_2 w-15w-9)}\Bigg)\nonumber \\[6pt]
&\qquad-\frac{4}{3}\left(\frac{2}{3w+1}\right)^{-\nu_2+\frac{3}{2}+\frac{2}{1+3w}}|k\eta_{\inf}|^{-\nu_2+\frac{5}{2}+\frac{2}{1+3w}}\Bigg(\frac{4^{\nu_2}\csc^2(\pi\nu_2)\left(\frac{2|k\eta_{\inf}|}{3w+1}\right)^{-2\nu_2}}{\Gamma(1\nu_2)^2(2\nu_2+6\nu_2 w-5w-3)}\nonumber \\[4pt]
&\qquad -\frac{3\times4^{-\nu_2}(\cot(\pi\nu_2)-i)^2\left(\frac{2|k\eta_{\inf}|}{3w+1}\right)^{2\nu_2}}{\Gamma(\nu_2+1)^2(2\nu_2+6\nu_2w+15w+9)}-\frac{6(\cot(\pi\nu_2)-i)\csc(\pi\nu_2)}{\Gamma(1-\nu_2)\Gamma(\nu_2+1)
(2\nu_2+6\nu_2 w-15w-9)}\Bigg)\Bigg],\\
\mathcal{I}_2 &\approx\frac{1}{H_{\rm inf}(k/k_{\rm inf})}\left(\frac{1+3w}{2k|\eta_{\rm inf}|}\right)^{\frac{2}{1+3w}}\Bigg[\frac{2}{3}(3w+1)(k\eta_{\rm reh})^{-\nu_2+\frac{2}{1+3w}+\frac{5}{2}}\Bigg(\frac{3\times4^{-\nu_2}\csc^2(\pi\nu_2)(k\eta_{\rm reh})^{2\nu_2}}{\Gamma(\nu_2+1)^2(2\nu_2+6\nu_2 w+15w+9)}\nonumber\\
&\qquad-\frac{4^{\nu_2}\csc^2(\pi\nu_2)(k\eta_{\rm reh})^{-2\nu_2}}{\Gamma(1-\nu_2)^2(2\nu_2+6\nu_2 w-5w-3)}+\frac{6\cot(\pi\nu_2)\csc(\pi\nu_2)}{\Gamma(1-\nu_2)\Gamma(\nu_2+1)(2\nu_2+6\nu_2 w-15w-9)}\Bigg)\nonumber\\[6pt]
&\qquad-\frac{4}{3}\left(\frac{2}{3w+1}\right)^{-\nu_2+\frac{2}{1+3w}+\frac{3}{2}}|k\eta_{\inf}|^{-\nu_2+\frac{2}{1+3w}+\frac{5}{2}}\Bigg(\frac{3\times4^{-\nu_2}\csc^2(\pi\nu_2)\left(\frac{2|k\eta_{\inf}|}{3w+1}\right)^{2\nu_2}}{\Gamma(\nu_2+1)^2(2\nu_2+6\nu_2 w+15w+9)}\nt[4pt]
&\qquad-\frac{4^{\nu_2}\csc^2(\pi\nu_2)\left(\frac{2|k\eta_{\inf}|}{3w+1}\right)^{-2\nu_2}}{\Gamma(1-\nu_2)^2(2\nu_2+6\nu_2 w-5w-3)}+\frac{6\cot(\pi\nu_2)\csc(\pi\nu_2)}{\Gamma(1-\nu_2)\Gamma(\nu_2+1)(2\nu_2+6\nu_2 w-15w-9)}\Bigg)\Bigg].
\end{align}
The full exact expression of $\mathcal{I}_{2,3}$ given in appendix \ref{app:reheating}.  The first term $\mathcal{I}_1$ is coming from the inflationary era and can be given by
\begin{align}\label{I1}
\mathcal{I}_1& \approx\frac{2}{3}\left(-k\eta_{\inf}\right)^{\frac{3}{2}-3\nu_1}\Bigg[\frac{3\times4^{-\nu_1}(\cot(\pi\nu_1)+i)^2\left(-k\eta_{\inf}\right)^{4\nu_1}}{(2\nu_1+3)\Gamma(\nu_1+1)^2}+\frac{6(\cot(\pi\nu_1)+i)\csc(\pi\nu_1)\left(-k\eta_{\inf}\right)^{2\nu_1}}{(2\nu_1-3)\Gamma(1-\nu_1)\Gamma(\nu_1+1)}\nt
&\qquad-\frac{4^{\nu_1}\csc^2(\pi\nu_1)}{(2\nu_1-1)\Gamma(1-\nu_1)^2}\Bigg]-\frac{\Gamma(-\frac{1}{4}+\frac{\nu_1}{2})}{2 \sqrt{\pi} \sin^2(\nu_1\pi) \Gamma(\frac{1}{4}-\frac{\nu_1}{2}) \Gamma(\frac{1}{4}+\frac{\nu_1}{2}) \Gamma(\frac{1}{4}+\frac{3\nu_1}{2})} \times \bigg[ \Gamma\left(\frac{3}{4}-\frac{3\nu_1}{2}\right) \Gamma\left(\frac{1}{4}+\frac{3\nu_1}{2}\right) \nt
&\qquad- 2 e^{-i\nu_1\pi} \Gamma\left(\frac{3}{4}-\frac{\nu_1}{2}\right) \Gamma\left(\frac{1}{4}+\frac{\nu_1}{2}\right) + e^{-2i\nu_1\pi} \Gamma\left(\frac{1}{4}-\frac{\nu_1}{2}\right) \Gamma\left(\frac{3}{4}+\frac{\nu_1}{2}\right) \bigg].
\end{align}
We note that the integral $\mathcal{I}_1$ requires special care for $\xi=0,\frac{1}{6}$ and for integer values of $\nu_1$, where the expression develops singular behavior. At $T = 10^{15}\,\mathrm{GeV}$ $(\eta_{\rm reh}=\eta_{\rm inf})$, and the bispectra reduces to the instantaneous reheating limit. As the reheating temperature is lowered ($\eta_{\rm reh}-\eta_{\rm inf}$ increases), contribution from $\mathcal{I}_3$ and $\mathcal{I}_2$ appears in the bispectra and, the resulting behavior is sensitive to the reheating history. In this regime, the shape of the bispectrum is affected by the reheating-era equation of state and the reheating temperature. 

\begin{figure}[t]
    \centering
    \begin{minipage}[b]{0.48\linewidth}
        \centering
        \includegraphics[width=\linewidth]{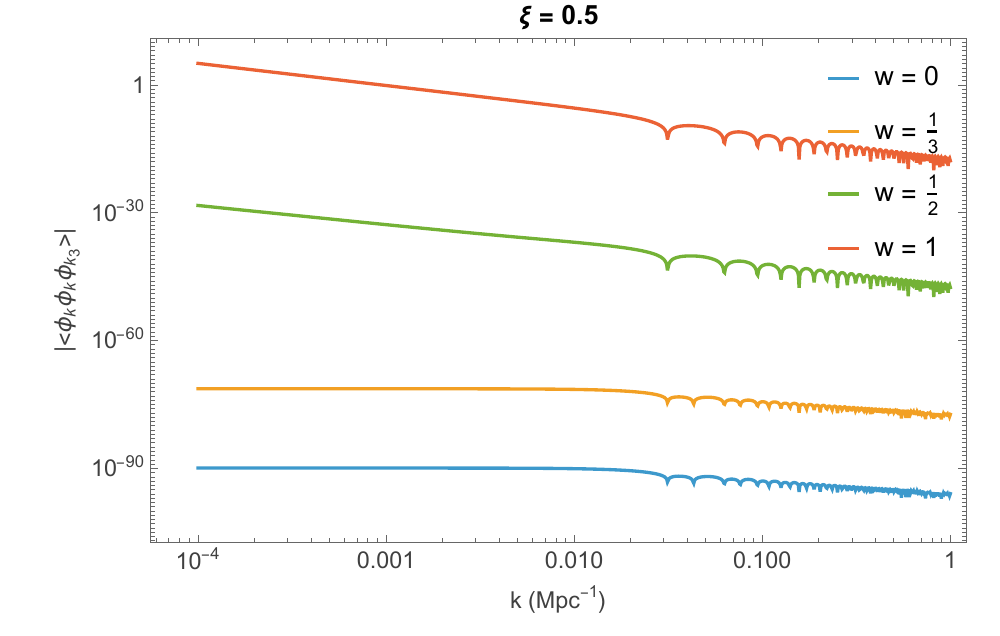}
        \captionsetup{style=centered_sub}\caption*{(a) Spectrum for $\xi=0.5$}
    \end{minipage}
    \hfill
    \begin{minipage}[b]{0.48\linewidth}
        \centering
        \includegraphics[width=\linewidth]{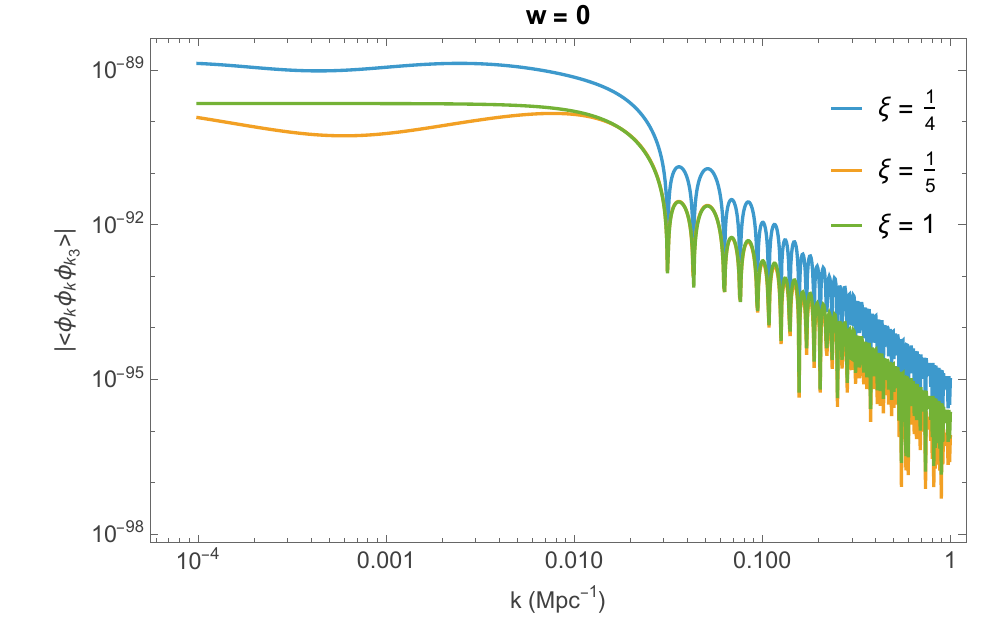}
        \captionsetup{style=centered_sub}\caption*{(b) Spectrum for  $w=0$}
    \end{minipage}
    \caption{The plot shows the bispectra in squeezed limit for scalar field, propagated to $\eta=\eta_{\rm CMB}$ hypersurface for the squeezed momenta of $k_3=10^{-8}\rm Mpc^{-1}$. (a) The bispectra is in persistent heavy field regime for $w=0$, and in transitional regime for $w\ge\frac{1}{3}$. (b) The bispectra is in persistent heavy field regime.}
    \label{fig:non-minimal-combined_spectra}
\end{figure}

From a phenomenological perspective, the observable signature of the bispectrum is dictated by the effective mass $(m_{\rm eff}^2=m^2+\xi\mathcal{R})$ of the spectator field across different cosmic epochs.  By parameterizing the field dynamics through the indices $\nu_1$ (inflation) and $\nu_2$ (reheating), we can rigorously classify the superhorizon evolution based on the scaling behavior of $\frac{k_3}{k}$ into three distinct physical regimes:
\begin{itemize}
\item \textbf{The Persistent Light-Field Regime ($\nu_1, \nu_2 \in \mathbb{R}$):} The field remains light throughout both inflation and reheating. In this scenario, the superhorizon modes can experience sustained tachyonic instability, leading to the enhancement of signal amplitude dictated by the specific values of the non-minimal coupling $\xi$ and the effective equation of state $w$. In this regime, we can drop the subleading term and thus the dominant behavior is dictated by:
    \begin{align}
    &\overline{\ev{\phi_{k}\phi_{k}\phi_{k_3}}}(\eta\ge\eta_{\rm reh})\nt
    &=\frac{g}{a(\eta)^3 k\sqrt{2k_3}}\Im\bigg\{ (\alpha_{k}^{(2)}e^{-ik\eta}+\beta_{k}^{(2)}e^{ik\eta})^2 (\alpha_{k_3}^{(2)}e^{-ik_3\eta}+\beta_{k_3}^{(2)}e^{ik_3\eta}) \bigg[\frac{i2^{\nu_1}\sqrt{\pi}\Gamma(\nu_1)e^{-3i(\nicefrac{\pi}{4}+\pi\nu_1/2)}}{8H_{\rm inf}k^{3/2}}\mathcal{I}_1(\nu_1)\left(\frac{k_3}{k}\right)^{-\nu_1}\nt
    &\qquad + \mathcal{U}_{k_3}(\nu_2) \frac{i2^{\nu_2}\sqrt{\pi}\Gamma(\nu_2)}{8k^{3/2}} \left(\frac{k_3}{k}\right)^{-\nu_2} \bigg( (\alpha_k^{(1)*})^2 e^{i(\frac{\pi}{4}+\frac{\pi\nu_2}{2}+6\mu k\eta_{\inf})} \mathcal{I}_3(\nu_2)  + 2\alpha_k^{(1)*}\beta_{k}^{(1)*}e^{-i(\frac{\pi}{4}+\frac{\pi\nu_2}{2})} \mathcal{I}_2(\nu_2) \nt
    &\qquad\qquad+ (\beta_{k}^{(1)*})^2 e^{-i(\frac{3\pi}{4}+\frac{3\pi\nu_2}{2}+6\mu k\eta_{\inf})} \mathcal{I}_3^*(\nu_2) \bigg)+\frac{1}{2k\sqrt{2k_3}}\bigg((\alpha_{k}^{(2)*})^2\bigg(\alpha_{k_3}^{(2)*}\mathcal{I}_4(2k+k_3)+\beta_{k_3}^{(2)*}\mathcal{I}_4(2k-k_3)\bigg)\nt
    &\qquad+2\alpha_k^{(2)*}\beta_{k}^{(2)*}(\beta_{k_3}^{(2)*}\mathcal{I}_4(-k_3)+\alpha_{k_3}^{(2)*}\mathcal{I}_4(k_3))+(\beta_k^{(2)*})^2(\alpha_{k_3}^{(2)}\mathcal{I}_4(k_3-2k)+\beta_{k_3}^{(2)*}\mathcal{I}_4(-2k-k_3))\bigg)\bigg]\bigg\},
    \end{align}
where
    \begin{equation}
        \mathcal{U}_{k_3}(\nu_2)=\beta_{k_3}^{(1)*}-ie^{i\pi\nu_2}\alpha_{k_3}^{(1)*}.
    \end{equation}
\item \textbf{The Persistent Heavy-Field Regime ($\nu_1 = i\mu_1$, $\nu_2 = i\mu_2$):} The spectator resides in the heavy-field regime until the end of reheating. In this limit, the bispectrum is purely oscillatory in momentum space with frequency set by $\nu_1$ and $\nu_2$
and the amplitude of the bispectrum suffers from a severe Boltzmann-like suppression active for all modes. This makes the bispectrum heavily suppressed and as a result it becomes challenging to detect them. The bispectrum has interesting dependence on non-minimal coupling for $w=0$, as changing $\xi$ introduces feature on both superhorizon and subhorizon scale, the expected bispectrum oscillation on longer wavelength is sensitive to non-minimal coupling $\xi$ as shown in Fig.~\ref{fig:non-minimal-combined_spectra}.
\item \textbf{The Transitional Regime ($\nu_1 \in i\mathbb{R}$, $\nu_2 \in \mathbb{R}$):} The field behaves as a heavy field during inflation $(\xi>\frac{3}{16})$ but transitions into a light field during the reheating epoch. While the inflationary contribution in this regime, scales as the purely oscillatory $(k/k_3)^{i\mu_1}$, the reheating part of the correlator contains the factor $(k/k_3)^{\nu_2}$, which experiences a late-time tachyonic enhancement for specific equations of state (e.g., $w > 1/3$). The Boltzmann-like suppression acts similarly on all modes whereas tachyonic instability is most effective at superhorizon scale. This transition implies that while non-Gaussianity is generated in the heavy regime during inflation, it is primarily produced in the ultralight (tachyonic) regime during reheating for $\xi>\frac{3}{16}$. Thus the bispectrum describes two overlapping signal originating from different dynamical behavior of the same field during two cosmological eras. This signal can not be produced by inflation alone, and it is only present from the inclusion of reheating era. The leading order behavior in this regime can be given as:
     \begin{align}
    &\overline{\ev{\phi_{k}\phi_{k}\phi_{k_3}}}(\eta\ge\eta_{\rm reh})\nt
    &=\frac{g}{a(\eta)^3k\sqrt{2k_3}}\Im\bigg\{(\alpha_{k}^{(2)}e^{-ik\eta}+\beta_{k}^{(2)}e^{ik\eta})^2(\alpha_{k_3}^{(2)}e^{-ik_3\eta}+\beta_{k_3}^{(2)}e^{ik_3\eta})\nt
    &\times\bigg[\frac{i\pi\,e^{-\frac{\pi\mu_1}{2}-3i\nicefrac{\pi}{4}}}{8H_{\rm inf}k^{3/2}\sqrt{\mu_1(1-e^{-2\pi\mu_1})}}\left(
    2^{-i\mu_1+\frac{1}{2}}\,e^{-3\pi\mu_1/2-i\delta}\mathcal{I}_1(-i\mu_1)e^{i\mu_1 \ln(k_3/k)}+2^{i\mu_1+\frac{1}{2}}\,e^{3\pi\mu_1/2+i\delta}\mathcal{I}_1(i\mu_1)e^{i\mu_1 \ln(k/k_3)}\right)\nt
    &\qquad+\mathcal{U}_{k_3}(\nu_2) \frac{i2^{\nu_2}\sqrt{\pi}\Gamma(\nu_2)}{8k^{3/2}} \left(\frac{k_3}{k}\right)^{-\nu_2} \bigg( (\alpha_k^{(1)*})^2 e^{i(\frac{\pi}{4}+\frac{\pi\nu_2}{2}+6\mu k\eta_{\inf})} \mathcal{I}_3(\nu_2)  + 2\alpha_k^{(1)*}\beta_{k}^{(1)*}e^{-i(\frac{\pi}{4}+\frac{\pi\nu_2}{2})} \mathcal{I}_2(\nu_2) \nt
    &\qquad+ (\beta_{k}^{(1)*})^2 e^{-i(\frac{3\pi}{4}+\frac{3\pi\nu_2}{2}+6\mu k\eta_{\inf})} \mathcal{I}_3^*(\nu_2) \bigg)+\frac{1}{2k\sqrt{2k_3}}\bigg((\alpha_{k}^{(2)*})^2\alpha_{k_3}^{(2)*}\mathcal{I}_4(2k+k_3)+(\alpha_{k}^{(2)*})^2\beta_{k_3}^{(2)*}\mathcal{I}_4(2k-k_3)\nt
    &\qquad+2\alpha_k^{(2)*}\beta_{k}^{(2)*}(\beta_{k_3}^{(2)*}\mathcal{I}_4(-k_3)+\alpha_{k_3}^{(2)*}\mathcal{I}_4(k_3))+(\beta_k^{(2)*})^2(\alpha_{k_3}^{(2)}\mathcal{I}_4(k_3-2k)+\beta_{k_3}^{(2)*}\mathcal{I}_4(-2k-k_3))\bigg)\bigg]\bigg\}.
\end{align}
Figure~\ref{fig:non-minimal-combined_spectra} shows that while the superhorizon bispectrum corresponding to $\xi=0.5$ is nearly flat for $w =0,1/3$, it exhibits decaying behavior for $w > 1/3$. On the subhorizon scale, the dynamics is universally oscillatory, but the scale dependence shows a clear dependence on the reheating equation of state. 
\end{itemize}

\subsection{Discussions and Outlook}
In the conventional treatment one computes the correlator at the end of inflation, then propagates it forward using a transfer function that encodes the subsequent linear evolution. 
In the present work we relaxed this assumption, to undertake full quantum treatment by defining adiabatic vacuum both in the asymptotic past and future, and finally demonstrate that correlation functions indeed carry imprints of different cosmological phases in a non-trivial manner. The correlators of an interacting field (such as $\phi^3$ theory presented through out) is not a linearly propagated quantity, unlike the curvature perturbation $\zeta$, which is conserved on superhorizon scales in single-field models. In the QFT framework, the non-adiabatic evolution through several cosmological phases leads a non-trivial contributions to the phase and amplitude of the cosmological correlator. A transfer function, which acts multiplicatively on the inflationary correlators, cannot reproduce those features by construction. Furthermore, one of our prime motivations of the present work is to explore the impact of reheating on the cosmological correlators. 

Note that the macroscopic background of reheating can be parameterized by a classical fluid equation of state $w$, however, the evolution of the perturbations across it may not be treated purely classically. The fluctuations are actively participating in scattering and decay processes governed by the non-linear interaction vertex. On subhorizon scales, these are fundamentally quantum mechanical phenomena that are entirely invisible to classical fluid equations of motion. Furthermore, the particle production across the transition, encoded in the Bogoliubov coefficients, is a quantum mechanical phenomenon arising from the non-adiabatic evolution of the mode functions. 
By rigorously tracking the Bogoliubov mode mixing and field-theoretic interaction history through the reheating phase, this work derives the modified, dynamically evolved various cosmological correlation function.

The implications of keeping interaction turned on during reheating are most transparently analyzed through the comoving correlator $a(\eta)^3\ev{\phi_{k_1}\phi_{k_2}\phi_{k_3}}$, which isolates the effects arising from interaction channel. The bispectrum is sensitive to the expansion history through the residual factor of $a(\eta)$ in the in-in integrand. For a quartic interaction this factor cancels exactly, leaving a Minkowski-like integral insensitive to the background, which is why we restricted attention to the cubic case. For the cubic interaction the residual $a(\eta)$ in the vertex ensures that the time integral during reheating is weighted differently from that of the radiation-era integral, and this difference is controlled by the reheating equation of state $w$. The equation of state is therefore not merely a kinematic parameter describing how fast the universe expands; it may actively controls the amplitude and phase of the correlators.

We have done two important cases studies. For conformally coupled fields, where the tachyonic instability is absent, the reheating correction to the superhorizon bispectrum is suppressed. Subhorizon modes, by contrast, oscillate freely throughout the reheating interval and accumulate a phase that is sensitive to the equation of state during that interval. For the non-minimally coupled case the physics is richer. As it allows for fields in heavy field regime during inflation (e.g. $\xi>\frac{3}{16}$) to experience light field tachyonic enhancement during reheating (for $w>\frac{1}{3}$). This shift in the dynamical behavior is captured by the reheating contribution and post reheating evolution of bispectra. The resulting behavior has interesting dependence on reheating equation of state as it affects the superhorizon behavior as well as enhance the overall signal strength depending on the reheating temperature and reheating EoS.
The sensitivity of this feature to the post-inflationary phases is the clearest illustration of the broader point: the quantum mechanical history of the mode functions through the post-inflationary era leaves imprints that are not accessible to conventional classical treatment of transporting correlation functions from inflation end to CMB sky.

There is a broader lesson for the program of using cosmological correlators to probe the reheating era. The standard approach to reheating constraints uses the spectral index $n_s$ and its running to constrain the number of e-folds of reheating through the inflationary consistency relations; this is sensitive to reheating only through the kinematic mapping between the inflationary scale and the CMB scale. The present paper demonstrates a different channel: the dynamics of the reheating phase directly modify the amplitude, shape, and spectral tilt of the correlators of fields that are active during reheating. This is a more direct probe of the expansion history, in the sense that it depends on the integral of the interaction over the reheating interval rather than just on the total duration. If the spectator field is identified with a dark matter candidate or a field responsible for isocurvature perturbations, this sensitivity translates into a direct connection between the CMB isocurvature spectrum and the microphysics of reheating.

The connection to the isocurvature power spectrum explored in Sec.~\ref{section 2} deserves particular emphasis in this context. The observation that the qualitative features of the spectator field correlators carry over to the entropy perturbation $\mathcal{S}$, with the power spectrum $\mathcal{P}_\mathcal{S}$ proportional to $\mathcal{P}_\phi$ and the bispectrum of $\mathcal{S}$ related to the four-point function of $\phi$, suggests that the reheating imprints computed in this paper are in principle observable, not through the primordial curvature perturbation, but through the isocurvature sector. Future CMB experiments with improved sensitivity to the isocurvature fraction and to bispectra in the squeezed limit will therefore probe exactly the regime where the reheating corrections computed  on large scales (small $k/k_\mathrm{inf}$) are most significant.

The present work is a first step in this direction, carried out within the controlled setting of a massless spectator field with a cubic interaction and a single-phase reheating epoch. Many of the most interesting physical questions, the role of preheating, the effect of inflaton--spectator couplings, the full triangle-shape dependence of the bispectrum for arbitrary $\xi$, and the precise connection to observable isocurvature statistics, remain open. What this paper establishes is an explicit QFT framework 
which requires treating post-inflationary epoch as an active phase that can also inject its own non-trivial imprints on the cosmological  correlators on the CMB sky.

\section{Appendix}
\appendix

\section{Estimation of $\eta_{\rm CMB}$}
To evaluate the power spectrum at physically relevant scales, we must determine the conformal time corresponding to the decoupling of the Cosmic Microwave Background (CMB). We approximate the post-inflationary history as a transition to a radiation-dominated (RD) fluid. Matching the de Sitter phase to the radiation phase at $\eta_{\text{inf}}$ imposes continuity on the scale factor $a(\eta)$ and the Hubble parameter, yielding the scale factor as:
\begin{equation}
    a(\eta) = -\frac{a_{\text{inf}}}{\eta_{\text{inf}}}\,(\eta - 2\eta_{\text{inf}}), \quad \text{for } \eta > \eta_{\text{inf}},
\end{equation}
where $\eta_{\text{inf}} < 0$ marks the end of inflation. The target epoch, $\eta_{\text{CMB}}$, corresponds to a redshift $z_{\text{CMB}} \approx 1100$, implying a scale factor of:
\begin{equation}
    \frac{a_{\text{CMB}}}{a_0} = \frac{1}{1 + z_{\text{CMB}}} \approx 9.1 \times 10^{-4}.
\end{equation}
where we have made $a_0$ explicit for this section. By inverting the scale factor solution $a(\eta_{\text{CMB}}) = a_{\text{CMB}}$, we can express the decoupling time in terms of inflationary parameters:
\begin{align}
    \eta_{\text{CMB}} &= 2\eta_{\text{inf}} - \frac{\eta_{\text{inf}}}{a_{\text{inf}}} a_{\text{CMB}} \nonumber \\
    &= \eta_{\text{inf}} \left( 2 - \frac{a_{\text{CMB}}/a_0}{a_{\text{inf}}/a_0} \right).
\end{align}
Assuming a high-scale inflation model ($H_{\text{inf}}=10^{-5}M_{\rm pl}\sim10^{51}\rm Mpc^{-1}$), the scale factor at the end of inflation is approximately $a_{\text{inf}} \sim a_010^{-28}$ (given $a_0\eta_{\text{inf}} \sim -10^{-23}$ Mpc). Substituting these values yields:
\begin{equation}
    a_0\eta_{\text{CMB}} \approx a_0\eta_{\text{inf}} \left( - \frac{9.1 \times 10^{-4}}{10^{-28}} \right) \approx 10^2 \, \text{Mpc}.
\end{equation}

We have kept $a_0$ implicit in the expression for $\eta_{\rm CMB}$ throughout the paper. To ensure this approximation is physically reasonable, we compute the coordinate time $t$ elapsed between the end of inflation and decoupling:
\begin{equation}
    t = \int_{\eta_{\text{inf}}}^{\eta_{\text{CMB}}} a(\eta) \, d\eta = \frac{a_{\text{inf}}/a_0}{2a_0|\eta_{\text{inf}}|} \left[ (a_0\eta_{\text{CMB}} - 2a_0\eta_{\text{inf}})^2 - a_0^2\eta_{\text{inf}}^2 \right].
\end{equation}
Using the derived values, we find $t \approx 0.041$ Mpc (approx. 133,000 years). While this underestimates the standard $\Lambda$CDM value ($\approx 380,000$ years) due to the neglect of the matter-dominated transition, it remains within the correct order of magnitude. This confirms that $\eta_{\text{CMB}} \sim 100$ Mpc is a sufficient approximation for evaluating the spectral shape at late times.

\section{Explicit Calculation of the Bispectrum Contributions}
In this appendix, we provide the detailed evaluation of the time integrals contributing to the bispectrum.
\subsection{Evaluation of the Radiation Contribution}
\label{app:radiation}
The radiation part of conformally coupled bispectrum given in \eqref{conformal:radiation_part} is calculated as:
\begin{align}
    &\overline{\ev{\phi_{k_1}\phi_{k_2}\phi_{k_3}}}_{\text{radiation part}}\nt
    &\quad=2g\Im{\int_{\eta_{\text{reh}}}^{\eta}D\eta' W_{k_1}(\eta, \eta')W_{k_2}(\eta,\eta')W_{k_3}(\eta,\eta')}\nt
    &\quad=2g\Im{\frac{e^{-iK\eta}}{8a(\eta)^3k_1k_2k_3}\int_{\eta_{\text{inf}}}^{\eta}d\eta' a(\eta')e^{iK\eta'}}\nt
    &\quad=\frac{2ga_{\text{reh}}}{8a(\eta)^3k_1k_2k_3}\Im{e^{-iK\eta}\int_{\eta_{\text{reh}}}^{\eta}d\eta' \left(1+\frac{2(\eta_{\text{reh}}-\eta)/(1+3w)}{3\mu\eta_{\text{inf}}-\eta_{\text{reh}}}\right) e^{iK\eta'}}\nt
    &\quad=\frac{2ga_{\text{reh}}}{8a(\eta)^3k_1k_2k_3K^2(1+3w)\left(\eta _{\text{reh}}-3 \mu  \eta _{\inf }\right)}\Im\bigg\{ie^{-iK\eta}  \bigg[e^{i K \eta _{\text{reh}}} \left(2 i+K(1+3w)(\eta_{\rm reh}-3\mu\eta_{\rm inf})\right)\nt
    &\qquad-e^{iK \eta} \big(2i+K\eta_{\rm reh}(-1+3w)+2K\eta-3\mu K(1+3w)\eta_{\rm inf})\big)\bigg]  \bigg\}\nt
    &\quad=\frac{2ga_{\text{reh}}}{8a(\eta)^3k_1k_2k_3K^2(1+3w)\left(\eta _{\text{reh}}-3 \mu  \eta _{\inf }\right)}\Im{2(1-e^{iK(\eta_{\rm reh}-\eta)})+iK(1+3w)(\eta_{\rm reh}-3\mu\eta_{\inf})e^{iK(\eta_{\rm reh}-\eta)}}\nt
    &\qquad+\frac{2ga_{\text{reh}}}{8a(\eta)^3k_1k_2k_3K^2(1+3w)\left(\eta _{\text{reh}}-3 \mu  \eta _{\inf }\right)}\Im{2-2iK\eta-iK\eta_{\rm reh}(-1+3w)+3iK(1+w)\eta_{\inf}}\nt
    &\quad=\frac{2ga_{\text{reh}}[-2\sin(K(\eta_{\rm reh}-\eta))+K\{(1+3w)\eta_{\rm reh}-3(1+w)\eta_{\rm inf}\}\cos{K(\eta_{\rm reh}-\eta)}]}{2a(\eta)^3k_1k_2k_3K^2(1+3w)\left(\eta _{\text{reh}}-3 \mu  \eta _{\inf }\right)}\nt
    &\qquad-\frac{2ga_{\text{reh}}K[(-1+3w)\eta_{\rm reh}+2\eta-3(1+w)\eta_{\inf}]}{8a(\eta)^3k_1k_2k_3K^2(1+3w)\left(\eta _{\text{reh}}-3 \mu  \eta _{\inf }\right)}.
\end{align}
The integral in the radiation part of non-minimally coupled scalar field bispectrum given in \eqref{I4} can be calculated in following manner:
\begin{align}
\mathcal{I}_4(K)&=\int_{\eta_{\text{reh}}}^{\eta}d\eta' a(\eta_{\text{reh}})\left(1+\frac{2(\eta_{\text{reh}}-\eta)/(1+3w)}{3\mu\eta_{\text{inf}}-\eta_{\text{reh}}}\right) e^{iK\eta'}\nt
    &=a(\eta_{\text{reh}})\int_{\eta_{\text{reh}}}^{\eta}d\eta' \left(1+\frac{2(\eta_{\text{reh}}-\eta)/(1+3w)}{3\mu\eta_{\text{inf}}-\eta_{\text{reh}}}\right) e^{iK\eta'}\nt
    &=a_{\text{reh}}\frac{i e^{i K \eta _{\text{reh}}} \left(K (3 w+1) \left(\eta _{\text{reh}}-3 \mu  \eta _{\inf }\right)+2 i\right)-i e^{i K\eta} \left(-3 K \mu  (3 w+1) \eta _{\inf }+K (3 w-1) \eta _{\text{reh}}+2 K \eta+2 i\right)}{K^2 (3 w+1) \left(\eta _{\text{reh}}-3 \mu  \eta _{\inf }\right)}.
\end{align}
\subsection{Evaluation of the Reheating Contribution}\label{app:reheating}
The reheating part of the correlator in \eqref{reheating-part-nonminimal} can be found as:
\begin{align}
\overline{\langle\phi_{k}\phi_{k}\phi_{k_3}\rangle}_{\text{reh}}&= 2g \, \text{Im} \Bigg\{ f_{k}^2 f_{k_3} \frac{i\sqrt{\pi}}{8} \nt
&\qquad\bigg[(\alpha_k^{(1)*})^2 e^{i(\frac{\pi}{2} + \pi\nu_2^* + 6\mu k\eta_{\inf})} \bigg(\beta_{k_3}^{(1)*}e^{-i(\frac{\pi}{4}+\frac{\pi\nu_2}{2})} 2^{\nu_2}\Gamma(\nu_2) k_3^{-\nu_2} \int_{\eta_{\text{inf}}}^{\eta_{\text{reh}}} d\eta' \, a(\eta') \bigl[H_{\nu_2^*}^{(1)}(k\tilde\eta')\bigr]^2 (\tilde\eta')^{-\nu_2+\frac{3}{2}} \nt
&\qquad\quad + \beta_{k_3}^{(1)*}e^{-i(\frac{\pi}{4}-\frac{\pi\nu_2}{2})} 2^{-\nu_2}\Gamma(-\nu_2) k_3^{\nu_2} \int_{\eta_{\text{inf}}}^{\eta_{\text{reh}}} d\eta' \, a(\eta') \bigl[H_{\nu_2^*}^{(1)}(k\tilde\eta')\bigr]^2 (\tilde\eta')^{\nu_2+\frac{3}{2}} \nt
&\qquad\quad - \alpha_{k_3}^{(1)*}e^{i(\frac{\pi}{4}+\frac{\pi\nu_2^*}{2})} 2^{\nu_2^*}\Gamma(\nu_2^*) k_3^{-\nu_2^*} \int_{\eta_{\text{inf}}}^{\eta_{\text{reh}}} d\eta' \, a(\eta') \bigl[H_{\nu_2^*}^{(1)}(k\tilde\eta')\bigr]^2 (\tilde\eta')^{-\nu_2^*+\frac{3}{2}} \nt
&\qquad\quad - \alpha_{k_3}^{(1)*}e^{i(\frac{\pi}{4}-\frac{\pi\nu_2^*}{2})} 2^{-\nu_2^*}\Gamma(-\nu_2^*) k_3^{\nu_2^*} \int_{\eta_{\text{inf}}}^{\eta_{\text{reh}}} d\eta' \, a(\eta') \bigl[H_{\nu_2^*}^{(1)}(k\tilde\eta')\bigr]^2 (\tilde\eta')^{\nu_2^*+\frac{3}{2}} \bigg) \nt
&\quad + 2\alpha_k^{(1)*}\beta_k^{(1)*} e^{i\frac{\pi}{2}(\nu_2^*-\nu_2)} \bigg( \beta_{k_3}^{(1)*}e^{-i(\frac{\pi}{4}+\frac{\pi\nu_2}{2})} 2^{\nu_2}\Gamma(\nu_2) k_3^{-\nu_2} \int_{\eta_{\text{inf}}}^{\eta_{\text{reh}}} d\eta' \, a(\eta') H_{\nu_2^*}^{(1)}(k\tilde\eta') H_{\nu_2}^{(2)}(k\tilde\eta') (\tilde\eta')^{-\nu_2+\frac{3}{2}} \nt
&\qquad\quad + \beta_{k_3}^{(1)*}e^{-i(\frac{\pi}{4}-\frac{\pi\nu_2}{2})} 2^{-\nu_2}\Gamma(-\nu_2) k_3^{\nu_2} \int_{\eta_{\text{inf}}}^{\eta_{\text{reh}}} d\eta' \, a(\eta') H_{\nu_2^*}^{(1)}(k\tilde\eta') H_{\nu_2}^{(2)}(k\tilde\eta') (\tilde\eta')^{\nu_2+\frac{3}{2}} \nt
&\qquad\quad - \alpha_{k_3}^{(1)*}e^{i(\frac{\pi}{4}+\frac{\pi\nu_2^*}{2})} 2^{\nu_2^*}\Gamma(\nu_2^*) k_3^{-\nu_2^*} \int_{\eta_{\text{inf}}}^{\eta_{\text{reh}}} d\eta' \, a(\eta') H_{\nu_2^*}^{(1)}(k\tilde\eta') H_{\nu_2}^{(2)}(k\tilde\eta') (\tilde\eta')^{-\nu_2^*+\frac{3}{2}} \nt
&\qquad\quad - \alpha_{k_3}^{(1)*}e^{i(\frac{\pi}{4}-\frac{\pi\nu_2^*}{2})} 2^{-\nu_2^*}\Gamma(-\nu_2^*) k_3^{\nu_2^*} \int_{\eta_{\text{inf}}}^{\eta_{\text{reh}}} d\eta' \, a(\eta') H_{\nu_2^*}^{(1)}(k\tilde\eta') H_{\nu_2}^{(2)}(k\tilde\eta') (\tilde\eta')^{\nu_2^*+\frac{3}{2}} \bigg) \nt
&\quad + (\beta_k^{(1)*})^2 e^{-i(\frac{\pi}{2} + \pi\nu_2 + 6\mu k\eta_{\inf})} \bigg( \beta_{k_3}^{(1)*}e^{-i(\frac{\pi}{4}+\frac{\pi\nu_2}{2})} 2^{\nu_2}\Gamma(\nu_2) k_3^{-\nu_2} \int_{\eta_{\text{inf}}}^{\eta_{\text{reh}}} d\eta' \, a(\eta') \bigl[H_{\nu_2}^{(2)}(k\tilde\eta')\bigr]^2 (\tilde\eta')^{-\nu_2+\frac{3}{2}} \nt
&\qquad\quad + \beta_{k_3}^{(1)*}e^{-i(\frac{\pi}{4}-\frac{\pi\nu_2}{2})} 2^{-\nu_2}\Gamma(-\nu_2) k_3^{\nu_2} \int_{\eta_{\text{inf}}}^{\eta_{\text{reh}}} d\eta' \, a(\eta') \bigl[H_{\nu_2}^{(2)}(k\tilde\eta')\bigr]^2 (\tilde\eta')^{\nu_2+\frac{3}{2}} \nt
&\qquad\quad - \alpha_{k_3}^{(1)*}e^{i(\frac{\pi}{4}+\frac{\pi\nu_2^*}{2})} 2^{\nu_2^*}\Gamma(\nu_2^*) k_3^{-\nu_2^*} \int_{\eta_{\text{inf}}}^{\eta_{\text{reh}}} d\eta' \, a(\eta') \bigl[H_{\nu_2}^{(2)}(k\tilde\eta')\bigr]^2 (\tilde\eta')^{-\nu_2^*+\frac{3}{2}} \nt
&\qquad\quad - \alpha_{k_3}^{(1)*}e^{i(\frac{\pi}{4}-\frac{\pi\nu_2^*}{2})} 2^{-\nu_2^*}\Gamma(-\nu_2^*) k_3^{\nu_2^*} \int_{\eta_{\text{inf}}}^{\eta_{\text{reh}}} d\eta' \, a(\eta') \bigl[H_{\nu_2}^{(2)}(k\tilde\eta')\bigr]^2 (\tilde\eta')^{\nu_2^*+\frac{3}{2}} \bigg) \bigg] \Bigg\}.
\end{align}
Depending on weather $\nu_2$ is real or imaginary, one can use:
\begin{align}
H^{(1)}_{\nu^*}(z)&=
\begin{cases}
H^{(1)}_\nu(z) & \nu\in\mathbb{R} \\
e^{i\pi\nu}H^{(1)}_\nu(z) & \nu\in i\mathbb{R}
\end{cases}\\
    H_{\nu}^{(1)}(z)&=e^{-i\pi\nu} H_{-\nu}^{(1)}(z)\\
    H_{\nu}^{(2)}(z)&=e^{i\pi\nu} H_{-\nu}^{(2)}(z)
\end{align}
This allows to us to simplify the reheating part of the bispectra. For real $\nu_2$, it takes the following form:
\begin{align*}
\overline{\langle\phi_{k}\phi_{k}\phi_{k_3}\rangle}_{\text{reh}} &= 2g \, \text{Im} \Bigg\{ f_{k}^2 f_{k_3} \frac{i\sqrt{\pi}}{8} \Bigg[ \\
&\quad (\alpha_k^{(1)*})^2 e^{i(\frac{\pi}{2} + \pi\nu_2 + 6\mu k\eta_{\inf})} \bigg(\beta_{k_3}^{(1)*}e^{-i(\frac{\pi}{4}+\frac{\pi\nu_2}{2})} 2^{\nu_2}\Gamma(\nu_2) k_3^{-\nu_2} \int_{\eta_{\text{inf}}}^{\eta_{\text{reh}}} d\eta' \, a(\eta') \bigl[H_{\nu_2}^{(1)}(k\tilde\eta')\bigr]^2 (\tilde\eta')^{-\nu_2+\frac{3}{2}} \\
&\qquad + \beta_{k_3}^{(1)*}e^{-i(\frac{\pi}{4}-\frac{\pi\nu_2}{2})} 2^{-\nu_2}\Gamma(-\nu_2) k_3^{\nu_2} \int_{\eta_{\text{inf}}}^{\eta_{\text{reh}}} d\eta' \, a(\eta')  e^{-2i\pi\nu_2}\bigl[H_{-\nu_2}^{(1)}(k\tilde\eta')\bigr]^2 (\tilde\eta')^{\nu_2+\frac{3}{2}} \\
&\qquad - \alpha_{k_3}^{(1)*}e^{i(\frac{\pi}{4}+\frac{\pi\nu_2}{2})} 2^{\nu_2}\Gamma(\nu_2) k_3^{-\nu_2} \int_{\eta_{\text{inf}}}^{\eta_{\text{reh}}} d\eta' \, a(\eta') \bigl[H_{\nu_2}^{(1)}(k\tilde\eta')\bigr]^2 (\tilde\eta')^{-\nu_2+\frac{3}{2}} \\
&\qquad - \alpha_{k_3}^{(1)*}e^{i(\frac{\pi}{4}-\frac{\pi\nu_2}{2})} 2^{-\nu_2}\Gamma(-\nu_2) k_3^{\nu_2} \int_{\eta_{\text{inf}}}^{\eta_{\text{reh}}} d\eta' \, a(\eta')e^{-2i\pi\nu_2} \bigl[H_{-\nu_2}^{(1)}(k\tilde\eta')\bigr]^2 (\tilde\eta')^{\nu_2+\frac{3}{2}} \bigg) \\
&\quad + 2\alpha_k^{(1)*}\beta_k^{(1)*} \bigg( \beta_{k_3}^{(1)*}e^{-i(\frac{\pi}{4}+\frac{\pi\nu_2}{2})} 2^{\nu_2}\Gamma(\nu_2) k_3^{-\nu_2} \int_{\eta_{\text{inf}}}^{\eta_{\text{reh}}} d\eta' \, a(\eta') H_{\nu_2}^{(1)}(k\tilde\eta') H_{\nu_2}^{(2)}(k\tilde\eta') (\tilde\eta')^{-\nu_2+\frac{3}{2}} \\
&\qquad + \beta_{k_3}^{(1)*}e^{-i(\frac{\pi}{4}-\frac{\pi\nu_2}{2})} 2^{-\nu_2}\Gamma(-\nu_2) k_3^{\nu_2} \int_{\eta_{\text{inf}}}^{\eta_{\text{reh}}} d\eta' \, a(\eta') H_{\nu_2}^{(1)}(k\tilde\eta') H_{\nu_2}^{(2)}(k\tilde\eta') (\tilde\eta')^{\nu_2+\frac{3}{2}} \\
&\qquad - \alpha_{k_3}^{(1)*}e^{i(\frac{\pi}{4}+\frac{\pi\nu_2}{2})} 2^{\nu_2}\Gamma(\nu_2) k_3^{-\nu_2} \int_{\eta_{\text{inf}}}^{\eta_{\text{reh}}} d\eta' \, a(\eta') H_{\nu_2}^{(1)}(k\tilde\eta') H_{\nu_2}^{(2)}(k\tilde\eta') (\tilde\eta')^{-\nu_2+\frac{3}{2}} \\
&\qquad - \alpha_{k_3}^{(1)*}e^{i(\frac{\pi}{4}-\frac{\pi\nu_2}{2})} 2^{-\nu_2}\Gamma(-\nu_2) k_3^{\nu_2} \int_{\eta_{\text{inf}}}^{\eta_{\text{reh}}} d\eta' \, a(\eta') H_{\nu_2}^{(1)}(k\tilde\eta') H_{\nu_2}^{(2)}(k\tilde\eta') (\tilde\eta')^{\nu_2+\frac{3}{2}} \bigg) \\
&\quad + (\beta_k^{(1)*})^2 e^{-i(\frac{\pi}{2} + \pi\nu_2 + 6\mu k\eta_{\inf})} \bigg( \beta_{k_3}^{(1)*}e^{-i(\frac{\pi}{4}+\frac{\pi\nu_2}{2})} 2^{\nu_2}\Gamma(\nu_2) k_3^{-\nu_2} \int_{\eta_{\text{inf}}}^{\eta_{\text{reh}}} d\eta' \, a(\eta') \bigl[H_{\nu_2}^{(2)}(k\tilde\eta')\bigr]^2 (\tilde\eta')^{-\nu_2+\frac{3}{2}} \\
&\qquad + \beta_{k_3}^{(1)*}e^{-i(\frac{\pi}{4}-\frac{\pi\nu_2}{2})} 2^{-\nu_2}\Gamma(-\nu_2) k_3^{\nu_2} \int_{\eta_{\text{inf}}}^{\eta_{\text{reh}}} d\eta' \, a(\eta') e^{2i\pi\nu_2}\bigl[H_{-\nu_2}^{(2)}(k\tilde\eta')\bigr]^2 (\tilde\eta')^{\nu_2+\frac{3}{2}} \\
&\qquad - \alpha_{k_3}^{(1)*}e^{i(\frac{\pi}{4}+\frac{\pi\nu_2}{2})} 2^{\nu_2}\Gamma(\nu_2) k_3^{-\nu_2} \int_{\eta_{\text{inf}}}^{\eta_{\text{reh}}} d\eta' \, a(\eta') \bigl[H_{\nu_2}^{(2)}(k\tilde\eta')\bigr]^2 (\tilde\eta')^{-\nu_2+\frac{3}{2}} \\
&\qquad - \alpha_{k_3}^{(1)*}e^{i(\frac{\pi}{4}-\frac{\pi\nu_2}{2})} 2^{-\nu_2}\Gamma(-\nu_2) k_3^{\nu_2} \int_{\eta_{\text{inf}}}^{\eta_{\text{reh}}} d\eta' \, a(\eta') e^{2i\pi\nu_2}\bigl[H_{-\nu_2}^{(2)}(k\tilde\eta')\bigr]^2 (\tilde\eta')^{\nu_2+\frac{3}{2}} \bigg) \Bigg] \Bigg\}\nt
\overline{\langle\phi_{k}\phi_{k}\phi_{k_3}\rangle}_{\text{reh}} &= 2g \, \text{Im} \Bigg\{ \frac{f_{k}^2 f_{k_3}}{k\sqrt{k}} \frac{i\sqrt{\pi}}{8} \Bigg[ \\
&\quad (\alpha_k^{(1)*})^2 e^{i(\frac{\pi}{2} + \pi\nu_2 + 6\mu k\eta_{\inf})} \bigg(\beta_{k_3}^{(1)*}e^{-i(\frac{\pi}{4}+\frac{\pi\nu_2}{2})} 2^{\nu_2}\Gamma(\nu_2) \left(\frac{k_3}{k}\right)^{-\nu_2} \mathcal{I}_{3}(\nu_2)\\
&\qquad + \beta_{k_3}^{(1)*}e^{-i(\frac{\pi}{4}-\frac{\pi\nu_2}{2})} 2^{-\nu_2}\Gamma(-\nu_2) e^{-2i\pi\nu_2} \left(\frac{k_3}{k}\right)^{\nu_2} \mathcal{I}_{3}(-\nu_2) \\
&\qquad - \alpha_{k_3}^{(1)*}e^{i(\frac{\pi}{4}+\frac{\pi\nu_2}{2})} 2^{\nu_2}\Gamma(\nu_2)  \left(\frac{k_3}{k}\right)^{-\nu_2} \mathcal{I}_{3}(\nu_2) \\
&\qquad - \alpha_{k_3}^{(1)*}e^{i(\frac{\pi}{4}-\frac{\pi\nu_2}{2})} 2^{-\nu_2}\Gamma(-\nu_2) e^{-2i\pi\nu_2} \left(\frac{k_3}{k}\right)^{\nu_2} \mathcal{I}_{3}(-\nu_2) \\
&\quad + 2\alpha_k^{(1)*}\beta_k^{(1)*} \bigg( \beta_{k_3}^{(1)*}e^{-i(\frac{\pi}{4}+\frac{\pi\nu_2}{2})} 2^{\nu_2}\Gamma(\nu_2)  \left(\frac{k_3}{k}\right)^{-\nu_2} \mathcal{I}_{2}(\nu_2)+ \beta_{k_3}^{(1)*}e^{-i(\frac{\pi}{4}-\frac{\pi\nu_2}{2})} 2^{-\nu_2}\Gamma(-\nu_2) \left(\frac{k_3}{k}\right)^{\nu_2} \mathcal{I}_{2}(-\nu_2)\bigg)  \\
&\qquad - \alpha_{k_3}^{(1)*}e^{i(\frac{\pi}{4}+\frac{\pi\nu_2}{2})} 2^{\nu_2}\Gamma(\nu_2) \left(\frac{k_3}{k}\right)^{-\nu_2} \mathcal{I}_{2}(\nu_2) - \alpha_{k_3}^{(1)*}e^{i(\frac{\pi}{4}-\frac{\pi\nu_2}{2})} 2^{-\nu_2}\Gamma(-\nu_2)  \left(\frac{k_3}{k}\right)^{\nu_2} \mathcal{I}_{2}(-\nu_2)\bigg) \\
&\quad + (\beta_k^{(1)*})^2 e^{-i(\frac{\pi}{2} + \pi\nu_2 + 6\mu k\eta_{\inf})} \bigg( \beta_{k_3}^{(1)*}e^{-i(\frac{\pi}{4}+\frac{\pi\nu_2}{2})} 2^{\nu_2}\Gamma(\nu_2)\left(\frac{k_3}{k}\right)^{-\nu_2} \mathcal{I}_{3}(\nu_2)^* \\
&\qquad + \beta_{k_3}^{(1)*}e^{-i(\frac{\pi}{4}-\frac{\pi\nu_2}{2})} 2^{-\nu_2}\Gamma(-\nu_2) e^{2i\pi\nu_2} \left(\frac{k_3}{k}\right)^{\nu_2} \mathcal{I}_{3}(-\nu_2)^* \\
&\qquad - \alpha_{k_3}^{(1)*}e^{i(\frac{\pi}{4}+\frac{\pi\nu_2}{2})} 2^{\nu_2}\Gamma(\nu_2) \left(\frac{k_3}{k}\right)^{-\nu_2} \mathcal{I}_{3}(\nu_2)^* - \alpha_{k_3}^{(1)*}e^{i(\frac{\pi}{4}-\frac{\pi\nu_2}{2})} 2^{-\nu_2}\Gamma(-\nu_2) e^{2i\pi\nu_2} \left(\frac{k_3}{k}\right)^{\nu_2} \mathcal{I}_{3}(-\nu_2)^*\bigg) \Bigg] \Bigg\}.
\end{align*}
For imaginary $\nu_2$:
\begin{align*}
\overline{\langle\phi_{k}\phi_{k}\phi_{k_3}\rangle}_{\text{reh}} &= 2g \, \text{Im} \Bigg\{ f_{k}^2 f_{k_3} \frac{i\sqrt{\pi}}{8} \Bigg[ \\
&\quad (\alpha_k^{(1)*})^2 e^{i(\frac{\pi}{2} + \pi\nu_2 + 6\mu k\eta_{\inf})} \bigg(\beta_{k_3}^{(1)*}e^{-i(\frac{\pi}{4}+\frac{\pi\nu_2}{2})} 2^{\nu_2}\Gamma(\nu_2) k_3^{-\nu_2} \int_{\eta_{\text{inf}}}^{\eta_{\text{reh}}} d\eta' \, a(\eta') e^{i2\pi\nu_2} \bigl[H_{\nu_2}^{(1)}(k\tilde\eta')\bigr]^2 (\tilde\eta')^{-\nu_2+\frac{3}{2}} \\
&\qquad + \beta_{k_3}^{(1)*}e^{-i(\frac{\pi}{4}-\frac{\pi\nu_2}{2})} 2^{-\nu_2}\Gamma(-\nu_2) k_3^{\nu_2} \int_{\eta_{\text{inf}}}^{\eta_{\text{reh}}} d\eta' \, a(\eta') \bigl[H_{-\nu_2}^{(1)}(k\tilde\eta')\bigr]^2 (\tilde\eta')^{\nu_2+\frac{3}{2}} \\
&\qquad - \alpha_{k_3}^{(1)*}e^{i(\frac{\pi}{4}-\frac{\pi\nu_2}{2})} 2^{-\nu_2}\Gamma(-\nu_2) k_3^{\nu_2} \int_{\eta_{\text{inf}}}^{\eta_{\text{reh}}} d\eta' \, a(\eta') \bigl[H_{-\nu_2}^{(1)}(k\tilde\eta')\bigr]^2 (\tilde\eta')^{\nu_2+\frac{3}{2}} \\
&\qquad - \alpha_{k_3}^{(1)*}e^{i(\frac{\pi}{4}+\frac{\pi\nu_2}{2})} 2^{\nu_2}\Gamma(\nu_2) k_3^{-\nu_2} \int_{\eta_{\text{inf}}}^{\eta_{\text{reh}}} d\eta' \, a(\eta') e^{i2\pi\nu_2} \bigl[H_{\nu_2}^{(1)}(k\tilde\eta')\bigr]^2 (\tilde\eta')^{-\nu_2+\frac{3}{2}} \bigg) \\
&\quad + 2\alpha_k^{(1)*}\beta_k^{(1)*} e^{-i\pi\nu_2} \bigg( \beta_{k_3}^{(1)*}e^{-i(\frac{\pi}{4}+\frac{\pi\nu_2}{2})} 2^{\nu_2}\Gamma(\nu_2) k_3^{-\nu_2} \int_{\eta_{\text{inf}}}^{\eta_{\text{reh}}} d\eta' \, a(\eta') e^{i\pi\nu_2} H_{\nu_2}^{(1)}(k\tilde\eta') H_{\nu_2}^{(2)}(k\tilde\eta') (\tilde\eta')^{-\nu_2+\frac{3}{2}} \\
&\qquad + \beta_{k_3}^{(1)*}e^{-i(\frac{\pi}{4}-\frac{\pi\nu_2}{2})} 2^{-\nu_2}\Gamma(-\nu_2) k_3^{\nu_2} \int_{\eta_{\text{inf}}}^{\eta_{\text{reh}}} d\eta' \, a(\eta') e^{i\pi\nu_2} H_{-\nu_2}^{(1)}(k\tilde\eta') H_{-\nu_2}^{(2)}(k\tilde\eta') (\tilde\eta')^{\nu_2+\frac{3}{2}} \\
&\qquad - \alpha_{k_3}^{(1)*}e^{i(\frac{\pi}{4}-\frac{\pi\nu_2}{2})} 2^{-\nu_2}\Gamma(-\nu_2) k_3^{\nu_2} \int_{\eta_{\text{inf}}}^{\eta_{\text{reh}}} d\eta' \, a(\eta') e^{i\pi\nu_2} H_{-\nu_2}^{(1)}(k\tilde\eta') H_{-\nu_2}^{(2)}(k\tilde\eta') (\tilde\eta')^{\nu_2+\frac{3}{2}} \\
&\qquad - \alpha_{k_3}^{(1)*}e^{i(\frac{\pi}{4}+\frac{\pi\nu_2}{2})} 2^{\nu_2}\Gamma(\nu_2) k_3^{-\nu_2} \int_{\eta_{\text{inf}}}^{\eta_{\text{reh}}} d\eta' \, a(\eta') e^{i\pi\nu_2} H_{\nu_2}^{(1)}(k\tilde\eta') H_{\nu_2}^{(2)}(k\tilde\eta') (\tilde\eta')^{-\nu_2+\frac{3}{2}} \bigg) \\
&\quad + (\beta_k^{(1)*})^2 e^{-i(\frac{\pi}{2} + \pi\nu_2 + 6\mu k\eta_{\inf})} \bigg( \beta_{k_3}^{(1)*}e^{-i(\frac{\pi}{4}+\frac{\pi\nu_2}{2})} 2^{\nu_2}\Gamma(\nu_2) k_3^{-\nu_2} \int_{\eta_{\text{inf}}}^{\eta_{\text{reh}}} d\eta' \, a(\eta') \bigl[H_{\nu_2}^{(2)}(k\tilde\eta')\bigr]^2 (\tilde\eta')^{-\nu_2+\frac{3}{2}} \\
&\qquad + \beta_{k_3}^{(1)*}e^{-i(\frac{\pi}{4}-\frac{\pi\nu_2}{2})} 2^{-\nu_2}\Gamma(-\nu_2) k_3^{\nu_2} \int_{\eta_{\text{inf}}}^{\eta_{\text{reh}}} d\eta' \, a(\eta') e^{i2\pi\nu_2} \bigl[H_{-\nu_2}^{(2)}(k\tilde\eta')\bigr]^2 (\tilde\eta')^{\nu_2+\frac{3}{2}} \\
&\qquad - \alpha_{k_3}^{(1)*}e^{i(\frac{\pi}{4}-\frac{\pi\nu_2}{2})} 2^{-\nu_2}\Gamma(-\nu_2) k_3^{\nu_2} \int_{\eta_{\text{inf}}}^{\eta_{\text{reh}}} d\eta' \, a(\eta') e^{i2\pi\nu_2} \bigl[H_{-\nu_2}^{(2)}(k\tilde\eta')\bigr]^2 (\tilde\eta')^{\nu_2+\frac{3}{2}} \\
&\qquad - \alpha_{k_3}^{(1)*}e^{i(\frac{\pi}{4}+\frac{\pi\nu_2}{2})} 2^{\nu_2}\Gamma(\nu_2) k_3^{-\nu_2} \int_{\eta_{\text{inf}}}^{\eta_{\text{reh}}} d\eta' \, a(\eta') \bigl[H_{\nu_2}^{(2)}(k\tilde\eta')\bigr]^2 (\tilde\eta')^{-\nu_2+\frac{3}{2}} \bigg) \Bigg] \Bigg\}\nt
    \overline{\langle\phi_{k}\phi_{k}\phi_{k_3}\rangle}_{\text{reh}} 
&= 2g \, \text{Im} \Bigg\{ \frac{f_{k}^2 f_{k_3}}{k\sqrt{k}} \frac{i\sqrt{\pi}}{8} \Bigg[  (\alpha_k^{(1)*})^2 e^{i(\frac{\pi}{2} + \pi\nu_2 + 6\mu k\eta_{\inf})} \bigg(  \beta_{k_3}^{(1)*}e^{-i(\frac{\pi}{4}+\frac{\pi\nu_2}{2})} 2^{\nu_2}\Gamma(\nu_2) e^{i2\pi\nu_2} \left(\frac{k_3}{k}\right)^{-\nu_2} \mathcal{I}_{3}(\nu_2) \\
    &\qquad + \beta_{k_3}^{(1)*}e^{-i(\frac{\pi}{4}-\frac{\pi\nu_2}{2})} 2^{-\nu_2}\Gamma(-\nu_2)\left(\frac{k_3}{k}\right)^{\nu_2} \mathcal{I}_{3}(-\nu_2)\bigg) - \alpha_{k_3}^{(1)*}e^{i(\frac{\pi}{4}-\frac{\pi\nu_2}{2})} 2^{-\nu_2}\Gamma(-\nu_2) \left(\frac{k_3}{k}\right)^{\nu_2} \mathcal{I}_{3}(-\nu_2)\bigg) \\
    &\qquad - \alpha_{k_3}^{(1)*}e^{i(\frac{\pi}{4}+\frac{\pi\nu_2}{2})} 2^{\nu_2}\Gamma(\nu_2) e^{i2\pi\nu_2} \left(\frac{k_3}{k}\right)^{-\nu_2} \mathcal{I}_{3}(\nu_2) \bigg) \\
    &\quad + 2\alpha_k^{(1)*}\beta_k^{(1)*} e^{-i\pi\nu_2} \bigg( \beta_{k_3}^{(1)*}e^{-i(\frac{\pi}{4}+\frac{\pi\nu_2}{2})} 2^{\nu_2}\Gamma(\nu_2) e^{i\pi\nu_2} \left(\frac{k_3}{k}\right)^{-\nu_2} \mathcal{I}_{2}(\nu_2) \\
    &\qquad + \beta_{k_3}^{(1)*}e^{-i(\frac{\pi}{4}-\frac{\pi\nu_2}{2})} 2^{-\nu_2}\Gamma(-\nu_2)e^{i\pi\nu_2} \left(\frac{k_3}{k}\right)^{\nu_2} \mathcal{I}_{2}(-\nu_2) - \alpha_{k_3}^{(1)*}e^{i(\frac{\pi}{4}-\frac{\pi\nu_2}{2})} 2^{-\nu_2}\Gamma(-\nu_2)e^{i\pi\nu_2}  \left(\frac{k_3}{k}\right)^{\nu_2} \mathcal{I}_{2}(-\nu_2)\\
    &\qquad - \alpha_{k_3}^{(1)*}e^{i(\frac{\pi}{4}+\frac{\pi\nu_2}{2})} 2^{\nu_2}\Gamma(\nu_2)e^{i\pi\nu_2}  \left(\frac{k_3}{k}\right)^{-\nu_2} \mathcal{I}_{2}(\nu_2) \bigg) \\
    &\quad + (\beta_k^{(1)*})^2 e^{-i(\frac{\pi}{2} + \pi\nu_2 + 6\mu k\eta_{\inf})} \bigg( \beta_{k_3}^{(1)*}e^{-i(\frac{\pi}{4}+\frac{\pi\nu_2}{2})} 2^{\nu_2}\Gamma(\nu_2) \left(\frac{k_3}{k}\right)^{-\nu_2} \mathcal{I}_{3}(-\nu_2)^* \\
    &\qquad + \beta_{k_3}^{(1)*}e^{-i(\frac{\pi}{4}-\frac{\pi\nu_2}{2})} 2^{-\nu_2}\Gamma(-\nu_2)e^{i2\pi\nu_2}\left(\frac{k_3}{k}\right)^{\nu_2} \mathcal{I}_{3}(\nu_2)^*\\
    &\qquad - \alpha_{k_3}^{(1)*}e^{i(\frac{\pi}{4}-\frac{\pi\nu_2}{2})} 2^{-\nu_2}\Gamma(-\nu_2)e^{i2\pi\nu_2}\left(\frac{k_3}{k}\right)^{\nu_2} \mathcal{I}_{3}(\nu_2)^* - \alpha_{k_3}^{(1)*}e^{i(\frac{\pi}{4}+\frac{\pi\nu_2}{2})} 2^{\nu_2}\Gamma(\nu_2) \left(\frac{k_3}{k}\right)^{-\nu_2} \mathcal{I}_{3}(-\nu_2)^* \bigg) \Bigg] \Bigg\}.
\end{align*}
Collectively,
\begin{align}
\overline{\langle\phi_{k}\phi_{k}\phi_{k_3}\rangle}_{\text{reh}} &= 2g \, \text{Im} \Bigg\{ \frac{f_{k}^2 f_{k_3}}{k\sqrt{k}} \frac{i\sqrt{\pi}}{8} \Bigg[ (\alpha_k^{(1)*})^2 e^{i(\frac{\pi}{2} + \pi\nu_2 + 6\mu k\eta_{\inf})} \bigg(  \beta_{k_3}^{(1)*}e^{-i(\frac{\pi}{4}+\frac{\pi\nu_2}{2})} 2^{\nu_2}\Gamma(\nu_2) e^{i2\pi\nu_2\delta_{I}} \left(\frac{k_3}{k}\right)^{-\nu_2} \mathcal{I}_{3}(\nu_2) \nt
&\qquad\qquad\qquad\qquad\qquad\qquad\qquad\quad+ \beta_{k_3}^{(1)*}e^{-i(\frac{\pi}{4}-\frac{\pi\nu_2}{2})} 2^{-\nu_2}\Gamma(-\nu_2)e^{-i2\pi\nu_2(1-\delta_{I})})\left(\frac{k_3}{k}\right)^{\nu_2} \mathcal{I}_{3}(-\nu_2)\bigg)  \nt
&\qquad\qquad\qquad\qquad\qquad\qquad\qquad\quad - \alpha_{k_3}^{(1)*}e^{i(\frac{\pi}{4}-\frac{\pi\nu_2}{2})} 2^{-\nu_2}\Gamma(-\nu_2) e^{-2i\pi\nu_2(1-\delta_I)}\left(\frac{k_3}{k}\right)^{\nu_2} \mathcal{I}_{3}(-\nu_2)\bigg) \nt
&\qquad\qquad\qquad\qquad\qquad\qquad\qquad\quad - \alpha_{k_3}^{(1)*}e^{i(\frac{\pi}{4}+\frac{\pi\nu_2}{2})} 2^{\nu_2}\Gamma(\nu_2) e^{i2\pi\nu_2\delta_{^I}} \left(\frac{k_3}{k}\right)^{-\nu_2} \mathcal{I}_{3}(\nu_2) \bigg) \nt
&\quad\qquad\qquad\qquad\qquad\qquad + 2\alpha_k^{(1)*}\beta_k^{(1)*} \bigg( \beta_{k_3}^{(1)*}e^{-i(\frac{\pi}{4}+\frac{\pi\nu_2}{2})} 2^{\nu_2}\Gamma(\nu_2)  \left(\frac{k_3}{k}\right)^{-\nu_2} \mathcal{I}_{2}(\nu_2) \nt
&\qquad\qquad\qquad\qquad\qquad\qquad\qquad\quad + \beta_{k_3}^{(1)*}e^{-i(\frac{\pi}{4}-\frac{\pi\nu_2}{2})} 2^{-\nu_2}\Gamma(-\nu_2) \left(\frac{k_3}{k}\right)^{\nu_2} \mathcal{I}_{2}(-\nu_2)\nt
&\qquad\qquad\qquad\qquad\qquad\qquad\qquad\quad - \alpha_{k_3}^{(1)*}e^{i(\frac{\pi}{4}-\frac{\pi\nu_2}{2})} 2^{-\nu_2}\Gamma(-\nu_2)  \left(\frac{k_3}{k}\right)^{\nu_2} \mathcal{I}_{2}((-1)^{\delta_I}\nu_2)\nt
&\qquad\qquad\qquad\qquad\qquad\qquad\qquad\quad - \alpha_{k_3}^{(1)*}e^{i(\frac{\pi}{4}+\frac{\pi\nu_2}{2})} 2^{\nu_2}\Gamma(\nu_2)  \left(\frac{k_3}{k}\right)^{-\nu_2} \mathcal{I}_{2}((-1)^{1-\delta_I}\nu_2) \bigg) \nt
&\quad\qquad\qquad\qquad\qquad + (\beta_k^{(1)*})^2 e^{-i(\frac{\pi}{2} + \pi\nu_2 + 6\mu k\eta_{\inf})} \bigg( \beta_{k_3}^{(1)*}e^{-i(\frac{\pi}{4}+\frac{\pi\nu_2}{2})} 2^{\nu_2}\Gamma(\nu_2) \left(\frac{k_3}{k}\right)^{-\nu_2} \mathcal{I}_{3}((-1)^{\delta_I}\nu_2)^* \nt
&\qquad\qquad\qquad\qquad\qquad\qquad\qquad\quad + \beta_{k_3}^{(1)*}e^{-i(\frac{\pi}{4}-\frac{\pi\nu_2}{2})} 2^{-\nu_2}\Gamma(-\nu_2)e^{i2\pi\nu_2}\left(\frac{k_3}{k}\right)^{\nu_2} \mathcal{I}_{3}((-1)^{1-\delta_I}\nu_2)^*\nt
&\qquad\qquad\qquad\qquad\qquad\qquad\qquad\quad - \alpha_{k_3}^{(1)*}e^{i(\frac{\pi}{4}-\frac{\pi\nu_2}{2})} 2^{-\nu_2}\Gamma(-\nu_2)e^{i2\pi\nu_2}\left(\frac{k_3}{k}\right)^{\nu_2} \mathcal{I}_{3}((-1)^{1-\delta_I}\nu_2)^*\nt
&\qquad\qquad\qquad\qquad\qquad\qquad\qquad\quad - \alpha_{k_3}^{(1)*}e^{i(\frac{\pi}{4}+\frac{\pi\nu_2}{2})} 2^{\nu_2}\Gamma(\nu_2) \left(\frac{k_3}{k}\right)^{-\nu_2}\mathcal{I}_{3}((-1)^{\delta_I}\nu_2)^* \bigg) \Bigg] \Bigg\}.
\end{align}

This involves several integral, of which, we evaluate \eqref{I2} in following manner:

\begin{align}
\mathcal{I}_3&=k^{-\nu_2+\frac{3}{2}}\int_{\eta_{\rm inf}}^{\eta_{\rm reh}}d\eta'\;a_{\rm inf}\left(\frac{1+3w}{2|\eta_{\rm inf}|}\right)^{\frac{2}{1+3w}}\left(\eta'-3\mu\eta_{\rm inf}\right)^{\frac{2}{1+3w}}\bigl[H_{\nu_2}^{(1)}(k\tilde\eta')\bigr]^2(\tilde\eta')^{-\nu_2+\frac32}\nonumber\\[1ex]
&=\frac{1}{H_{\rm inf}(k/k_{\rm inf})}\left(\frac{1+3w}{2k|\eta_{\rm inf}|}\right)^{\frac{2}{1+3w}}\int_{\frac{-2k\eta_{\rm inf}}{1+3w}}^{k\eta_{\rm reh}-3\mu\,k\eta_{\rm inf}}dx\;[H_{\nu_2}^{(1)}(x)]^2x^{-\nu_2+\frac32+\frac{2}{1+3w}}\nonumber\\[2ex]
&=\frac{1}{H_{\rm inf}(k/k_{\rm inf})}\left(\frac{1+3w}{2k|\eta_{\rm inf}|}\right)^{\frac{2}{1+3w}}
\frac{2}{3}(3w+1)\,x^{-\nu_2+\frac{2}{1+3w}+1+\frac{3}{2}}\nonumber\\
&\times\Bigg[\frac{4^{\nu_2}\csc^2(\pi\nu_2)\,(k\eta')^{-2\nu_2}}{\Gamma(1-\nu_2)^2\,\nu_2(6w+2)-5w-3]}\nonumber\\
&\qquad\times{}_2F_3\!\left(\frac12-\nu_2,\,\frac{-18\nu_2 w-6\nu_2+15w+9}{12w+4};1-2\nu_2,\,
1-\nu_2,\,\frac{-18\nu_2 w-6\nu_2+27w+13}{12w+4};-x^2\right)\nonumber\\[1.2ex]
&\quad+\frac{\cot(\pi\nu_2)-i}{\Gamma(\nu_2+1)^2}\Bigg(-\frac{2\csc(\pi\nu_2)\Gamma(\nu_2+1)}{\Gamma(1-\nu_2)\,[\nu_2(6w+2)-15w-9]}\nonumber\\
&\qquad\times{}_2F_3\!\left(\frac12,\,\frac{-6\nu_2 w-2\nu_2+15w+9}{12w+4};1-\nu_2,\,\nu_2+1,\,
\frac{-6\nu_2 w-2\nu_2+27w+13}{12w+4};-x^2\right)\nonumber\\[1.2ex]
&\quad\qquad-\frac{4^{-\nu_2}(k\eta')^{2\nu_2}}{\nu_2(6w+2)+15w+9}\nonumber\\
&\qquad\times{}_2F_3\!\left(\nu_2+\frac12,\,\frac{6\nu_2 w+2\nu_2+15w+9}{12w+4};\nu_2+1,\,2\nu_2+1,\,\frac{6\nu_2 w+2\nu_2+27w+13}{12w+4};-x^2\right)\Bigg)\Bigg]\Bigg|_{\eta'=\frac{2|k\eta_{\rm inf}|}{1+3w}}^{x=k\eta_{\rm reh}-3\mu\,k\eta_{\rm inf}}\nt
&=\frac{1}{H_{\rm inf}(k/k_{\rm inf})}\left(\frac{1+3w}{2k|\eta_{\rm inf}|}\right)^{\frac{2}{1+3w}}
\frac{2}{3}(3w+1)\left(k\eta_{\rm reh}-3\mu\,k\eta_{\rm inf}\right)^{-\nu_2+\frac{2}{1+3w}+1+\frac{3}{2}}\nonumber\\
&\times\Bigg[\frac{4^{\nu_2}\csc^2(\pi\nu_2)\,\left(k\eta_{\rm reh}-3\mu\,k\eta_{\rm inf}\right)^{-2\nu_2}}{\Gamma(1-\nu_2)^2\,\nu_2(6w+2)-5w-3]}\nonumber\\
&\times{}_2F_3\!\left(\frac12-\nu_2,\,\frac{-18\nu_2 w-6\nu_2+15w+9}{12w+4};1-2\nu_2,\,
1-\nu_2,\,\frac{-18\nu_2 w-6\nu_2+27w+13}{12w+4};-k^2\left(\eta_{\rm reh}-3\mu\eta_{\rm inf}\right)^2\right)\nonumber\\[1.2ex]
&\quad+\frac{\cot(\pi\nu_2)-i}{\Gamma(\nu_2+1)^2}\Bigg(-\frac{2\csc(\pi\nu_2)\Gamma(\nu_2+1)}{\Gamma(1-\nu_2)\,[\nu_2(6w+2)-15w-9]}\nonumber\\
&\times{}_2F_3\!\left(\frac12,\,\frac{-6\nu_2 w-2\nu_2+15w+9}{12w+4};1-\nu_2,\,\nu_2+1,\,
\frac{-6\nu_2 w-2\nu_2+27w+13}{12w+4};-k^2\left(\eta_{\rm reh}-3\mu\eta_{\rm inf}\right)^2\right)\nonumber\\[1.2ex]
&\quad\qquad-\frac{4^{-\nu_2}\left(k\eta_{\rm reh}-3\mu\,k\eta_{\rm inf}\right)^{2\nu_2}}{\nu_2(6w+2)+15w+9}\nonumber\\
&\times{}_2F_3\!\left(\nu_2+\frac12,\,\frac{6\nu_2 w+2\nu_2+15w+9}{12w+4};\nu_2+1,\,2\nu_2+1,\,\frac{6\nu_2 w+2\nu_2+27w+13}{12w+4};-k^2\left(\eta_{\rm reh}-3\mu\eta_{\rm inf}\right)^2\right)\Bigg)\Bigg]\nt
&-\frac{1}{H_{\rm inf}(k/k_{\rm inf})}\left(\frac{1+3w}{2k|\eta_{\rm inf}|}\right)^{\frac{2}{1+3w}}
\frac{2}{3}(3w+1)\,\left(\frac{2|k\eta_{\rm inf}|}{1+3w}\right)^{-\nu_2+\frac{2}{1+3w}+1+\frac{3}{2}}\nonumber\\
&\times\Bigg[\frac{4^{\nu_2}\csc^2(\pi\nu_2)\,\left(\frac{2k|\eta_{\rm inf}|}{1+3w}\right)^{-2\nu_2}}{\Gamma(1-\nu_2)^2\,\nu_2(6w+2)-5w-3]}\nonumber\\
&\qquad\times{}_2F_3\!\left(\frac12-\nu_2,\,\frac{-18\nu_2 w-6\nu_2+15w+9}{12w+4};1-2\nu_2,\,
1-\nu_2,\,\frac{-18\nu_2 w-6\nu_2+27w+13}{12w+4};-k^2\left(\frac{2|\eta_{\rm inf}|}{1+3w}\right)^2\right)\nonumber\\[1.2ex]
&\quad+\frac{\cot(\pi\nu_2)-i}{\Gamma(\nu_2+1)^2}\Bigg(-\frac{2\csc(\pi\nu_2)\Gamma(\nu_2+1)}{\Gamma(1-\nu_2)\,[\nu_2(6w+2)-15w-9]}\nonumber\\
&\qquad\times{}_2F_3\!\left(\frac12,\,\frac{-6\nu_2 w-2\nu_2+15w+9}{12w+4};1-\nu_2,\,\nu_2+1,\,
\frac{-6\nu_2 w-2\nu_2+27w+13}{12w+4};-k^2\left(\frac{2|\eta_{\rm inf}|}{1+3w}\right)^2\right)\nonumber\\[1.2ex]
&\quad\qquad-\frac{4^{-\nu_2}\left(\frac{2k|\eta_{\rm inf}|}{1+3w}\right)^{2\nu_2}}{\nu_2(6w+2)+15w+9}\nonumber\\
&\qquad\times{}_2F_3\!\left(\nu_2+\frac12,\,\frac{6\nu_2 w+2\nu_2+15w+9}{12w+4};\nu_2+1,\,2\nu_2+1,\,\frac{6\nu_2 w+2\nu_2+27w+13}{12w+4};-k^2\left(\frac{2|\eta_{\rm inf}|}{1+3w}\right)^2\right)\Bigg)\Bigg].
\end{align}
On superhorizon scale:
\begin{align}
\mathcal{I}_3&=\frac{1}{H_{\rm inf}(k/k_{\rm inf})}
\left(\frac{1+3w}{2k|\eta_{\rm inf}|}\right)^{\frac{2}{1+3w}}
\Bigg[\frac{2}{3}(3w+1)(k\eta_{\rm reh})^{-\nu_2+\frac{5}{2}+\frac{2}{1+3w}}
\nonumber\Bigg((k\eta_{\rm reh})^{-2\nu_2}\left[\frac{4^{\nu_2}\csc^2(\pi\nu_2)}{\Gamma(1-\nu_2)^2(2\nu_2+6\nu_2 w-5w-3)}\right.\nonumber \\[4pt]
&\qquad\left.-\frac{3\times4^{-\nu_2}(\cot(\pi\nu_2)-i)^2(k\eta_{\rm reh})^{4\nu_2}}
{\Gamma(\nu_2+1)^2(2\nu_2+6\nu_2 w+15w+9)}\right]\nonumber -\frac{6(\cot(\pi\nu_2)-i)\csc(\pi\nu_2)}{\Gamma(1-\nu_2)\Gamma(\nu_2+1)(2\nu_2+6\nu_2 w-15w-9)}\Bigg)\nonumber \\[6pt]
&\quad-\frac{4}{3}\left(\frac{2}{3w+1}\right)^{-\nu_2+\frac{3}{2}+\frac{2}{1+3w}}|k\eta_{\inf}|^{-\nu_2+\frac{5}{2}+\frac{2}{1+3w}}\Bigg(2^{-2\nu_2}\left(\frac{1}{3w+1}\right)^{-2\nu_2}|k\eta_{\inf}|^{-2\nu_2}\nonumber \\[4pt]
&\qquad \times\left[\frac{4^{\nu_2}\csc^2(\pi\nu_2)}{\Gamma(1\nu_2)^2(2\nu_2+6\nu_2 w-5w-3)}\right.\left.-\frac{3\times2^{2\nu_2}(\cot(\pi\nu_2)-i)^2\left(\frac{1}{3w+1}\right)^{4\nu_2}|k\eta_{\inf}|^{4\nu_2}}{\Gamma(\nu_2+1)^2(2\nu_2+6\nu_2w+15w+9)}\right]\nonumber \\[4pt]
&\qquad-\frac{6(\cot(\pi\nu_2)-i)\csc(\pi\nu_2)}{\Gamma(1-\nu_2)\Gamma(\nu_2+1)
(2\nu_2+6\nu_2 w-15w-9)}\Bigg)\Bigg].
\end{align}
\begin{align}
\mathcal{I}_2&=\frac{1}{H_{\rm inf}|\eta_{\rm inf}|}\left(\frac{1+3w}{2|\eta_{\rm inf}|}\right)^{\frac{2}{1+3w}}k^{-\nu_2+\frac{3}{2}}\int_{\frac{-2\eta_{\rm inf}}{1+3w}}^{\eta_{\rm reh}-3\mu\eta_{\rm inf}}\eta'^{-\nu_2+\frac{2}{3w+1}+\frac32}H_{\nu_2}^{(1)}(k\eta')\,H_{\nu_2}^{(2)}(k\eta')\,d\eta'\nonumber\\[1ex]
&=\frac{1}{H_{\rm inf}(k/k_{\rm inf})}\left(\frac{1+3w}{2k|\eta_{\rm inf}|}\right)^{\frac{2}{1+3w}}\int_{\frac{-2k\eta_{\rm inf}}{1+3w}}^{k\eta_{\rm reh}-3\mu\,k\eta_{\rm inf}}x^{-\nu_2+\frac{2}{3w+1}+\frac32}H_{\nu_2}^{(1)}(x)\,H_{\nu_2}^{(2)}(x)\,dx\nt
&=\frac{1}{H_{\rm inf}(k/k_{\rm inf})}\left(\frac{1+3w}{2k|\eta_{\rm inf}|}\right)^{\frac{2}{1+3w}}\frac{2}{3}(3w+1)\,\csc(\pi\nu_2)\,x^{-\nu_2+\frac{2}{1+3w}+1+\frac{3}{2}}\nonumber\\
&\times\Bigg[\frac{3}{\Gamma(\nu_2+1)^2}\Bigg(\frac{4^{-\nu_2}\csc(\pi\nu_2)\,(k\eta')^{2\nu_2}}{\nu_2(6w+2)+15w+9}\nonumber\\
&\qquad\times{}_2F_3\!\left(\nu_2+\frac12,\,\frac{6\nu_2 w+2\nu_2+15w+9}{12w+4};\nu_2+1,\,2\nu_2+1,\,\frac{6\nu_2 w+2\nu_2+27w+13}{12w+4};-x^2\right)\nonumber\\[1.2ex]
&\qquad+\frac{2\cot(\pi\nu_2)\Gamma(\nu_2+1)}{\Gamma(1-\nu_2)\,[\nu_2(6w+2)-3(5w+3)]}\nonumber\\
&\qquad\times{}_2F_3\!\left(\frac12,\,\frac{-6\nu_2 w-2\nu_2+15w+9}{12w+4};1-\nu_2,\,\nu_2+1,\,\frac{-6\nu_2 w-2\nu_2+27w+13}{12w+4};-x^2\right)\Bigg)\nonumber\\[1.5ex]
&\quad-\frac{4^{\nu_2}\csc(\pi\nu_2)\,(k\eta')^{-2\nu_2}}{\Gamma(1-\nu_2)^2\,[\nu_2(6w+2)-5w-3]}\nonumber\\
&\times{}_2F_3\!\left(\frac12-\nu_2,\,\frac{-18\nu_2 w-6\nu_2+15w+9}{12w+4};1-2\nu_2,\,1-\nu_2,\,\frac{-18\nu_2 w-6\nu_2+27w+13}{12w+4};-x^2\right)\Bigg]\Bigg|_{x=\frac{2|k\eta_{\rm inf}|}{1+3w}}^{x=k\eta_{\rm reh}-3\mu\,k\eta_{\rm inf}}\nt
&=\frac{1}{H_{\rm inf}(k/k_{\rm inf})}\left(\frac{1+3w}{2k|\eta_{\rm inf}|}\right)^{\frac{2}{1+3w}}\frac{2}{3}(3w+1)\,\csc(\pi\nu_2)\left(k\eta_{\rm reh}-3\mu\,k\eta_{\rm inf}\right)^{-\nu_2+\frac{2}{1+3w}+1+\frac{3}{2}}\nonumber\\
&\times\Bigg[\frac{3}{\Gamma(\nu_2+1)^2}\Bigg(\frac{4^{-\nu_2}\csc(\pi\nu_2)\,\left(k\eta_{\rm reh}-3\mu\,k\eta_{\rm inf}\right)^{2\nu_2}}{\nu_2(6w+2)+15w+9}\nonumber\\
&\times{}_2F_3\!\left(\nu_2+\frac12,\,\frac{6\nu_2 w+2\nu_2+15w+9}{12w+4};\nu_2+1,\,2\nu_2+1,\,\frac{6\nu_2 w+2\nu_2+27w+13}{12w+4};-k^2\left(\eta_{\rm reh}-3\mu\eta_{\rm inf}\right)^2\right)\nonumber\\[1.2ex]
&+\frac{2\cot(\pi\nu_2)\Gamma(\nu_2+1)}{\Gamma(1-\nu_2)\,[\nu_2(6w+2)-3(5w+3)]}\nonumber\\
&\times{}_2F_3\!\left(\frac12,\,\frac{-6\nu_2 w-2\nu_2+15w+9}{12w+4};1-\nu_2,\,\nu_2+1,\,\frac{-6\nu_2 w-2\nu_2+27w+13}{12w+4};-k^2\left(\eta_{\rm reh}-3\mu\eta_{\rm inf}\right)^2\right)\Bigg)\nonumber\\[1.5ex]
&-\frac{4^{\nu_2}\csc(\pi\nu_2)\,\left(k\eta_{\rm reh}-3\mu\,k\eta_{\rm inf}\right)^{-2\nu_2}}{\Gamma(1-\nu_2)^2\,[\nu_2(6w+2)-5w-3]}\nonumber\\
&\times{}_2F_3\!\left(\frac12-\nu_2,\,\frac{-18\nu_2 w-6\nu_2+15w+9}{12w+4};1-2\nu_2,\,1-\nu_2,\,\frac{-18\nu_2 w-6\nu_2+27w+13}{12w+4};-k^2\left(\eta_{\rm reh}-3\mu\eta_{\rm inf}\right)^2\right)\Bigg]\nt
&-\frac{1}{H_{\rm inf}(k/k_{\rm inf})}\left(\frac{1+3w}{2k|\eta_{\rm inf}|}\right)^{\frac{2}{1+3w}}\frac{2}{3}(3w+1)\,\csc(\pi\nu_2)\left(\frac{2|k\eta_{\rm inf}|}{1+3w}\right)^{-\nu_2+\frac{2}{1+3w}+1+\frac{3}{2}}\nt
&\times\Bigg[\frac{3}{\Gamma(\nu_2+1)^2}\Bigg(\frac{4^{-\nu_2}\csc(\pi\nu_2)\,\left(\frac{2k|\eta_{\rm inf}|}{1+3w}\right)^{2\nu_2}}{\nu_2(6w+2)+15w+9}\nt
&\times{}_2F_3\!\left(\nu_2+\frac12,\,\frac{6\nu_2 w+2\nu_2+15w+9}{12w+4};\nu_2+1,\,2\nu_2+1,\,\frac{6\nu_2 w+2\nu_2+27w+13}{12w+4};-k^2\left(\frac{2|\eta_{\rm inf}|}{1+3w}\right)^2\right)\nonumber\\[1.2ex]
&+\frac{2\cot(\pi\nu_2)\Gamma(\nu_2+1)}{\Gamma(1-\nu_2)\,[\nu_2(6w+2)-3(5w+3)]}\nonumber\\
&\times{}_2F_3\!\left(\frac12,\,\frac{-6\nu_2 w-2\nu_2+15w+9}{12w+4};1-\nu_2,\,\nu_2+1,\,\frac{-6\nu_2 w-2\nu_2+27w+13}{12w+4};-k^2\left(\frac{2|\eta_{\rm inf}|}{1+3w}\right)^2\right)\Bigg)\nonumber\\[1.5ex]
&-\frac{4^{\nu_2}\csc(\pi\nu_2)\,\left(\frac{2k|\eta_{\rm inf}|}{1+3w}\right)^{-2\nu_2}}{\Gamma(1-\nu_2)^2\,[\nu_2(6w+2)-5w-3]}\nonumber\\
&\times{}_2F_3\!\left(\frac12-\nu_2,\,\frac{-18\nu_2 w-6\nu_2+15w+9}{12w+4};1-2\nu_2,\,1-\nu_2,\,\frac{-18\nu_2 w-6\nu_2+27w+13}{12w+4};-k^2\left(\frac{2|\eta_{\rm inf}|}{1+3w}\right)^2\right)\Bigg].
\end{align}
On superhorizon scale:
\begin{align}
\mathcal{I}_2 
&=\frac{1}{H_{\rm inf}(k/k_{\rm inf})}\left(\frac{1+3w}{2k|\eta_{\rm inf}|}\right)^{\frac{2}{1+3w}}\Bigg[\frac{2}{3}(3w+1)(k\eta_{\rm reh})^{-\nu_2+\frac{2}{1+3w}+\frac{5}{2}}\Bigg((k\eta_{\rm reh})^{-2\nu_2}
\Bigg[\frac{3\times4^{-\nu_2}\csc^2(\pi\nu_2)(k\eta_{\rm reh})^{4\nu_2}}{\Gamma(\nu_2+1)^2(2\nu_2+6\nu_2 w+15w+9)}\nonumber\\
&\qquad-\frac{4^{\nu_2}\csc^2(\pi\nu_2)}{\Gamma(1-\nu_2)^2(2\nu_2+6\nu_2 w-5w-3)}\Bigg]+\frac{6\cot(\pi\nu_2)\csc(\pi\nu_2)}{\Gamma(1-\nu_2)\Gamma(\nu_2+1)(2\nu_2+6\nu_2 w-15w-9)}\Bigg)\nonumber\\[6pt]
&-\frac{4}{3}\left(\frac{2}{3w+1}\right)^{-\nu_2+\frac{2}{1+3w}+\frac{3}{2}}|k\eta_{\inf}|^{-\nu_2+\frac{2}{1+3w}+\frac{5}{2}}\Bigg(2^{-2\nu_2}\left(\frac{1}{3w+1}\right)^{-2\nu_2}|k\eta_{\inf}|^{-2\nu_2}\nonumber\\
&\qquad\times\Bigg[\frac{3\times2^{2\nu_2}\csc^2(\pi\nu_2)\left(\frac{1}{3w+1}\right)^{4\nu_2}|k\eta_{\inf}|^{4\nu_2}}{\Gamma(\nu_2+1)^2(2\nu_2+6\nu_2 w+15w+9)}-\frac{4^{\nu_2}\csc^2(\pi\nu_2)}{\Gamma(1-\nu_2)^2(2\nu_2+6\nu_2 w-5w-3)}\Bigg]\nonumber\\[4pt]
&\qquad+\frac{6\cot(\pi\nu_2)\csc(\pi\nu_2)}{\Gamma(1-\nu_2)\Gamma(\nu_2+1)(2\nu_2+6\nu_2 w-15w-9)}\Bigg)\Bigg].
\end{align}
\subsection{Evaluation of the Inflation Contribution}
The Integral \eqref{I1} is calculated as follows:
\begin{align}
\mathcal{I}_1 &=\int_{-\infty}^{k\eta_{\text{inf}}}dx [H_{\nu_1}^{(2)}(-x)]^2(-x)^{-\nu_1+1/2}\nt
&=\frac{2^{1-2\nu_1}\,|x|^{\frac32-3\nu_1}\,(\cot(\pi\nu_1)-i)}{3(2\nu_1-3)(2\nu_1+3)(2\nu_1-1)\,\Gamma(1-\nu_1)^2\,\Gamma(\nu_1+1)^2}\Bigg[-2^{2\nu_1+1}(12\nu_1^2+12\nu_1-9)\,\csc(\pi\nu_1)\,\Gamma(1-\nu_1)\Gamma(\nu_1+1)\,
x^{2\nu_1}\nonumber\\
&\qquad\times{}_2F_3\!\left(\frac12,\frac34-\frac{\nu_1}{2};1-\nu_1,\frac74-\frac{\nu_1}{2},\nu_1+1;-x^2\right)+(2\nu_1-3)\Bigg(16^{\nu_1}(2\nu_1+3)(\cot(\pi\nu_1)+i)\,\Gamma(\nu_1+1)^2\nonumber\\
&\qquad\qquad\times{}_2F_3\!\left(\frac34-\frac{3\nu_1}{2},\frac12-\nu_1;1-2\nu_1,\frac74-\frac{3\nu_1}{2},1-\nu_1;-x^2\right)-3(2\nu_1-1)(\cot(\pi\nu_1)-i)\,\Gamma(1-\nu_1)^2(k|\eta|)^{4\nu_1}\nonumber\\
&\qquad\qquad\times{}_2F_3\!\left(\frac{\nu_1}{2}+\frac34,\nu_1+\frac12;\frac{\nu_1}{2}+\frac74,\nu_1+1,2\nu_1+1;-x^2\right)\Bigg)\Bigg]\bigg\lvert_{x=-\infty}^{x=k\eta_{\rm inf}}\nt
&=\frac{2^{1-2\nu_1}\,|k\eta_{\rm inf}|^{\frac32-3\nu_1}\,(\cot(\pi\nu_1)-i)}{3(2\nu_1-3)(2\nu_1+3)(2\nu_1-1)\,\Gamma(1-\nu_1)^2\,\Gamma(\nu_1+1)^2}\nonumber\\[1ex]
&\times\Bigg[-2^{2\nu_1+1}(12\nu_1^2+12\nu_1-9)\,\csc(\pi\nu_1)\,\Gamma(1-\nu_1)\Gamma(\nu_1+1)\,
(k|\eta_{\rm inf}|)^{2\nu_1}{}_2F_3\!\left(\frac12,\frac34-\frac{\nu_1}{2};1-\nu_1,\frac74-\frac{\nu_1}{2},\nu_1+1;-k^2\eta^2_{\rm inf}\right)\nonumber\\[1ex]
&\quad+(2\nu_1-3)\Bigg(16^{\nu_1}(2\nu_1+3)(\cot(\pi\nu_1)+i)\,\Gamma(\nu_1+1)^2{}_2F_3\!\left(\frac34-\frac{3\nu_1}{2},\frac12-\nu_1;1-2\nu_1,\frac74-\frac{3\nu_1}{2},1-\nu_1;-k^2\eta^2_{\rm inf}\right)\nonumber\\[1ex]
&\qquad-3(2\nu_1-1)(\cot(\pi\nu_1)-i)\,\Gamma(1-\nu_1)^2(k|\eta_{\rm inf}|)^{4\nu_1}{}_2F_3\!\left(\frac{\nu_1}{2}+\frac34,\nu_1+\frac12;\frac{\nu_1}{2}+\frac74,\nu_1+1,2\nu_1+1;-k^2\eta^2_{\rm inf}\right)\Bigg)\Bigg]\nt
&\qquad-\frac{\Gamma(-\frac{1}{4}+\frac{\nu_1}{2})}{2 \sqrt{\pi} \sin^2(\nu_1\pi) \Gamma(\frac{1}{4}-\frac{\nu_1}{2}) \Gamma(\frac{1}{4}+\frac{\nu_1}{2}) \Gamma(\frac{1}{4}+\frac{3\nu_1}{2})} \times \bigg[ \Gamma(\frac{3}{4}-\frac{3\nu_1}{2}) \Gamma(\frac{1}{4}+\frac{3\nu_1}{2}) - 2 e^{-i\nu_1\pi} \Gamma(\frac{3}{4}-\frac{\nu_1}{2}) \Gamma(\frac{1}{4}+\frac{\nu_1}{2}) \nt
&\qquad+ e^{-2i\nu_1\pi} \Gamma(\frac{1}{4}-\frac{\nu_1}{2}) \Gamma(\frac{3}{4}+\frac{\nu_1}{2}) \bigg]\nt
&=\frac{2}{3}\left(-k\eta_{\inf}\right)^{\frac{3}{2}-3\nu_1}\Bigg[\frac{3\times4^{-\nu_1}(\cot(\pi\nu_1)+i)^2\left(-k\eta_{\inf}\right)^{4\nu_1}}{(2\nu_1+3)\Gamma(\nu_1+1)^2}+\frac{6(\cot(\pi\nu_1)+i)\csc(\pi\nu_1)\left(-k\eta_{\inf}\right)^{2\nu_1}}{(2\nu_1-3)\Gamma(1-\nu_1)\Gamma(\nu_1+1)}\nt
&\qquad-\frac{4^{\nu_1}\csc^2(\pi\nu_1)}{(2\nu_1-1)\Gamma(1-\nu_1)^2}\Bigg]+\mathcal{C}_{UV}.
\end{align}
Here $\mathcal{C}_{\rm UV}$ denotes a finite constant arising from the early-time contribution $\eta\to-\infty$, whose precise value depends on the choice of $\xi$.
\begin{align}
   \mathcal{C}_{UV} &\approx -\frac{\Gamma(-\frac{1}{4}+\frac{\nu_1}{2})}{2 \sqrt{\pi} \sin^2(\nu_1\pi) \Gamma(\frac{1}{4}-\frac{\nu_1}{2}) \Gamma(\frac{1}{4}+\frac{\nu_1}{2}) \Gamma(\frac{1}{4}+\frac{3\nu_1}{2})} \times \bigg[ \Gamma(\frac{3}{4}-\frac{3\nu_1}{2}) \Gamma(\frac{1}{4}+\frac{3\nu_1}{2}) - 2 e^{-i\nu_1\pi} \Gamma(\frac{3}{4}-\frac{\nu_1}{2}) \Gamma(\frac{1}{4}+\frac{\nu_1}{2}) \nt
   &\qquad+ e^{-2i\nu_1\pi} \Gamma(\frac{1}{4}-\frac{\nu_1}{2}) \Gamma(\frac{3}{4}+\frac{\nu_1}{2}) \bigg].
\end{align}
\section{Four-Point Correlator for the Conformally Coupled Scalar}\label{app:contact}\label{app:contact_interaction}
In this appendix, we detail the calculation of the contact interaction diagram. 
Consistent with the main text, we assume a sudden transition to a radiation-dominated era where the scale factor behaves as $a(\eta) \propto \eta$. 
For a massless field with conformal coupling, the interaction vertex $\sqrt{-g} \phi^4$ reduces to the Minkowski form due to the cancellation between the metric determinant and the field rescaling—regardless of the reheating dynamics.
\begin{figure}[h]
        \centering	
        \begin{tikzpicture}
		\def\TopY{0}       
		\def\VertexY{-2.5} 
		\def\Width{2.5}    
		
		\draw[gray, line width=2pt] (-\Width, \TopY) -- (\Width, \TopY) node[right, black] {$\eta$};
		
		\coordinate (V) at (0, \VertexY);
		
		\draw[thick] (V) -- (-1.5, \TopY);
		\draw[thick] (V) -- (-0.5, \TopY);
		\draw[thick] (V) -- (0.5, \TopY);
		\draw[thick] (V) -- (1.5, \TopY);
		
		\fill[black] (V) circle (3pt);
		
		 \node[above] at (-1.5, \TopY) {\small $k_1$};
		 \node[above] at (-0.5, \TopY) {\small $k_2$};
		 \node[above] at (0.5, \TopY) {\small $k_3$};
		 \node[above] at (1.5, \TopY) {\small $k_4$};
		
	\end{tikzpicture}
        \captionsetup{style=centered_sub}
        \caption{Contact diagram}
        \label{fig:four-point-contact}
    \end{figure}
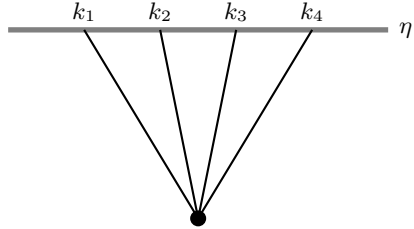
\begin{align}\label{contact interaction-four point}
    \overline{\ev{\phi_{k_1}\phi_{k_2}\phi_{k_3}\phi_{k_4}}}(\eta)
    &=2g\Im{\int_{-\infty}^{\eta}D\eta' W_{k_1}(\eta, \eta')W_{k_2}(\eta,\eta')W_{k_3}(\eta,\eta')W_{k_4}(\eta,\eta')}\nt
    &=2g\Im{\frac{e^{-iK\eta}}{a(\eta)^4}\int_{-\infty}^{\eta}d\eta' a(\eta')^4\frac{1}{a(\eta')^4}\frac{e^{i(k_1+k_2+k_3+k_4)\eta'}}{16k_1k_2k_3k_4}}\nt
    &=2g\Im{\frac{e^{-iK\eta}}{16a(\eta)^4k_1k_2k_3k_4}\int_{-\infty}^{\eta}d\eta' e^{iK\eta'}}\nt
    &=2g\Im{\frac{-i}{16a(\eta)^4k_1k_2k_3k_4 K}}=-\frac{2g}{16a(\eta)^4k_1k_2k_3k_4 K}.
\end{align}
We essentially factor out this trivial "Minkowski" evolution and focus on the exchange diagrams, where the breaking of time-translation invariance leads to distinct cosmological signatures.
We first consider the sudden transition model, where the universe transitions instantaneously from inflation to radiation. In this limit, the four-point correlator during the radiation era is given by:
\begin{equation}
    \overline{\langle\phi_{k_1}\phi_{k_2}\phi_{k_3}\phi_{k_4}\rangle}(\eta) = -\frac{g H_{\mathrm{inf}}^4 \eta^8_{\mathrm{inf}}}{8(\eta+2\lvert\eta_{\mathrm{inf}}\rvert)^4 k_1 k_2 k_3 k_4 K}.
\end{equation}

When we include a finite reheating era parametrized by the equation of state $w$, the result generalizes to:
\begin{align}
    \overline{\langle\phi_{k_1}\phi_{k_2}\phi_{k_3}\phi_{k_4}\rangle}(\eta) &= -\frac{gH_{\mathrm{inf}}^4\eta_{\mathrm{inf}}^4}{8 k_1 k_2 k_3 k_4 K}  \frac{1}{\left(\frac{1+3w}{2\lvert\eta_{\mathrm{inf}}\rvert}\right)^{\frac{8}{1+3w}} \left(\eta_{\mathrm{reh}} + \lvert 3\eta_{\mathrm{inf}}\rvert \mu\right)^{\frac{8}{1+3w}}}\frac{1}{\left[1 + \frac{2(\eta_{\mathrm{reh}}-\eta)}{3(1+w)\eta_{\mathrm{inf}}-(1+3w)\eta_{\mathrm{reh}}}\right]^4}.
\end{align}

\bibliography{biblography}
\end{document}